\documentclass[letterpaper]{article}

\usepackage{geometry}
\usepackage{subfigure}
\usepackage{graphicx}
\usepackage{subcaption}
\usepackage{booktabs}
\usepackage{tabularx}

\usepackage[dvipsnames]{xcolor}
\usepackage[ruled,vlined,linesnumbered]{algorithm2e}
\DontPrintSemicolon
\newcommand{\Hdr}[1]{\hspace{0.5em}\textcolor{blue}{\ttfamily // #1}}
\newcommand{\RC}[1]{\tcp*[r]{\textcolor{blue}{\ttfamily #1}}}

\SetKwComment{LineC}{\textcolor{blue}{// }}{}
\newcommand{\LC}[1]{\LineC*[l]{\textcolor{blue}{\ttfamily #1}}} 

\AtBeginEnvironment{algorithm}{%
  \SetAlFnt{\fontsize{7.8pt}{8.5pt}\selectfont}% body
  \SetAlCapFnt{\fontsize{7.8pt}{8.5pt}\selectfont}% caption text
  \SetAlCapNameFnt{\fontsize{7.8pt}{8.5pt}\selectfont}% “Algorithm” label
}

\usepackage{amsmath}
\usepackage{amsfonts}
\usepackage{url}
\usepackage[colorlinks=true, allcolors=BrickRed]{hyperref}

\usepackage{lineno}
\usepackage{authblk}
\author[1]{Ho Fung Tsoi*}
\author[1]{Dylan Rankin}
\author[2]{Vladimir Loncar}
\author[3,4]{Philip Harris}
\affil[1]{University of Pennsylvania, USA}
\affil[2]{Institute of Physics Belgrade, Serbia}
\affil[3]{Massachusetts Institute of Technology, USA}
\affil[4]{Institute for Artificial Intelligence and Fundamental Interactions, USA}
\title{\vspace{-2cm}\includegraphics[width=0.28\textwidth]{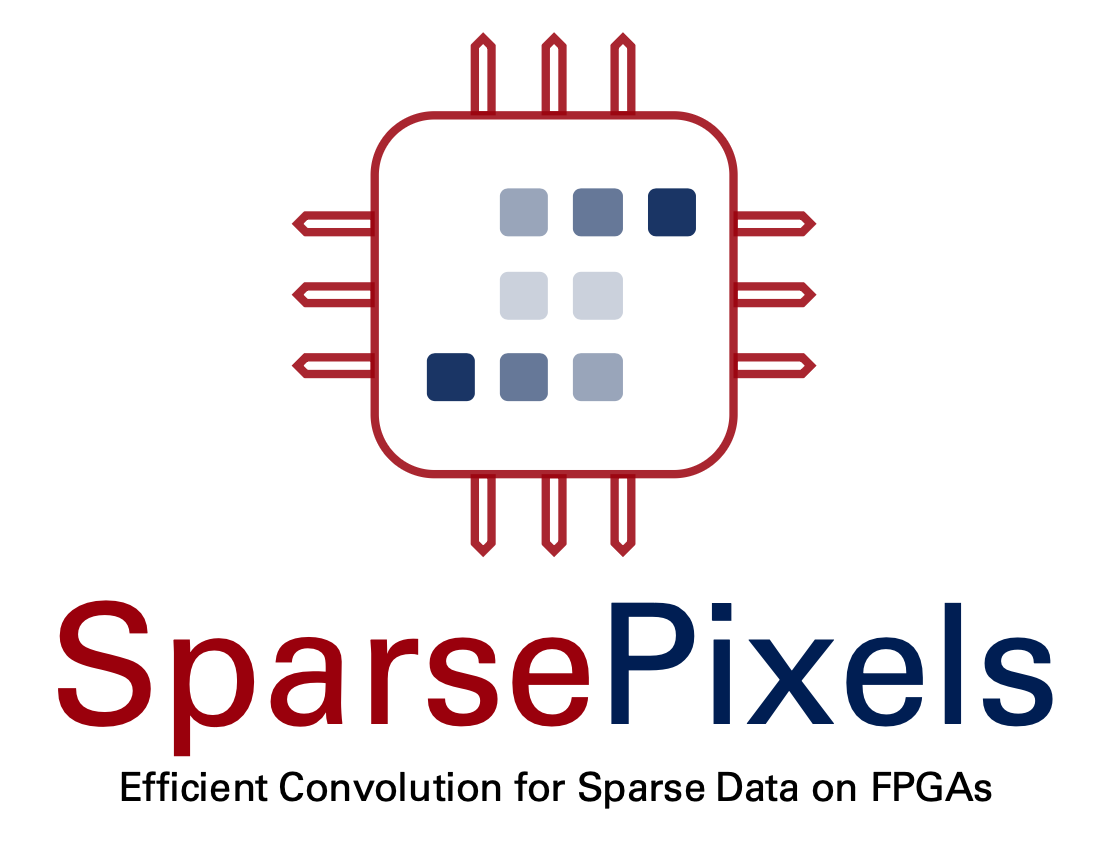}\\ \vspace{0.3cm}\textbf{\Large SparsePixels: Efficient Convolution for Sparse Data on FPGAs}}
\date{*Email: ho.fung.tsoi@cern.ch}
\begin{document}

\maketitle

\begin{abstract}
Inference of standard convolutional neural networks (CNNs) on FPGAs often incurs high latency and a long initiation interval due to the deep nested loops required to densely convolve every input pixel regardless of its feature value.
However, input features can be spatially sparse in some image data, where semantic information may occupy only a small fraction of the pixels and most computation would be wasted on empty regions.
In this work, we introduce SparsePixels, a framework that implements sparse convolution on FPGAs by selectively retaining and computing on a small subset of active input pixels while ignoring the rest, reducing the per-layer compute from $\mathcal{O}(HWK^2C_{\text{in}}C_{\text{out}})$ to $\mathcal{O}(n^2C_{\text{in}}C_{\text{out}})$ and the storage from $\mathcal{O}(HWC)$ to $\mathcal{O}(nC)$ when $n^2\ll HW$.
Because computation always runs over a single pre-specified pixel budget that is frozen at synthesis, the inference latency is data-independent and constant at runtime.
We show that, for identifying neutrino interactions in naturally sparse LArTPC images with 4k pixels, a standard CNN with a compact size of 4k parameters incurs an inference latency of 48.665 $\mu$s on an FPGA, whereas a sparse CNN of the same base architecture, computing on less than 1\% of the input pixels, achieves a $\times 73$ speedup to 0.665 $\mu$s with resource utilization well within on-chip budgets, trading only a small percent-level performance loss.
This work aims to benefit future algorithm development for efficient data readout in modern experiments with strict latency requirements of microseconds or below.
%Code is publicly released at \url{https://github.com/hftsoi/sparse-pixels}.
\end{abstract}

\section{Introduction}
\label{sec:introduction}

Convolutional neural networks (CNNs) have gained wide interest and usage for image-based recognition tasks in particle physics experiments, including muon tracking in gas chambers~\cite{Francescato:2021ezq,Coccaro:2023nol,Maglianella:2023inw}, jet flavor tagging in calorimeters~\cite{deOliveira:2015xxd,Kasieczka:2017nvn,Kasieczka:2019dbj,Moreno:2019bmu,Pol:2021iqw}, anomaly detection for new physics searches at colliders~\cite{Kasieczka:2021xcg,Aarrestad:2021oeb,Govorkova:2021utb,CMS-DP-2023-086,CMS-DP-2024-121,Gandrakota:2024yqs}, and neutrino identification in liquid argon time projection chambers (LArTPCs)~\cite{Aurisano:2016jvx,MicroBooNE:2016dpb,Domine:2019zhm,DUNE:2020gpm,Chung:2025dag}.
This is because experimental data are often encoded in geometrical formats originating from detector layouts and can be represented in fixed-grid views.
Hence, it is natural to use CNNs to process and extract signal from this type of data.
However, some of these images are spatially very sparse because certain signals leave small traces within large detector volumes.
For example, in the dataset presented in \cite{Maglianella:2023inw}, a muon traversing gas chambers leaves a track in a 2D image covering the full muon station, but a single muon track may occupy only 10 out of $9\times 384$ pixels ($<0.3\%$ of the input).
Similarly, in neutrino experiments, particle tracks in large LArTPC volumes can have, on average, only 0.01\% nonzero pixels~\cite{Domine:2019zhm}.
Because these signatures can have irregular shapes and appear anywhere in the image, CNNs must scan the entire input to capture them.
Implementing large standard CNNs can therefore incur high inference latency even on high-speed custom processors such as FPGAs~\cite{Aarrestad:2021zos}, posing a challenge for data readout in modern experiments that enforce ultra-low latency.

In modern particle experiments, data processing is typically divided into two domains: online (low-latency) processing and offline analysis.
Taking the CERN Large Hadron Collider (LHC) as an example~\cite{ATLAS:2020esi,2137107,CMS:2020cmk,Zabi:2020gjd}, proton bunches collide at 40 MHz, and each collision is an event.
The detector acts like a high-speed camera, recording trajectories and energy deposits across subdetectors.
The resulting raw information is about 1 MB per event, leading to a raw data rate of $\mathcal{O}(10)$ TB/s.
Storing all data is impossible due to limited bandwidth and storage capacity, hence a fast filtering system (the trigger) is required to select potentially interesting events for permanent storage in real time, concurrent with the 40 MHz collisions.
Because the input streams in at 40 MHz, the Level-1 trigger must decide whether to store an event within a few microseconds, set by buffer capacity, motivating the use of FPGAs/ASICs for real-time processing.
The latency constraint is strict because the system can buffer each data event only for a fixed duration, so a trigger algorithm must make the decision within the set time before the event is lost permanently.
It is crucial and challenging to design trigger algorithms for online use that are sensitive enough to select interesting events (as unselected events are discarded permanently) while remaining compact enough to meet low-latency constraints.
With the forthcoming High-Luminosity LHC upgrade~\cite{ZurbanoFernandez:2020cco}, developing more sophisticated and hardware-efficient online algorithms is essential for handling unprecedented rates and bandwidth.
After trigger selections, data are stored and analyzed offline without strict latency limits.

In this paper, we focus on the low-latency domain, targeting inference latencies on the order of microseconds and below for trigger applications in modern particle experiments such as the CERN LHC and MicroBooNE.
The general workflow for deploying machine learning (ML) models on FPGAs in this regime is well established in the community.
The $\tt{hls4ml}$ library~\cite{fastml_hls4ml,Duarte:2018ite} is widely used due to its flexible hardware-aware configuration, which supports translating trained models from common Python libraries into High-level Synthesis (HLS) firmware for FPGA deployment.
CNNs are supported and have been benchmarked~\cite{Aarrestad:2021zos,Tarafdar:2021pqm,Ghielmetti:2022ndm}.
For large images (e.g., with $\mathcal{O}(1000)$ or more pixels), inputs typically cannot be fully partitioned to enable parallel processing on an FPGA, so designs fall back to a streaming approach, processing one pixel at a time.
While this keeps resource usage small, it often yields high latencies.
When the image is sparse, say only a few percent or fewer pixels are nonzero on average, the standard convolution becomes highly inefficient because it must be computed on many empty regions.
Therefore, the sparsity exploited is in the input data instead of the model weights.
We investigate a class of CNNs that adaptively select and compute only on active input pixels, targeting large images with extreme spatial sparsity.

\textbf{Problem statement.}
\begin{itemize}
    \item Can we improve the processing efficiency of CNNs for spatially sparse images on FPGAs to allow for accelerated real-time inference in experiments that require microsecond-level latency or below?
\end{itemize}
\textbf{Our contributions.}
\begin{enumerate}
    \item Development of a sparse convolution method designed for efficient FPGA implementation for applications in the broader particle physics experiments, targeting low inference latency that is independent of input data and is constant at runtime. This contrasts with existing methods with input-dependent runtime, which would fail to meet the strict latency bound required by trigger systems. To our knowledge, this is the first implementation of fixed-latency sparse convolution in the sub-microsecond domain.
    \item Proof-of-concept on spatially sparse image data, demonstrating significant FPGA inference speedups with comparable model performance.
    \item A Python library for quantization-aware training (QAT) of sparse CNNs and C++ HLS modules for FPGA deployment. Code is publicly released at \url{https://github.com/hftsoi/sparse-pixels}.
\end{enumerate}

In the rest of the paper, Sec.~\ref{sec:related} reviews related developments that motivate this work, Sec.~\ref{sec:method} introduces the $\tt{SparsePixels}$ framework for sparse CNN training and FPGA implementation, Sec.~\ref{sec:exp} presents the experimental setup and validation results, and Sec.~\ref{sec:future} discusses current limitations and the planned improvements.

\section{Related work}
\label{sec:related}

Our work aims to improve upon the standard 2D convolution in $\tt{hls4ml}$~\cite{Aarrestad:2021zos} by designing an approach specifically for sparse image data on FPGAs in the low-latency regime, using a sparse convolution method inspired by works on acceleration of sparse 3D point cloud inference.

\textbf{Low-latency standard convolution on FPGAs.}
The $\tt{hls4ml}$~\cite{fastml_hls4ml,Duarte:2018ite} library is open-sourced and developed to facilitate deployment of ML models on FPGAs for fast inference in strictly constrained environments such as particle physics experiments at the LHC.
The library provides a Python interface for translating models trained in common libraries such as $\tt{Keras}$~\cite{chollet2015keras,Coelho:2020zfu} and $\tt{PyTorch}$~\cite{paszke2019pytorch} into high-level C/C++ via HLS tools, which can be synthesized into register-transfer level designs, enabling rapid customization and optimization of resources and latency for FPGA firmware.
$\tt{hls4ml}$ has integrated QAT for fixed-precision deployment~\cite{Coelho:2020zfu,Sun:2024soe}.
Implementations of various ML architectures have been demonstrated~\cite{Duarte:2018ite,Summers:2020xiy,Khoda:2022dwz,Tsoi:2023isc,Tsoi:2024ypg,Jiang:2024tkg}, including CNNs~\cite{Aarrestad:2021zos,Tarafdar:2021pqm,Ghielmetti:2022ndm}, which serve as the baseline in this paper.
Other tools such as $\tt{FINN}$~\cite{finn,blott2018finn} and $\tt{Chisel4ml}$~\cite{vrevca2025generating}, with slightly different optimization targets and workflows, also support low-latency inference of standard convolution on FPGAs.
However, across all these tools, the underlying challenge is similar: the amount of computation required by standard convolution on large inputs makes full parallel processing prohibitive, forcing designs toward streaming implementations (inputs are processed serially) that can incur too high a latency for real-time systems.
For example, $\tt{hls4ml}$ has demonstrated an inference latency of 5 $\mu$s for CNNs on $32\times 32$ images~\cite{Aarrestad:2021zos}, and this latency grows rapidly with input size.

\textbf{Accelerating convolution with sparsity.}
A direction to reduce CNN computation is to exploit sparsity to skip unnecessary operations.
\cite{7298681} explored model compression via parameter pruning, reducing redundancy in the parameter space and mapping convolutions to sparse matrix multiplications for speedup.
Other works~\cite{8735526,10.1109/TVLSI.2020.3002779,xu2024sparse} explored efficient data-flow schemes on FPGAs for CNNs with sparse structure to skip multiplications involving zeros.
These methods focus on weight sparsity or structured sparsity, and the computation still scales with the full image size.
Moreover, directly skipping computation based on irregular input sparsity often leads to input-dependent computations with the runtime varying with each input~\cite{8421093}, which is problematic in systems with strict latency bounds.
Sparse representation theory~\cite{7102696} exploits sparsity in a different sense, where a signal is compressed into a sparse latent representation, but such transformations could introduce data-dependency in the algorithm, whereas the sparsity we exploit is already present in the input pixels and is extracted directly without any feature transformation.
A more relevant direction is to focus on selective computation only on a subset of pixels from the input tensor and ignore the rest.
\cite{graham2015sparse3dconvolutionalneural} proposed a type of sparse convolution that computes an output pixel only when any of the input pixels within the kernel receptive field is nonzero, so the computational cost scales with the number of nonzero input pixels rather than the entire image/volume.
However, this approach can rapidly increase the number of nonzero pixels across convolutional layers: an output pixel can become nonzero if its receptive field includes any neighboring nonzero input, effectively dilating the nonzero set.
To address this, \cite{8579059} constrained the operation so that an output pixel is computed only when the same location at the input is nonzero, keeping the sparsity pattern fixed across layers and trading little or no performance loss for significant speedups.
$\tt{MinkowskiEngine}$~\cite{choy20194d}, $\tt{SpConv}$~\cite{yan2018second}, $\tt{TorchSparse}$~\cite{tang2022torchsparse}, and \texttt{TorchSparse++}~\cite{tangandyang2023torchsparse} are some 3D point cloud engines built upon such sparse convolution for high-performance inference on GPUs.

Despite these advances, such techniques have largely remained unexplored in the FPGA domain, which is critical for certain scientific experiments that demand especially low latencies.
Our proposed method addresses this with an implementation of sparse convolution that selectively retains and computes on a subset of pixels only, and achieves constant runtime for every input regardless of sparsity, enabling parallel processing on FPGAs to achieve large inference speedups.
This property is required for trigger systems in modern particle experiments, where the latency is strictly bounded and so algorithms with input-dependent runtime cannot be deployed.

\section{Method}
\label{sec:method}

We first present the background and define sparse convolution in Sec.~\ref{sec:method-sparsecnn}, then lay out the algorithmic details of the HLS implementation in Sec.~\ref{sec:method-hls}.

\subsection{Background and sparse convolution}
\label{sec:method-sparsecnn}

For simplicity, \textit{same} padding and unit stride are assumed in the rest of the paper.
In a standard 2D convolutional layer, for an input tensor $I$ with dimensions ($H$, $W$, $C_{\text{in}}$), an output tensor $O$ with dimensions ($H$, $W$, $C_{\text{out}}$), and a kernel weight tensor $T$ with dimensions ($K$, $K$, $C_{\text{in}}$, $C_{\text{out}}$), three nested loops are required to accumulate input-weight multiplications for each output pixel and channel~\cite{deeplearning}:
\begin{equation}
    \label{eq:conv2d}
    O[i,j,c_{\text{out}}]=b[c_{\text{out}}]+\sum_{c_{\text{in}}=0}^{C_{\text{in}}-1}\sum_{k_h=0}^{K-1}\sum_{k_w=0}^{K-1}I[i+k_h,j+k_w,c_{\text{in}}]T[k_h,k_w,c_{\text{in}},c_{\text{out}}],
\end{equation}
where $b$ denotes the $C_{\text{out}}$ bias terms, and the input indices are shifted to the proper spatial patch.
Eq.~\ref{eq:conv2d} is computed for each output pixel ($0\leq i<H$, $0\leq j<W$) and channel ($0\leq c_{\text{out}}<C_{\text{out}}$).
Thus, a standard Conv2D performs $HWC_{\text{out}}$ inner products, each with $C_{\text{in}}K^2$ multiplications, totaling $HWC_{\text{out}}C_{\text{in}}K^2$ multiplications (six nested loops).
Directly implementing these deep nested loops in HLS leads to high latency because each transition across loop levels, say exiting an inner loop to its enclosing outer loop or re-entering for the next iteration, costs one clock cycle.
A common workaround is to pipeline an outer loop and unroll all inner loops so iterations execute in parallel, but this is infeasible for typical image sizes due to resource usage and memory bandwidth.
Because of these limitations, $\tt{hls4ml}$ uses stream I/O to implement convolutional architectures~\cite{Aarrestad:2021zos}.
In stream I/O, inputs are processed sequentially via ``first in, first out'' (FIFO) buffers, which use fewer resources than parallel I/O.
For an input of size $H\times W\times C_{\text{in}}$, the $H\times W$ pixels are fed one-by-one, each carrying $C_{\text{in}}$ features.
Pixels are buffered until a full sliding window is available, then the multiply-accumulate operation is performed.
Since streaming reads one pixel per clock, the latency is at least $H\times W$ cycles to scan the image, plus additional cycles to fill and drain the sliding window, which scale with $W$ and $K$.
This approach enables large inputs but yields latencies too high for constrained systems.

\begin{figure}[!t]
     \centering
    \includegraphics[width=0.5\textwidth]{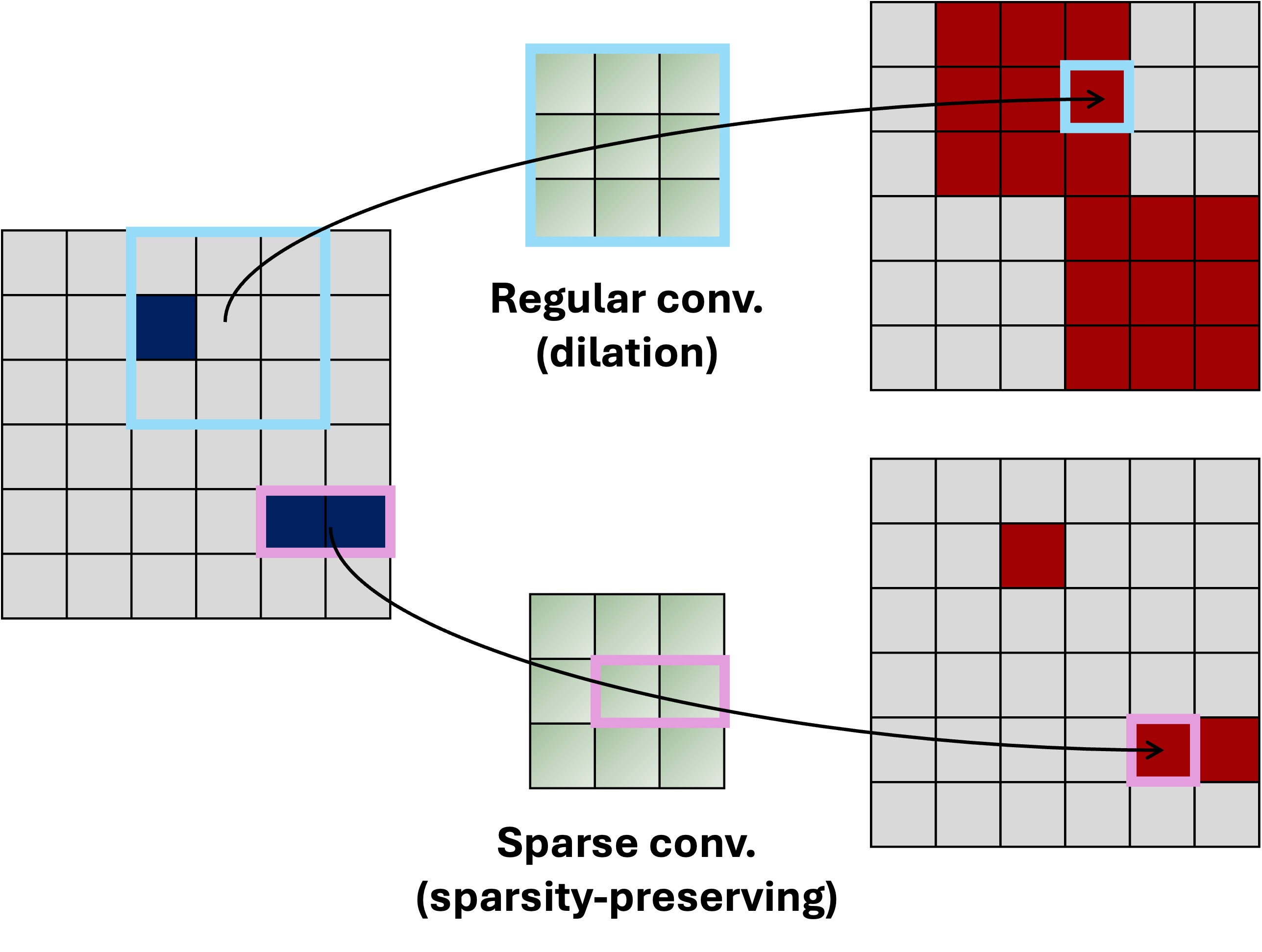}
    \caption{Dilation of active pixels under standard convolution versus sparsity-preserving convolution, where the active set is fixed.}
    \label{fig:dilation}
\end{figure}

For large images with spatially sparse features, implementing standard convolution is often prohibitive, or, if feasible, wastes most computation on empty regions that contain little or no semantic information and thus contribute little to feature extraction.
In contrast, sparse convolution is a special (constrained) type of convolution that selectively computes on a subset of input pixels instead of densely convolving the entire image.
Which pixels are selected can depend on the dataset or task, but in general, selection is based on pixel activity.
We consider an input pixel \textit{active} if its feature value exceeds a predefined threshold (often zero or a small noise value).
Inspired by \cite{8579059}, we constrain the operation so that: \textit{an output pixel is active if and only if the input at the same pixel location is active}, and only active pixels are computed.
This preserves the sparsity pattern between input and output and prevents dilation, as illustrated in Fig.~\ref{fig:dilation}.

\begin{figure}[!t]
     \centering
    \includegraphics[width=\textwidth]{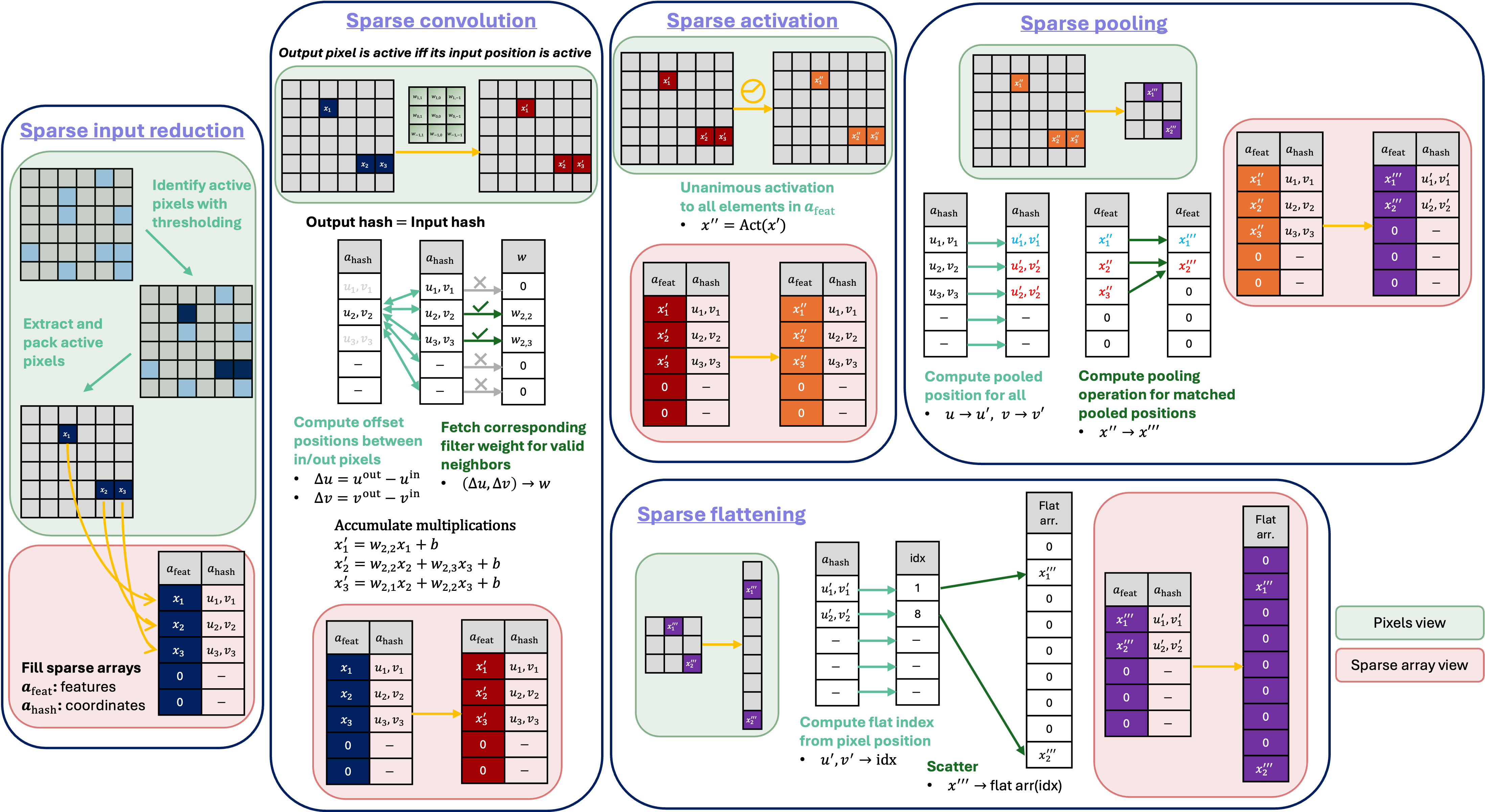}
    \caption{Schematic diagram of $\tt{SparsePixels}$. The framework consists of five main layers sufficient to build sparse CNNs that selectively compute on active pixels only. Sparse arrays are produced and consumed within the HLS implementation. The sparse input reduction layer dynamically retains up to $N_{\text{active}}^{\text{max}}$ (a hyperparameter) active pixels into two sparse arrays: the feature array $a_{\text{feat}}$ (features) and the hash array $a_{\text{hash}}$ (height and width coordinates). The sparse convolution layer performs a sparsity-preserving convolution, updating $a_{\text{feat}}$ while keeping $a_{\text{hash}}$ unchanged. The sparse activation layer applies an element-wise nonlinearity to $a_{\text{feat}}$. The sparse pooling layer applies downsampling, updating both $a_{\text{feat}}$ and $a_{\text{hash}}$. The sparse flattening layer transforms the sparse arrays into a flat dense array that can be connected to conventional dense layers for downstream tasks such as classification and regression.}
    \label{fig:schematic}
\end{figure}

Building on this sparsity-preserving convolution, we introduce the $\tt{SparsePixels}$ framework, which consists of five core sparse layers sufficient to build a sparse CNN: (1) sparse input reduction, (2) sparse convolution, (3) sparse activation, (4) sparse pooling, and (5) sparse flattening, illustrated in Fig.~\ref{fig:schematic}.
The framework is designed around implementing CNNs with sparse convolution in HLS.
The sparse input reduction is the first layer acting on the input, which extracts up to a fixed number of active pixels from the input images and groups them for subsequent processing.
The sparse convolutional layer performs a sparsity-preserving convolution that fixes the input-output sparsity.
The other sparse layers retain their usual functionality but operate on the stored active pixels.
For training, we have developed a Python library to build and train sparse CNNs that mirror the HLS behavior.
The training library uses $\tt{QKeras}$ as a backend for QAT~\cite{Coelho:2020zfu}, and will support high-granularity quantization (HGQ)~\cite{Sun:2024soe} in the future.
The HLS implementation of the sparse layers is detailed in the following section.

\subsection{HLS implementation}
\label{sec:method-hls}

In HLS, the core idea is to store only the active pixels in compact arrays and perform all subsequent computation on these arrays.

\subsubsection{Sparse input reduction}
\label{sec:method-hls-input}

\begin{figure}[!t]
    \centering
    \includegraphics[width=0.8\textwidth]{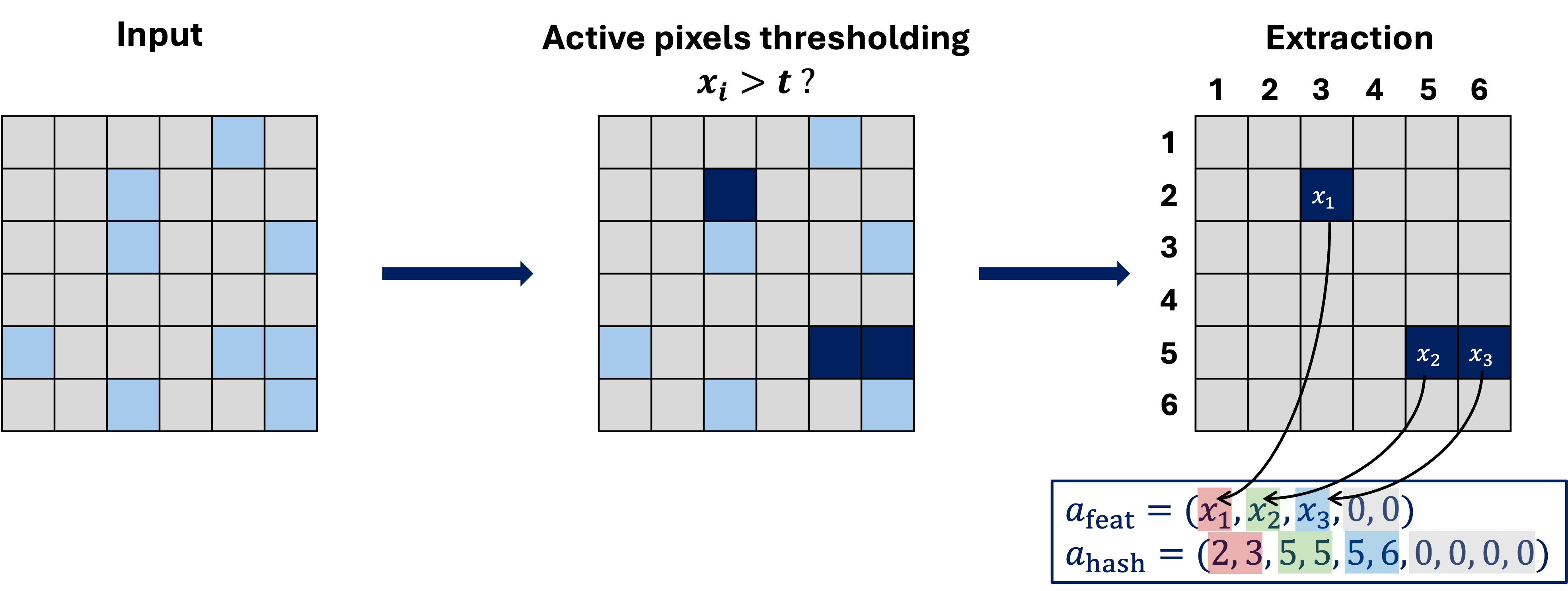}
    \caption{Identification of active pixels through thresholding. Only active pixels are stored in the sparse arrays: $a_{\text{feat}}$ holds features, and $a_{\text{hash}}$ holds the corresponding height and width coordinates. The example uses $N_{\text{active}}^{\text{max}}=5$.}
    \label{fig:input_reduce}
\end{figure}

The first step is to read out the active pixels from the input, as illustrated in Fig.~\ref{fig:input_reduce}.
For an input image with $H\times W$ pixels and $C$ channels, the active pixels are identified as those whose (channel-0 by convention) feature value exceeds a predefined threshold, and are extracted into two compact sparse arrays: the $\textit{feature array}$ $a_{\text{feat}}$ (storing all $C$ channel values for each retained pixel) and the $\textit{hash array}$ $a_{\text{hash}}$ (storing the corresponding height and width coordinates).
We fix a maximum number of active pixels per model, $N_{\text{active}}^{\text{max}}$, which sets the array sizes: $a_{\text{feat}}$ has $N_{\text{active}}^{\text{max}}\times C$ elements and $a_{\text{hash}}$ has $N_{\text{active}}^{\text{max}}\times 2$ elements.
Thus, instead of computing over all $H\times W$ pixels as in standard convolution, we operate on arrays whose size scales with $N_{\text{active}}^{\text{max}}\ll H\times W$ for sparse data.
In particular, when an input has fewer than $N_{\text{active}}^{\text{max}}$ active pixels, the unused slots in $a_{\text{feat}}$ and $a_{\text{hash}}$ are padded and flagged as invalid, so they are still computed but do not contribute to the output.
Because the active pixels are stored and processed entirely in $a_{\text{feat}}$ and $a_{\text{hash}}$, whose sizes are fixed by $N_{\text{active}}^{\text{max}}$, and every sparse layer always computes over all elements in the sparse arrays without skipping the padded ones, the total operation count stays identical for every input, so the inference latency is independent of the input sparsity and is frozen at synthesis.
The threshold only decides which input pixels are active and has no effect on latency and resources.
A naive implementation, say scanning all input pixels once and ``pushing'' active pixels into an array via an ``if-else'', creates data-dependent write addresses, where the next write index depends on how many active pixels were seen.
In HLS this induces loop-carried dependencies that can inflate scheduling and synthesis cost even when the array size is small.

\begin{figure}[!t]
     \centering
    \includegraphics[width=0.95\textwidth]{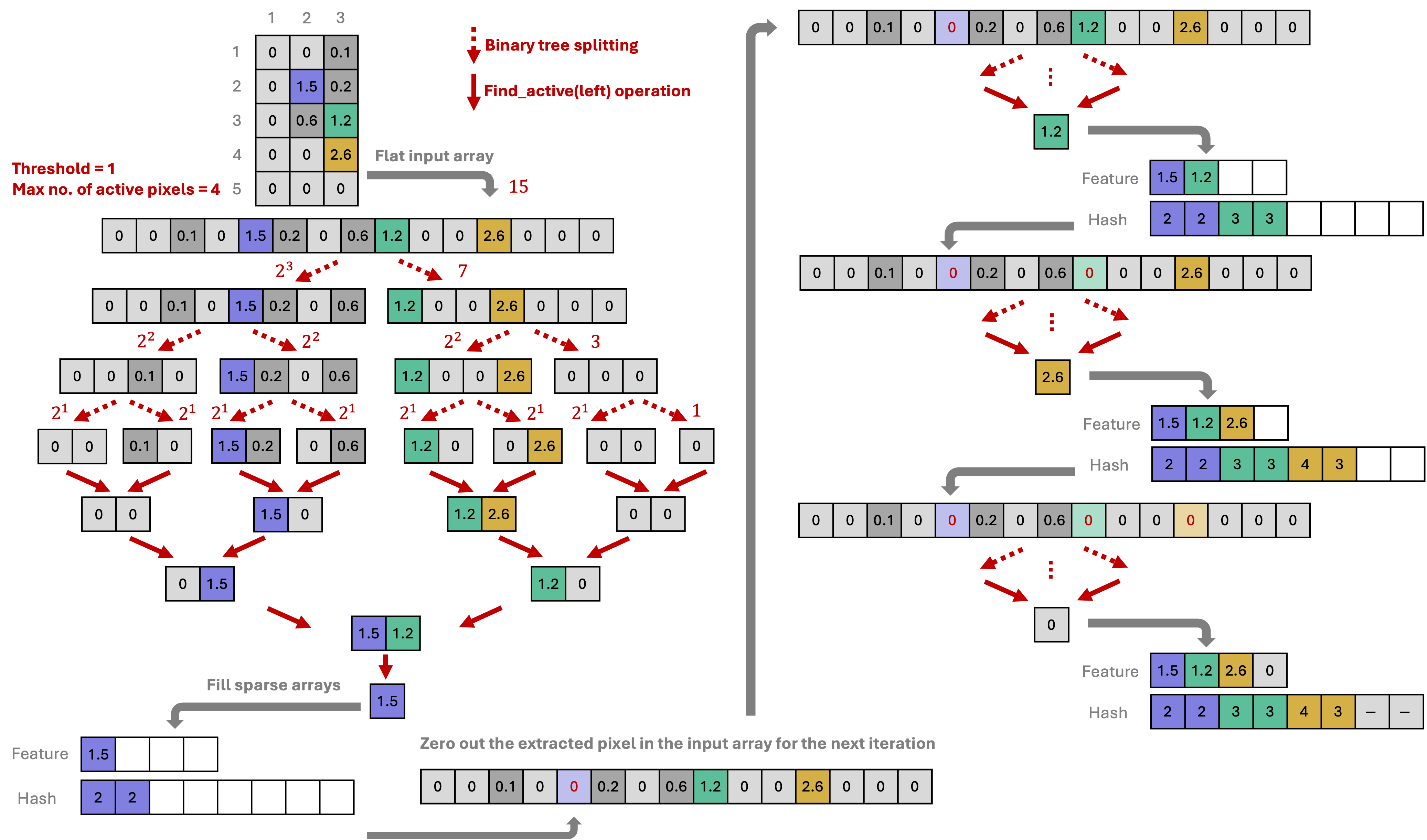}
    \caption{Schematic diagram of the recursive tree splitting and pairwise combiner used in the sparse input reduction layer.}
    \label{fig:input_reduce_steps}
\end{figure}

To avoid data-dependent writes, we use a recursive binary reduction scheme to read out the active pixels, as illustrated in Fig.~\ref{fig:input_reduce_steps}.
Let $N=H\times W$ and flatten the first channel into a length-$N$ array.
We then split it into two partitions where the left partition has size $2^{\lfloor \log_2(N-1)\rfloor}$ (largest power of two less than $N$) and the right holds the remainder.
The split recurses until we reach pairs.
The reduction tree depth is therefore
\begin{equation}
    \label{eq:reduction-depth}
    \text{reduction tree depth}=\lceil \log_{2}(H\times W)\rceil.
\end{equation}
Using a simple pairwise combiner that returns the left element if it is above threshold and otherwise the right, we reduce each level until one element remains, which results in the leftmost active pixel from the flat input array.
We write this pixel's features and coordinates into the first slots of $a_{\text{feat}}$ and $a_{\text{hash}}$, set its value to zero in the input array, and repeat the reduction to obtain the next leftmost active.
This continues until all $N_{\text{active}}^{\text{max}}$ slots are filled.
By convention, the ordering is row-major (top-left to bottom-right).
If an input has fewer active pixels than $N_{\text{active}}^{\text{max}}$, the remaining slots are padded with an invalid flag; otherwise the first $N_{\text{active}}^{\text{max}}$ in this scan ordering are retained and the rest are discarded.
This approach is hardware-efficient as it replaces the data-dependent ``push-back'' with a fixed-shape reduction tree.
The algorithmic details are given in Alg.~\ref{alg:sparse_input_reduce} (Appendix~\ref{sec:apd-input-conv}).

\subsubsection{Sparse convolution}
\label{sec:method-hls-conv}

\begin{figure}[!t]
     \centering
    \includegraphics[width=0.95\textwidth]{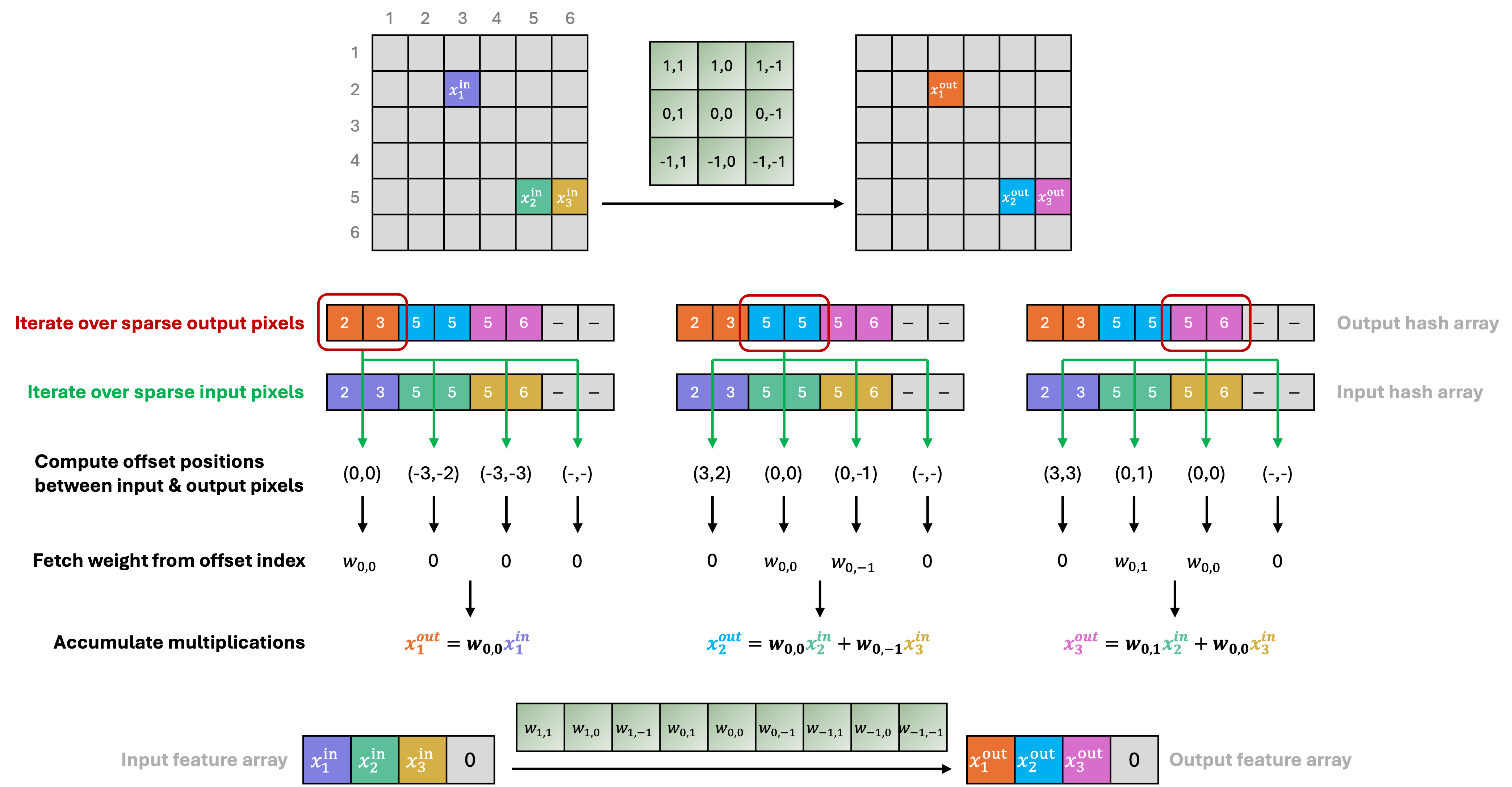}
    \caption{Schematic diagram of sparse convolution implemented in HLS. Active pixel locations are unchanged. Only the feature array is updated.}
    \label{fig:conv}
\end{figure}

Next, we perform sparse convolution that takes as inputs the two sparse arrays, as illustrated in Fig.~\ref{fig:conv}.
Because the operation preserves the sparsity pattern, the active pixel locations are identical at input and output.
Consequently, only the feature array is updated while the hash array remains unchanged.

Since only active pixels can contribute and the input and output share the same active locations, no information from inactive pixels is used.
On the sparse arrays, we loop over output pixels; for each output pixel we loop over input pixels and compute the offsets $(\Delta h,\Delta w)$ between their coordinates from the hash array.
If $(\Delta h,\Delta w)$ lies within the $K\times K$ receptive field, we map that offset to the corresponding kernel position and accumulate a channel-wise dot product between the input features and the appropriate weights; otherwise, the pair contributes nothing.
By construction, if an output location is invalid (padded when the input has fewer than $N_{\text{active}}^{\text{max}}$ active pixels), it will never get activated in subsequent layers.

For each output channel, the sparse convolution involves two nested loops over active pixels (outer: output active pixels; inner: input active pixels).
Because multiplications are based on an offset check and weight lookup, we do not loop over all $K^2$ kernel positions explicitly, hence the iteration count does not scale with the kernel size.
This is advantageous for sparse data, since active pixels can be far apart, and increasing $K$ to allow longer-range interactions does not increase the number of loop iterations.
The relative multiply-accumulate counts between standard and sparse convolutions can be seen as:
\begin{equation}
    \frac{N_{\text{active}}^{\text{max}}\times N_{\text{active}}^{\text{max}}\times C_{\text{in}}\times C_{\text{out}}\text{ (sparse conv.)}}{H\times W\times C_{\text{in}}\times C_{\text{out}}\times K\times K\text{ (standard conv.)}}\ll 1,
\end{equation}
whenever $N_{\text{active}}^{\text{max}}\times N_{\text{active}}^{\text{max}}\ll H\times W$.
Therefore, the compute is reduced from $\mathcal{O}(HWK^2C_{\text{in}}C_{\text{out}})$ to $\mathcal{O}((N_{\text{active}}^{\text{max}})^2C_{\text{in}}C_{\text{out}})$, and the storage from $\mathcal{O}(HWC)$ to $\mathcal{O}(N_{\text{active}}^{\text{max}}C)$.
Taking the neutrino image in Sec.~\ref{sec:exp} as an example (e.g., $H=W=63$, $N_{\text{active}}^{\text{max}}=20$, $K=3$, $C_{\text{in}}=C_{\text{out}}=1$), this corresponds to 400 vs. 35,721 multiplications per input channel per output filter, and storing 20 vs. 3969 features per channel, plus $2 N_{\text{active}}^{\text{max}}=40$ hash coordinates shared across all channels.
The algorithmic details are given in Alg.~\ref{alg:sparse_conv} (Appendix~\ref{sec:apd-input-conv}).

\subsubsection{Sparse activation}
\label{sec:method-hls-act}

As in standard activation, sparse activation applies an element-wise nonlinearity to the sparse feature array while keeping the hash array unchanged.
The number of elements processed is greatly reduced:
\begin{equation}
    \frac{N_{\text{active}}^{\text{max}}\times C\text{ (sparse act.)}}{H\times W\times C\text{ (standard act.)}}\ll 1,
\end{equation}
whenever $N_{\text{active}}^{\text{max}}\ll H\times W$.
The algorithmic details are given in Alg.~\ref{alg:sparse_activation} (Appendix~\ref{sec:apd-activation}).

\subsubsection{Sparse pooling}
\label{sec:method-hls-pool}

For sparse pooling, we first map every retained pixel's coordinates in the sparse array to its pooled coordinates and write into a new hash array to reflect the downsampled sparsity pattern.
We then loop over the pooled pixel positions and accumulate features from pixels that land in the same pool, and finally apply the pooling operation.
The algorithmic details are given in Alg.~\ref{alg:sparse_pool} (Appendix~\ref{sec:apd-pooling}).

\subsubsection{Sparse flattening}
\label{sec:method-hls-flatten}

All sparse layers above operate on sparse arrays that encode a reduced representation of the input image.
Before connecting to conventional dense layers for downstream tasks, we need to transform the sparse representation back to the dense representation.
We first initialize a flat dense array with zeros, then map the coordinates of each retained pixel to the corresponding index and scatter its $C$ channel features to the corresponding slots.
The algorithmic details are given in Alg.~\ref{alg:sparse_flatten} (Appendix~\ref{sec:apd-flatten}).

\section{Evaluations}
\label{sec:exp}

We perform scaling studies for the five sparse layers to understand their latency and resource utilization on an FPGA by scanning key hyperparameters, as presented in Sec.~\ref{sec:exp-scaling}.
We then perform experiments to evaluate sparse CNNs on an FPGA comparing against standard CNN baselines on a realistic neutrino dataset in Sec.~\ref{sec:exp-eval}: the dataset is described in Sec.~\ref{sec:exp-eval-datasets}, the experimental setup in Sec.~\ref{sec:exp-eval-setup}, and the results in Sec.~\ref{sec:exp-eval-results}.

\subsection{Scaling studies}
\label{sec:exp-scaling}

In the scaling studies, layers are synthesized with Vitis HLS (2023.1)~\cite{vitis}, targeting an AMD/Xilinx Alveo FPGA with part number xcu250-figd2104-2L-e.
We report FPGA resource utilization from the HLS compilation step (C-synthesis results), which typically provides conservative resource estimates.
We quote C-synthesis numbers here rather than logic synthesis because we are characterizing individual layers: logic synthesis can allocate a non-negligible fraction of resources into shared FIFO buffers across layers, which could lead to double-counting if quoted per layer.
We report latency in clock cycles, using a 200 MHz clock (a standard benchmark for $\tt{hls4ml}$ and deployments at the LHC), and utilization for four resource types: block random access memory (BRAM), digital signal processor (DSP), flip-flop (FF), and look-up table (LUT).
Utilization is given as a percentage relative to the total available on the target device: BRAMs$=$5,376, DSPs$=$12,288, FFs$=$3,456,000, and LUTs$=$1,728,000.

\begin{figure}[!t]
    \centering
    \includegraphics[width=0.35\textwidth]{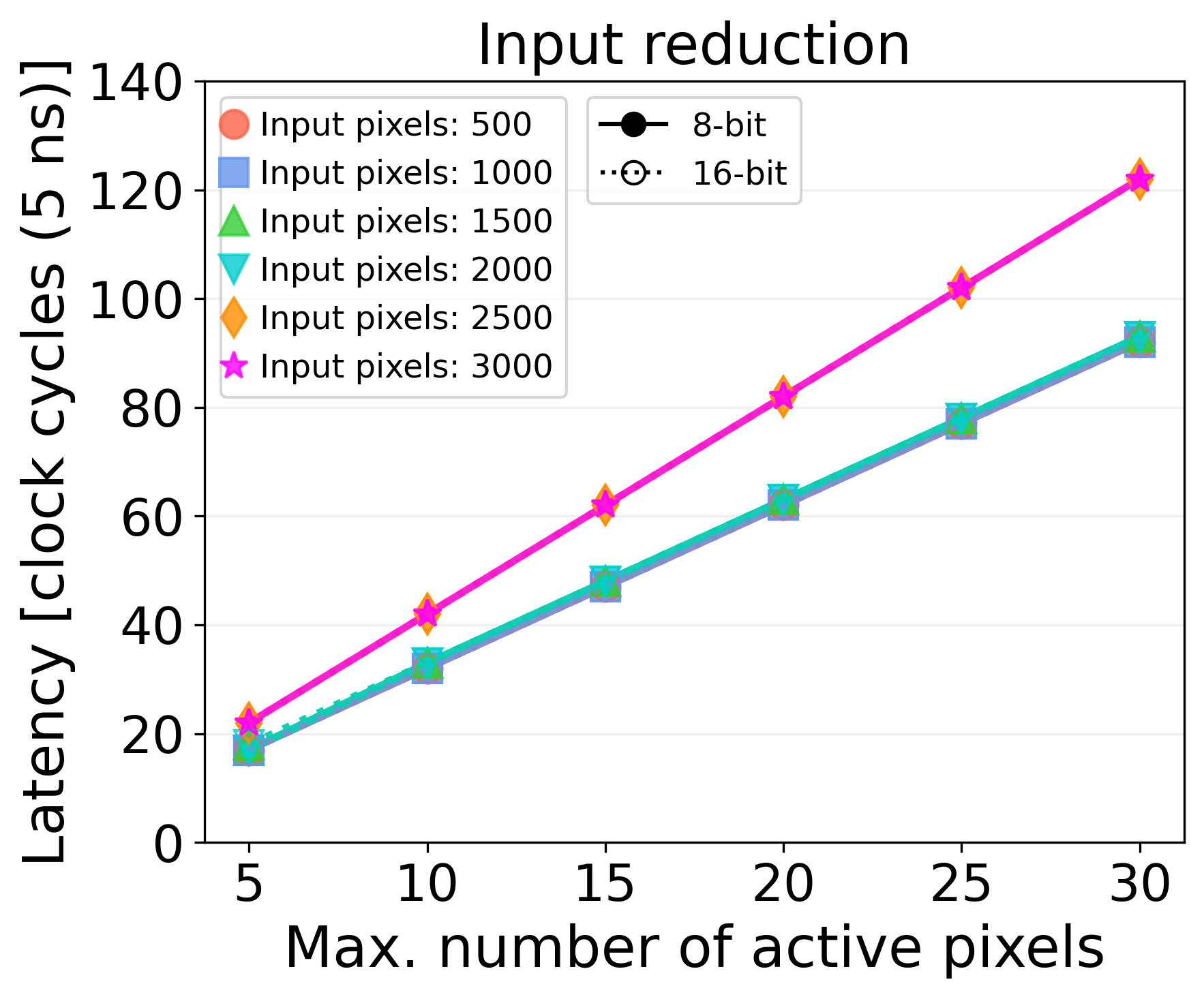}
    \includegraphics[width=0.35\textwidth]{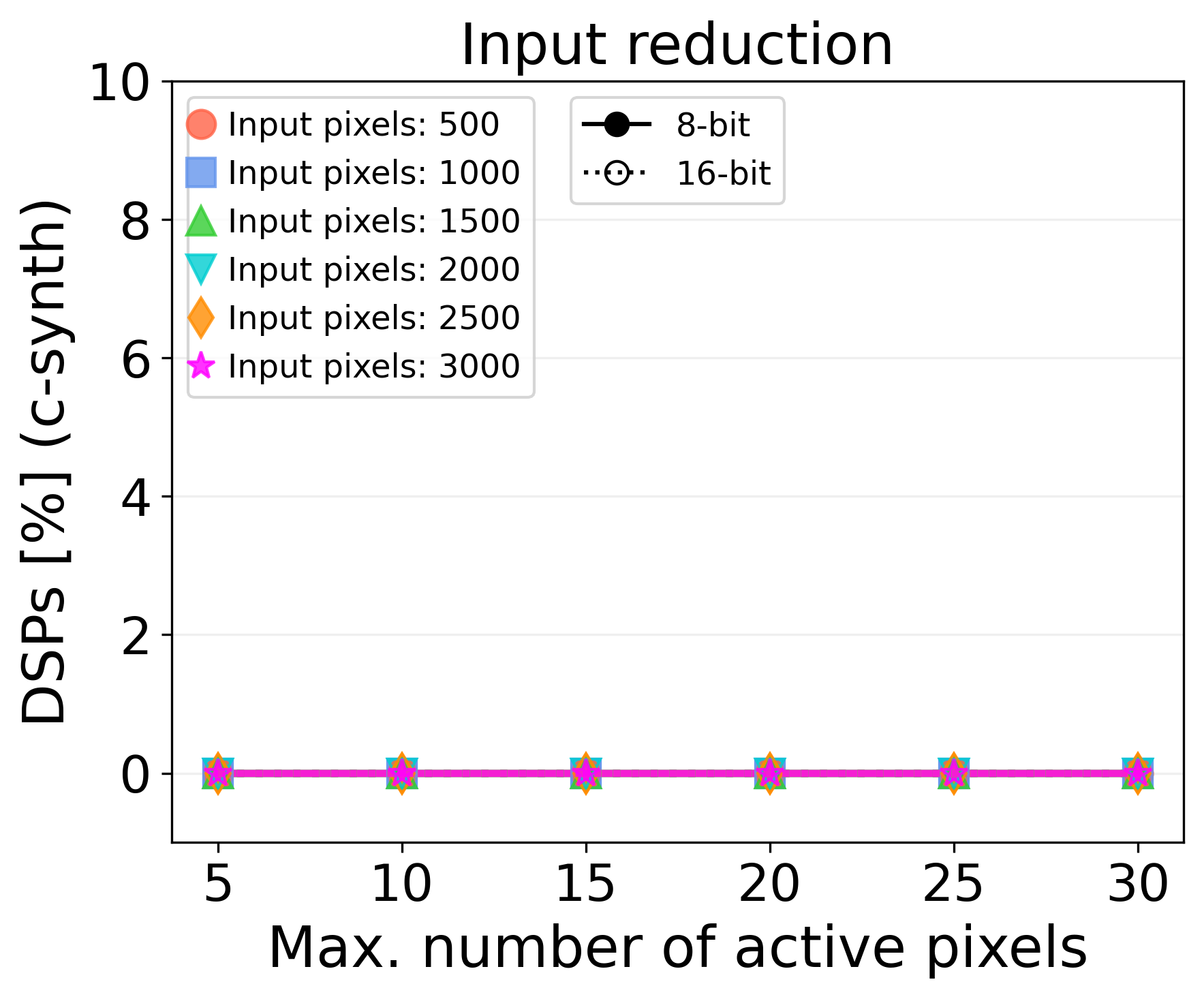}\\
    \includegraphics[width=0.35\textwidth]{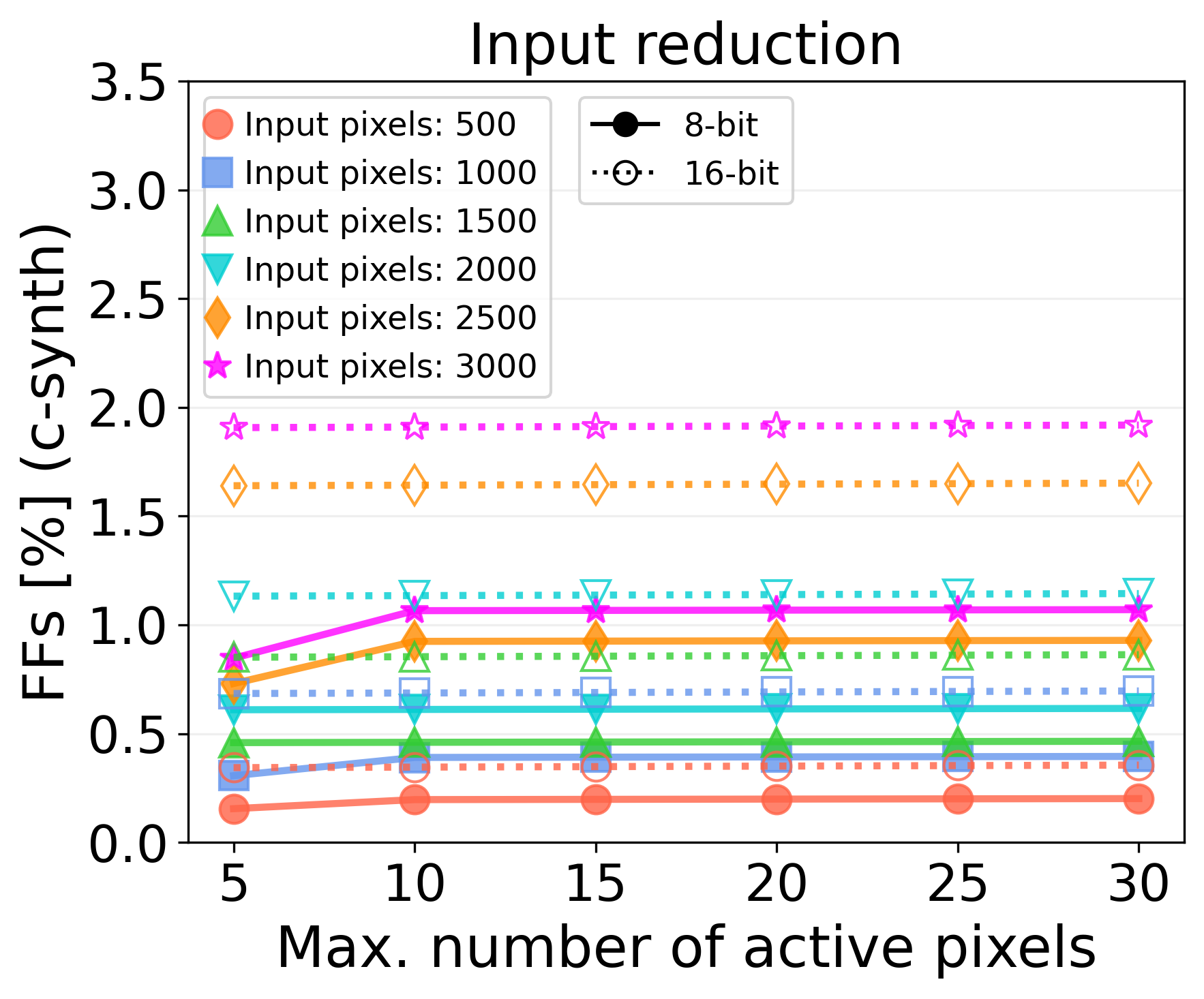}
    \includegraphics[width=0.35\textwidth]{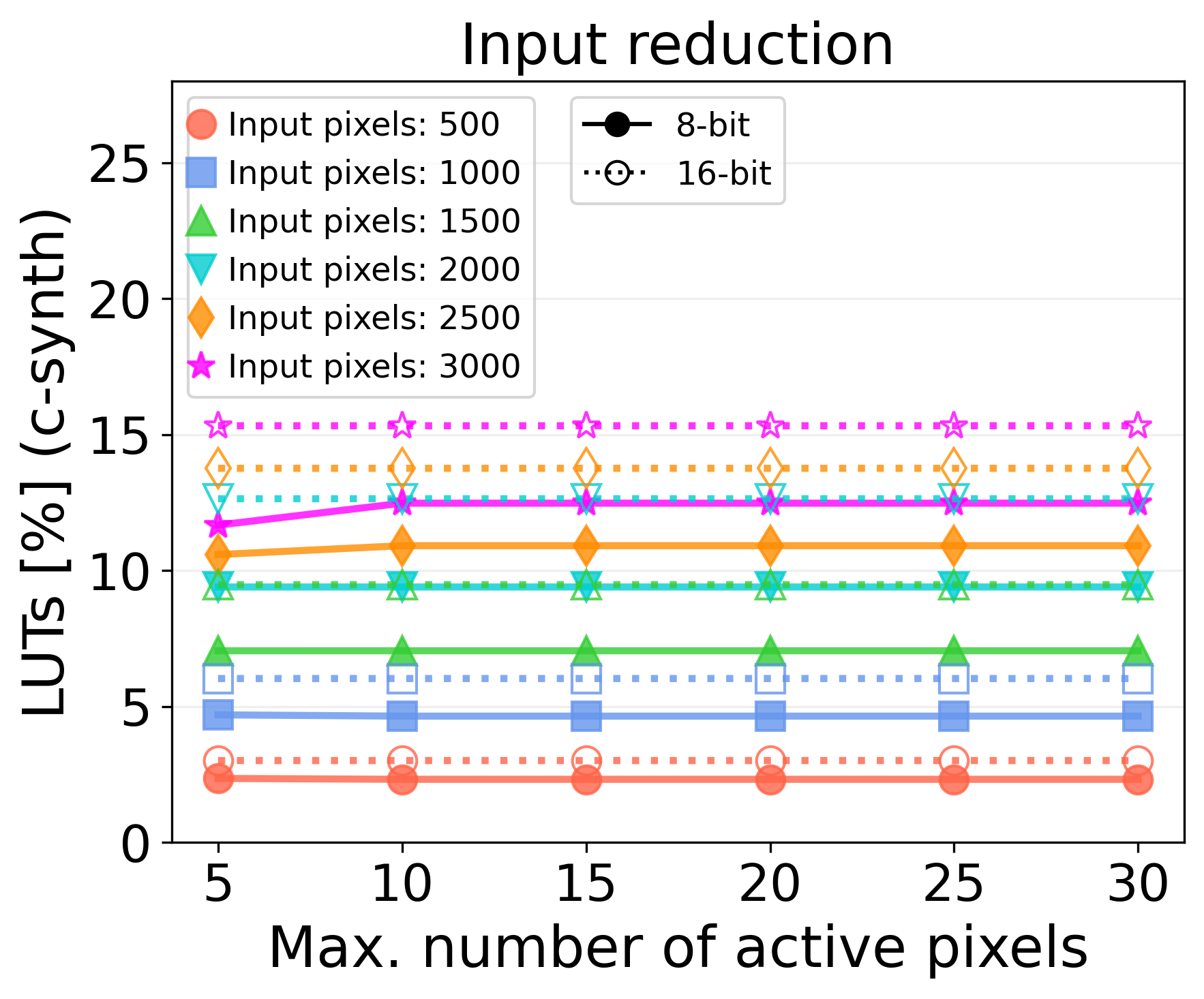}
    \caption{Scaling of the sparse input reduction layer for different input sizes and $N_{\text{active}}^{\text{max}}$: latency (upper left), DSPs (upper right), FFs (lower left), and LUTs (lower right). For latency, 1 clock cycle is equivalent to 5 ns. Resource utilization is obtained from the HLS C-synthesis step.}
    \label{fig:scaling_reduction}
\end{figure}

For the sparse input reduction layer, we test six input sizes: 500, 1000, 1500, 2000, 2500, and 3000 pixels.
For each input size, we scan $N_{\text{active}}^{\text{max}}$ from 5 to 30 in steps of 5.
The latency and resource utilization are shown in Fig.~\ref{fig:scaling_reduction}.
Latency scales linearly with $N_{\text{active}}^{\text{max}}$ across input sizes because the outer loop iterates over the retained active pixels.
The latency curves reflect the stepwise behavior of the reduction tree depth shown in Eq.~\ref{eq:reduction-depth} (Sec.~\ref{sec:method-hls-input}).
No DSPs are required because the readout involves only comparisons and data movements.
FF/LUT usage is nearly flat versus $N_{\text{active}}^{\text{max}}$ for a fixed input size, dominated by the fixed reduction-tree width, and increases with input size or bit-width.

\begin{figure}[!t]
    \centering
    \includegraphics[width=0.35\textwidth]{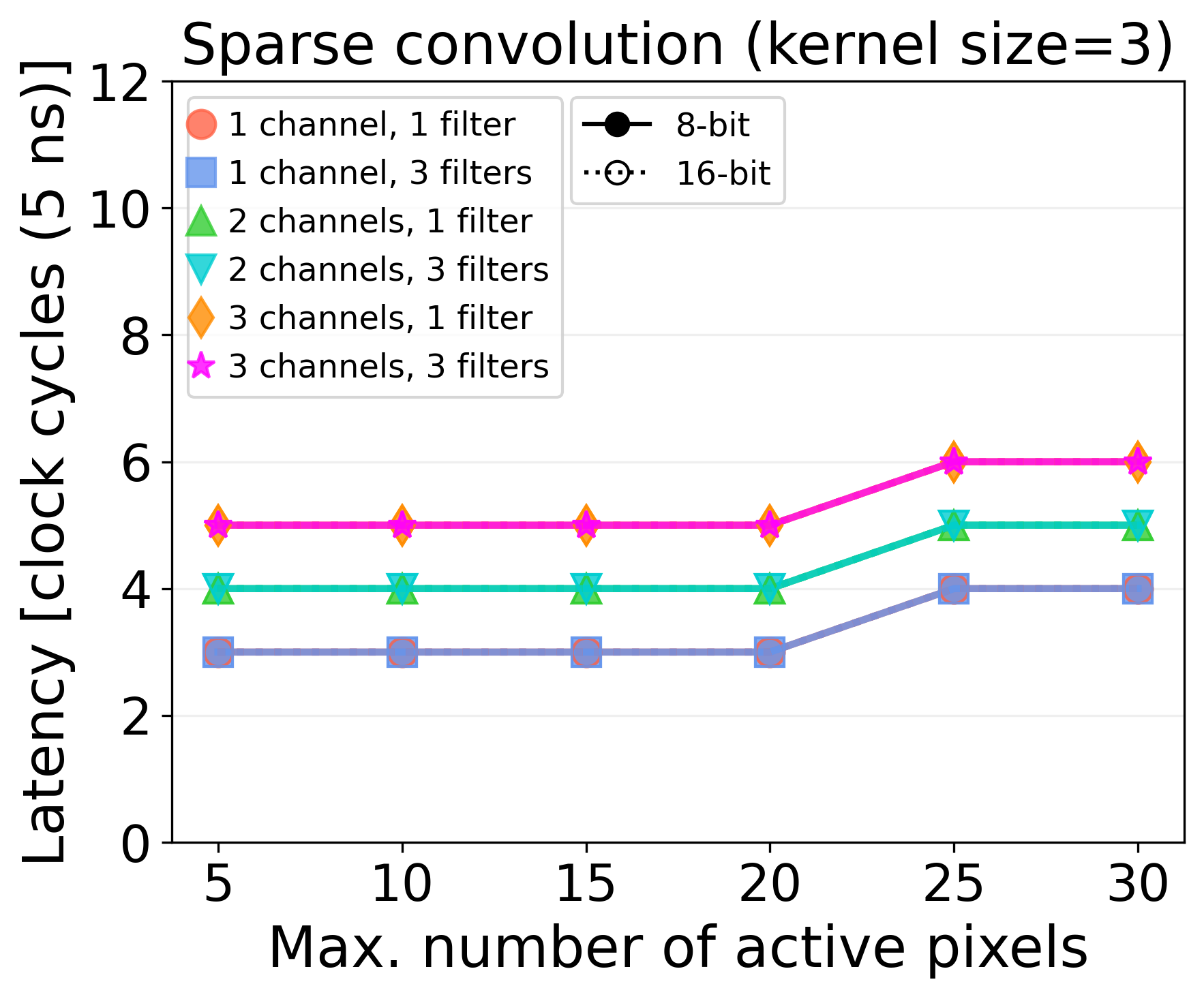}
    \includegraphics[width=0.35\textwidth]{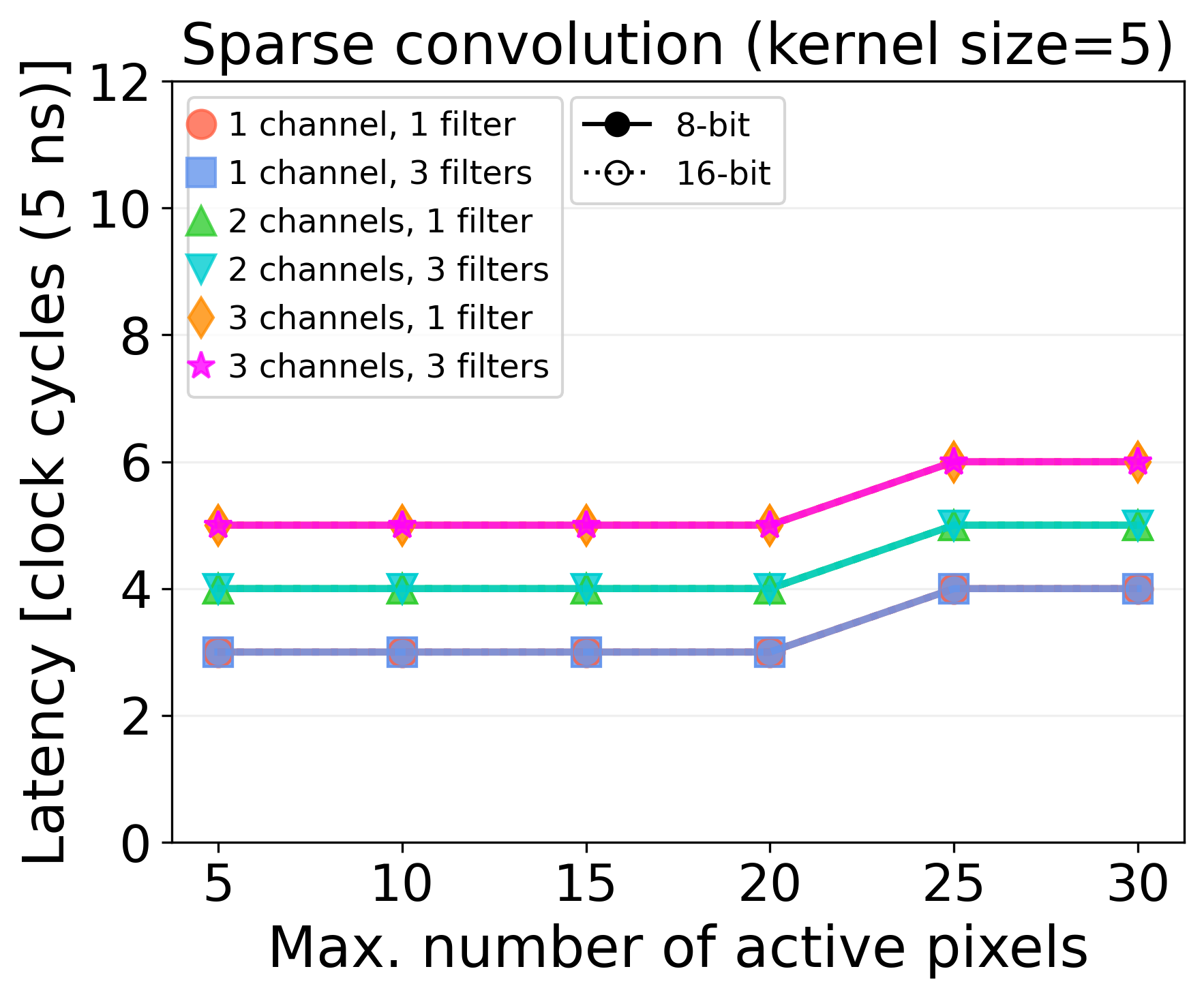}
    \caption{Latency scaling of the sparse convolutional layer for varying channels and $N_{\text{active}}^{\text{max}}$. Two kernel sizes are shown: $K=3$ (left) and $K=5$ (right). Loops are fully unrolled. For latency, 1 clock cycle is equivalent to 5 ns.}
    \label{fig:scaling_conv_latency}
\end{figure}

\begin{figure}[!t]
    \centering
    \includegraphics[width=0.33\textwidth]{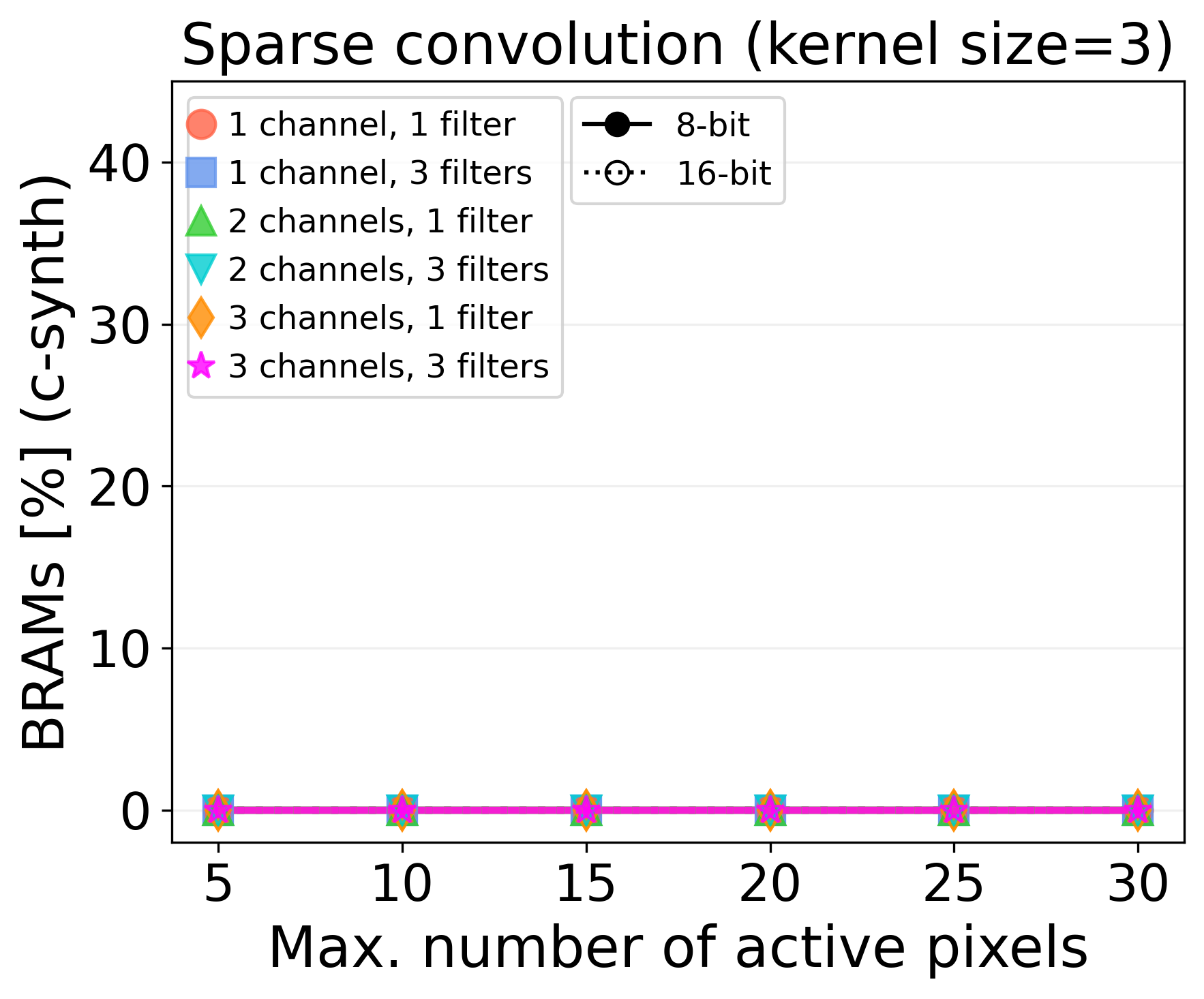}
    \includegraphics[width=0.33\textwidth]{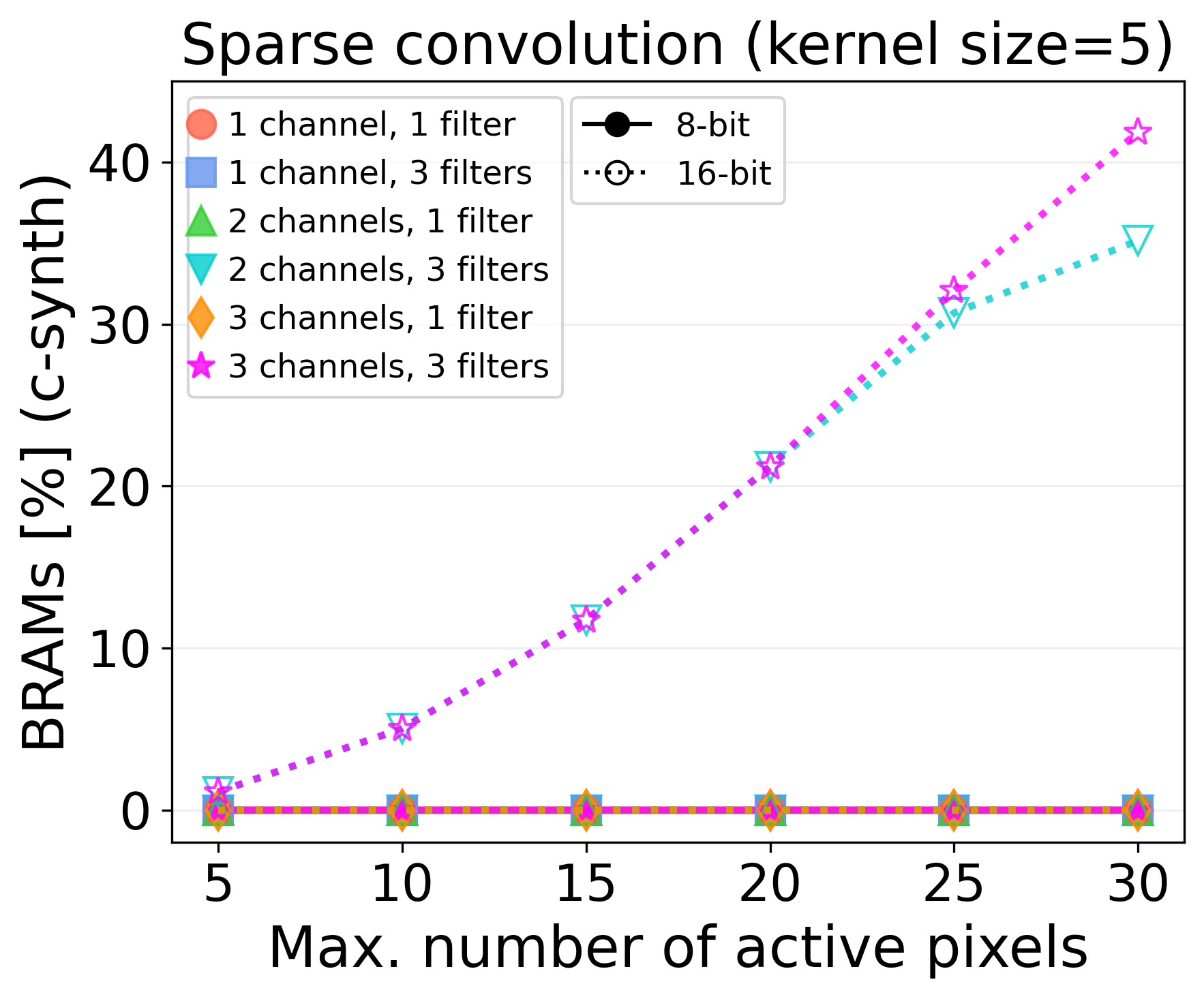}\\
    \includegraphics[width=0.33\textwidth]{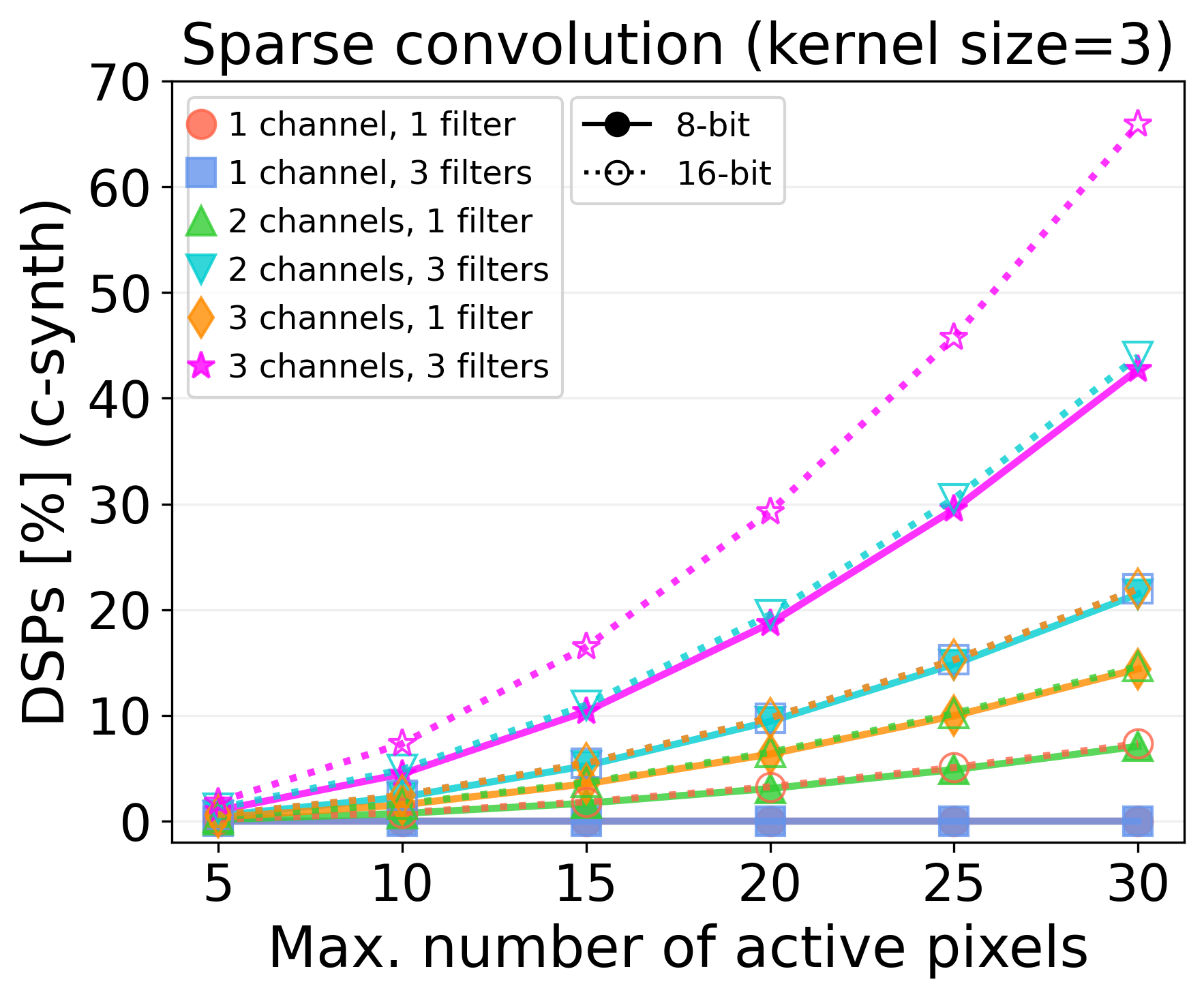}
    \includegraphics[width=0.33\textwidth]{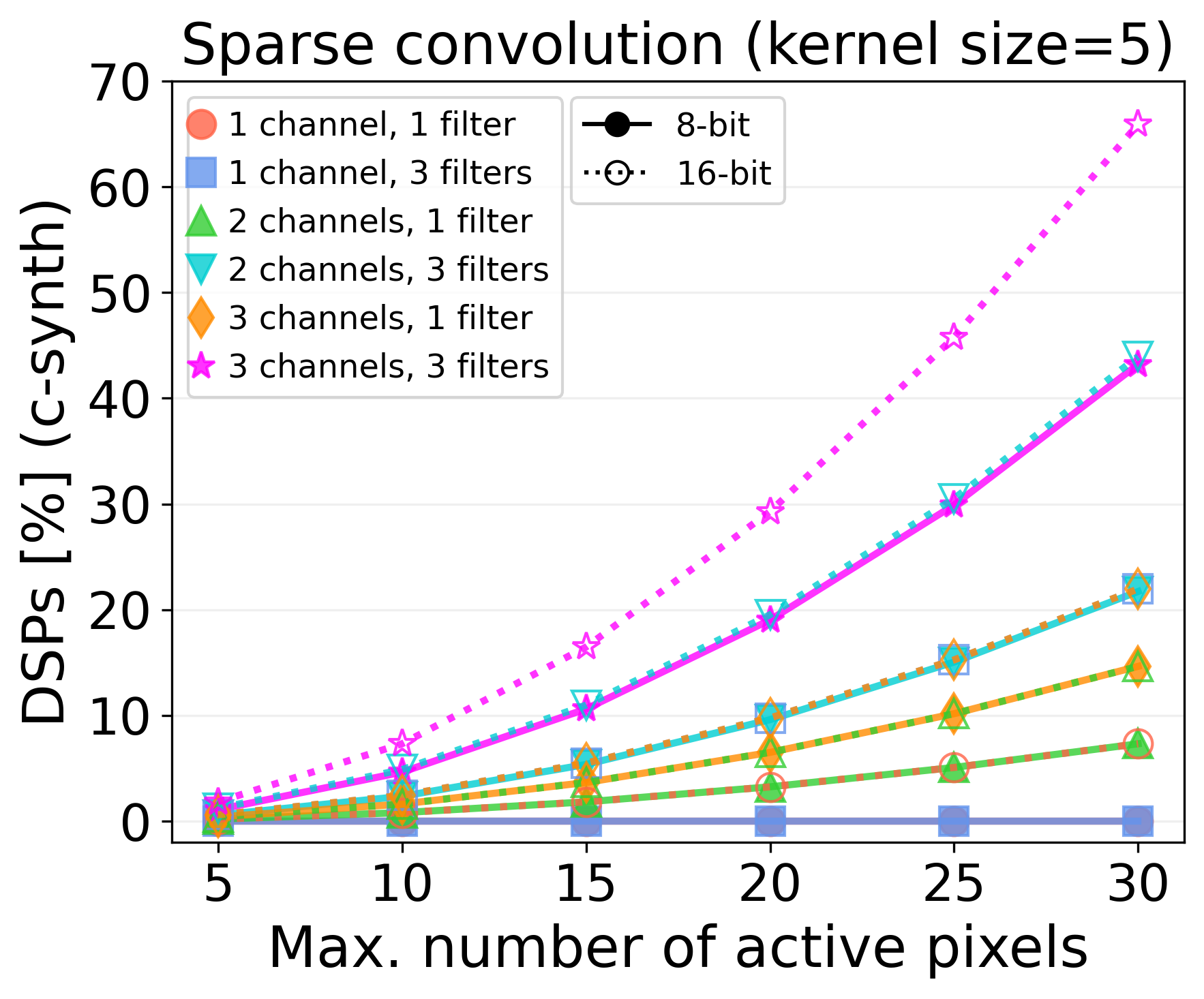}\\
    \includegraphics[width=0.33\textwidth]{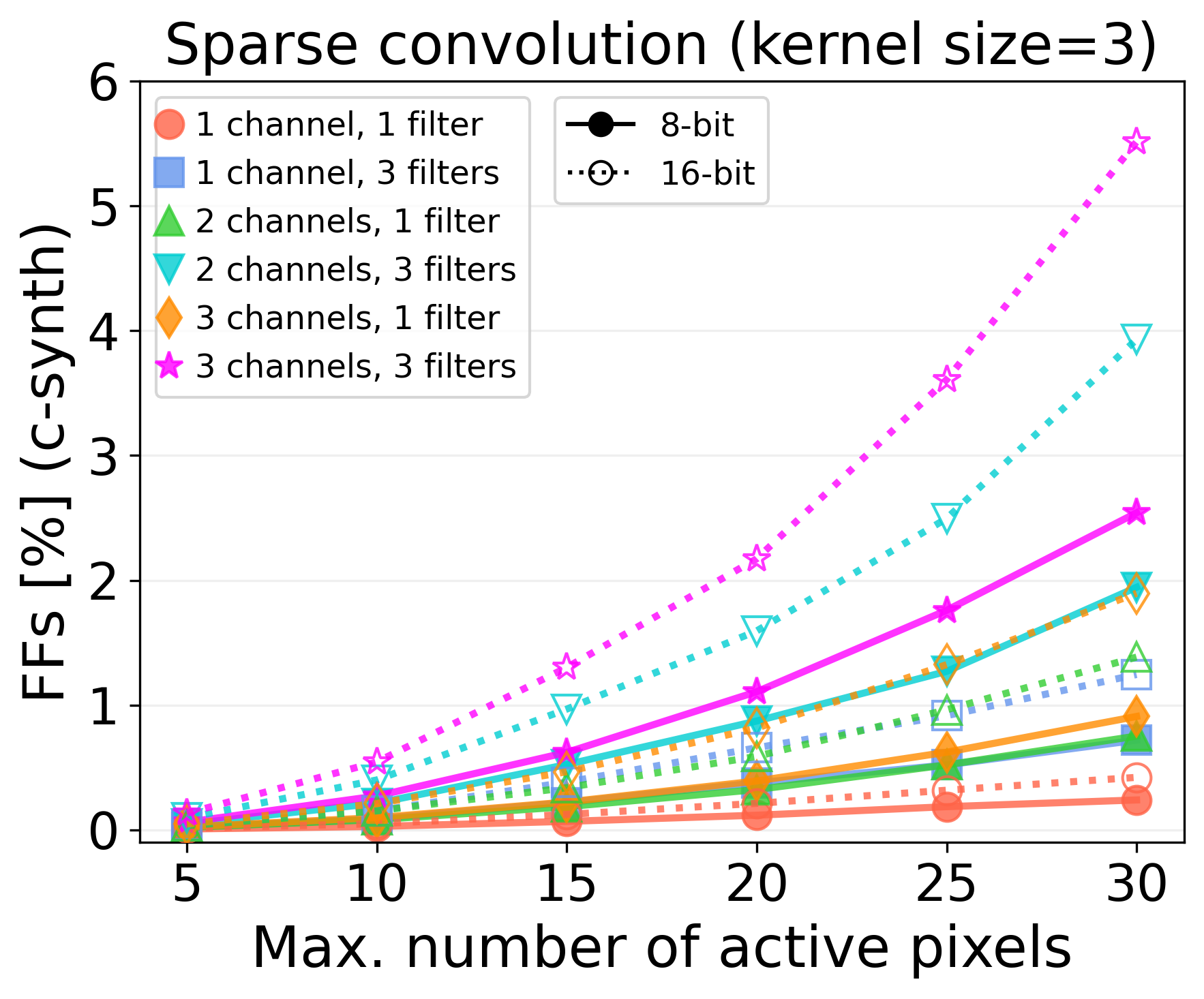}
    \includegraphics[width=0.33\textwidth]{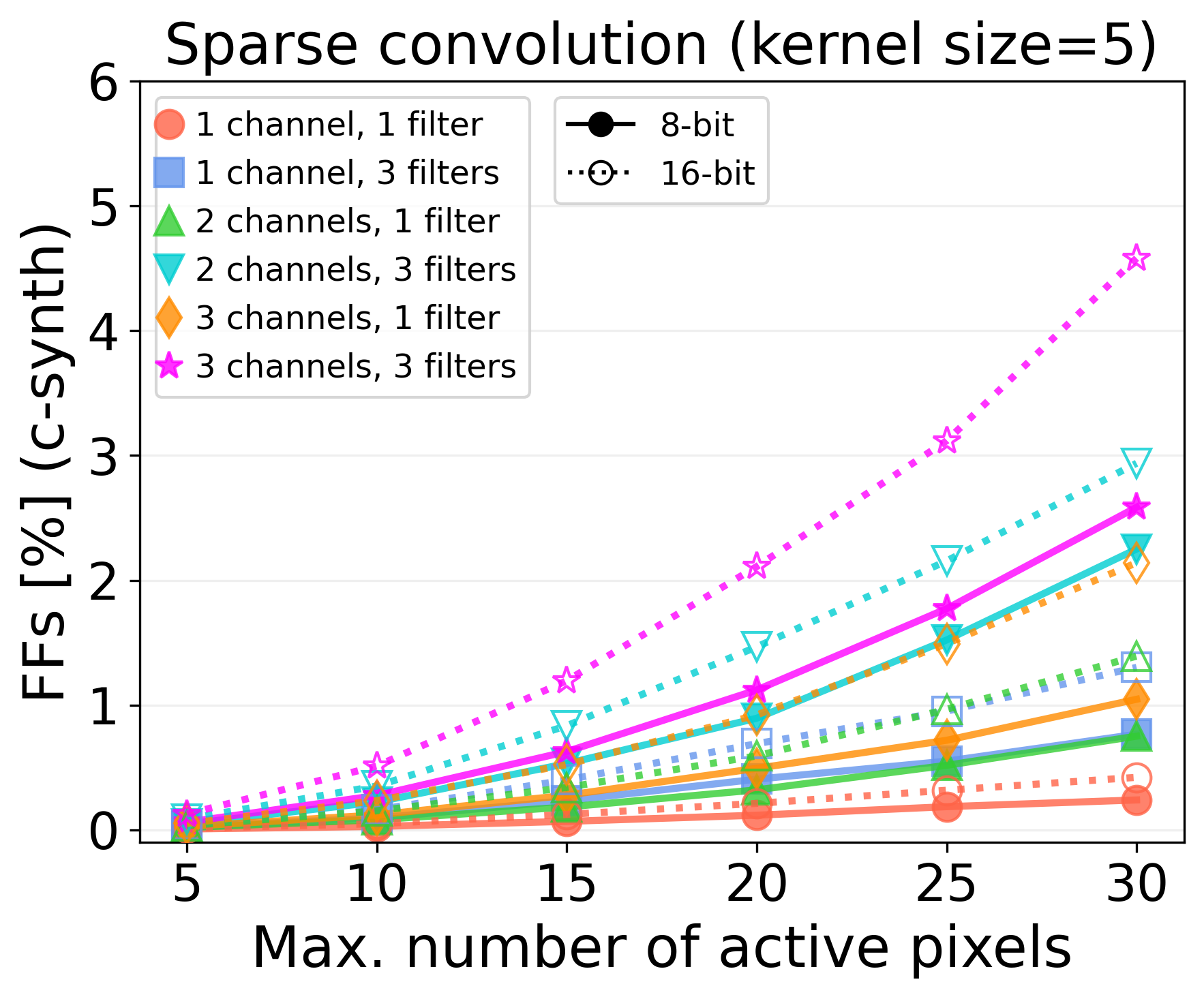}\\
    \includegraphics[width=0.33\textwidth]{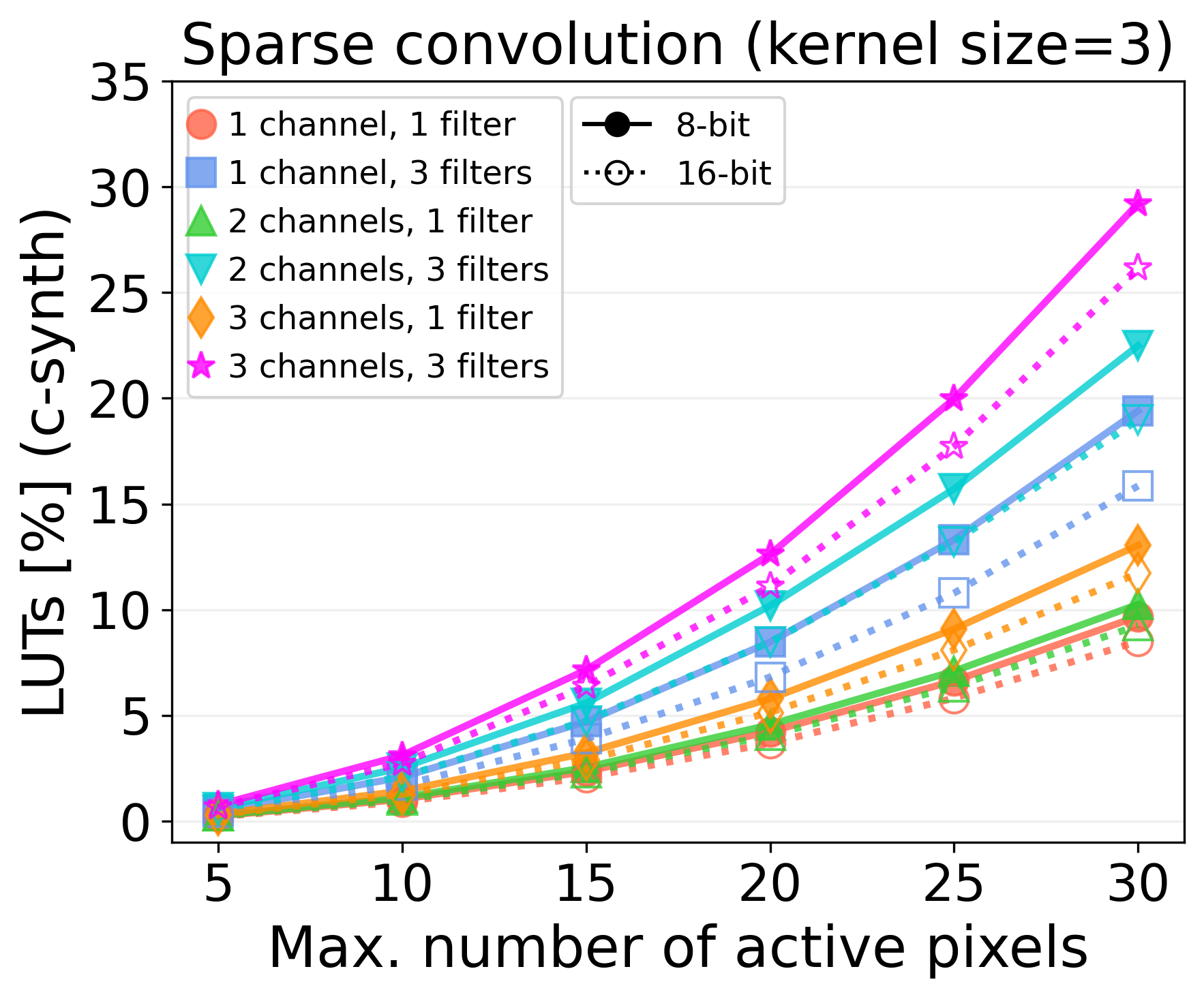}
    \includegraphics[width=0.33\textwidth]{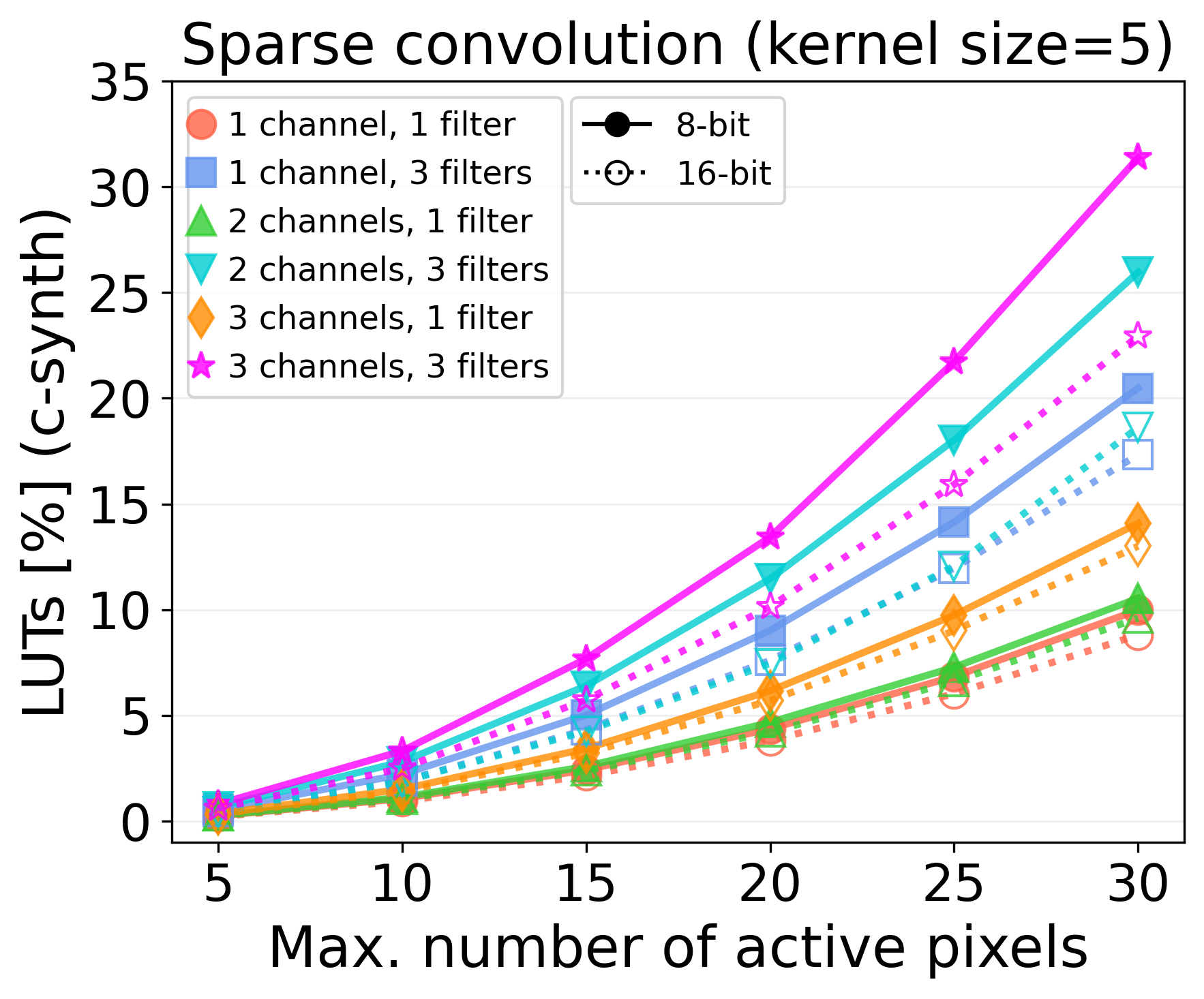}
    \caption{Resource scaling of the sparse convolutional layer for varying channels and $N_{\text{active}}^{\text{max}}$: BRAMs (row 1), DSPs (row 2), FFs (row 3), and LUTs (row 4). Two kernel sizes are shown: $K=3$ (left) and $K=5$ (right). Loops are fully unrolled. Resource utilization is obtained from the HLS C-synthesis step.}
    \label{fig:scaling_conv_resource}
\end{figure}

For the sparse convolutional layer, we evaluate $(C_{\text{in}},C_{\text{out}})\in\{(1,1),(1,3),(2,1),(2,2),(3,1),(3,3)\}$, kernel sizes $K\in\{3,5\}$, and $N_{\text{active}}^{\text{max}}$ from 5 to 30 in steps of 5.
We choose a small number of input/output channels because spatially sparse inputs typically exhibit simpler semantic patterns, so compact models suffice for low-latency applications and can achieve performance comparable to standard CNNs (see Sec.~\ref{sec:exp-eval-results}).
In addition, the current HLS design fully parallelizes computations wherever possible, and increasing the number of filters can significantly increase resource usage.
Exploring larger numbers of filters becomes feasible once the framework is extended with a flexible control over parallelization in the future (see Sec.~\ref{sec:future}).
The latency and resource utilization are shown in Fig.~\ref{fig:scaling_conv_latency} and Fig.~\ref{fig:scaling_conv_resource}, respectively.
Full or partial parallelization of standard convolution at these input sizes is beyond the capabilities of the current $\tt{hls4ml}$ implementation, so we do not include it.
Latency and resource usage are broadly similar for $K=3$ and $K=5$, except that the latter can require more BRAMs to store larger weight tensors as channels increase.
This matches the design in Sec.~\ref{sec:method-hls-conv}: the number of iterations does not scale with $K$ because kernel positions are accessed via offset checks rather than explicit loops over all $K^2$ positions.
In our tested configurations, the total number of iterations is small enough to fully unroll the loops, yielding very low latency at the cost of higher resource usage.
As $C_{\text{in}}$, $C_{\text{out}}$, or $N_{\text{active}}^{\text{max}}$ grows, the degree of parallelism increases and so do resources.

Scaling results for sparse activation are presented in Sec.~\ref{sec:apd-activation}, sparse pooling in Sec.~\ref{sec:apd-pooling}, and sparse flattening in Sec.~\ref{sec:apd-flatten}.

\subsection{Evaluation on relevant datasets}
\label{sec:exp-eval}

\subsubsection{Datasets}
\label{sec:exp-eval-datasets}

We perform experiments on a realistic neutrino dataset in this section.
For demonstration on different sparsity patterns, we also evaluate on two synthetic datasets (MNIST and LHC jet tagging), presented in Appendix~\ref{sec:apd-mnist-jet}.

\begin{figure}[!t]
    \centering
    \includegraphics[width=0.7\textwidth]{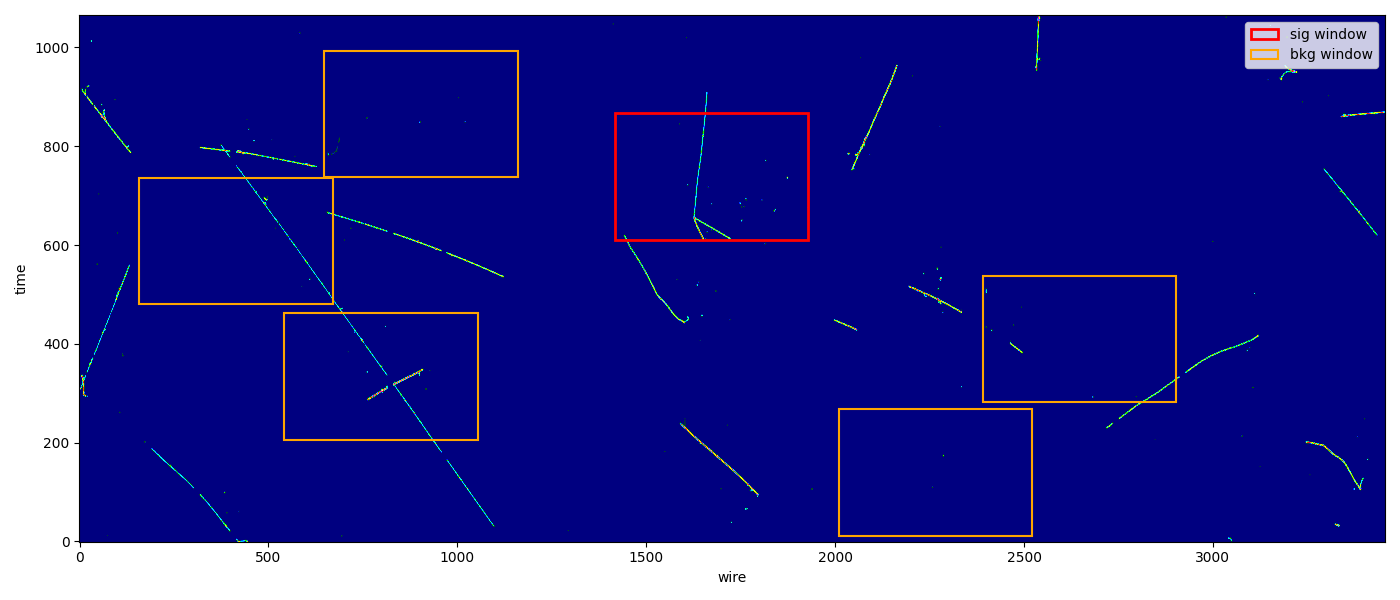}\\
    \includegraphics[width=0.3\textwidth]{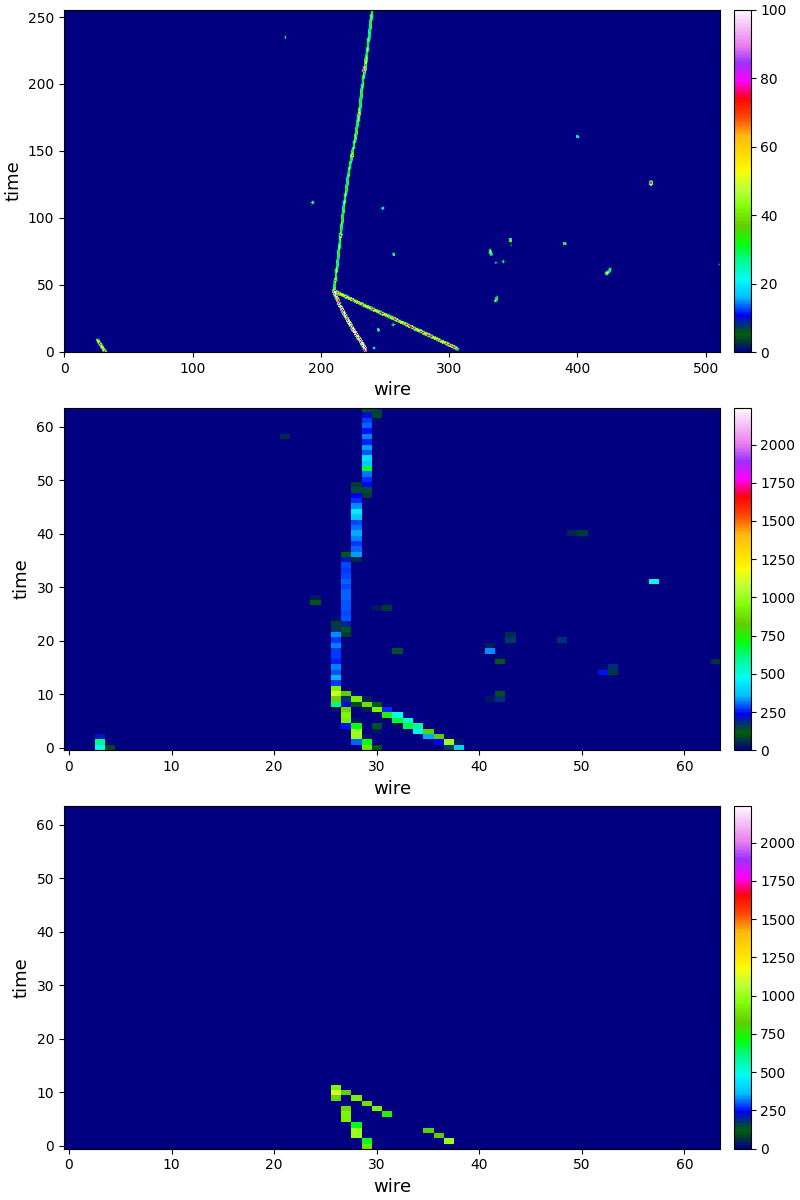}
    \includegraphics[width=0.3\textwidth]{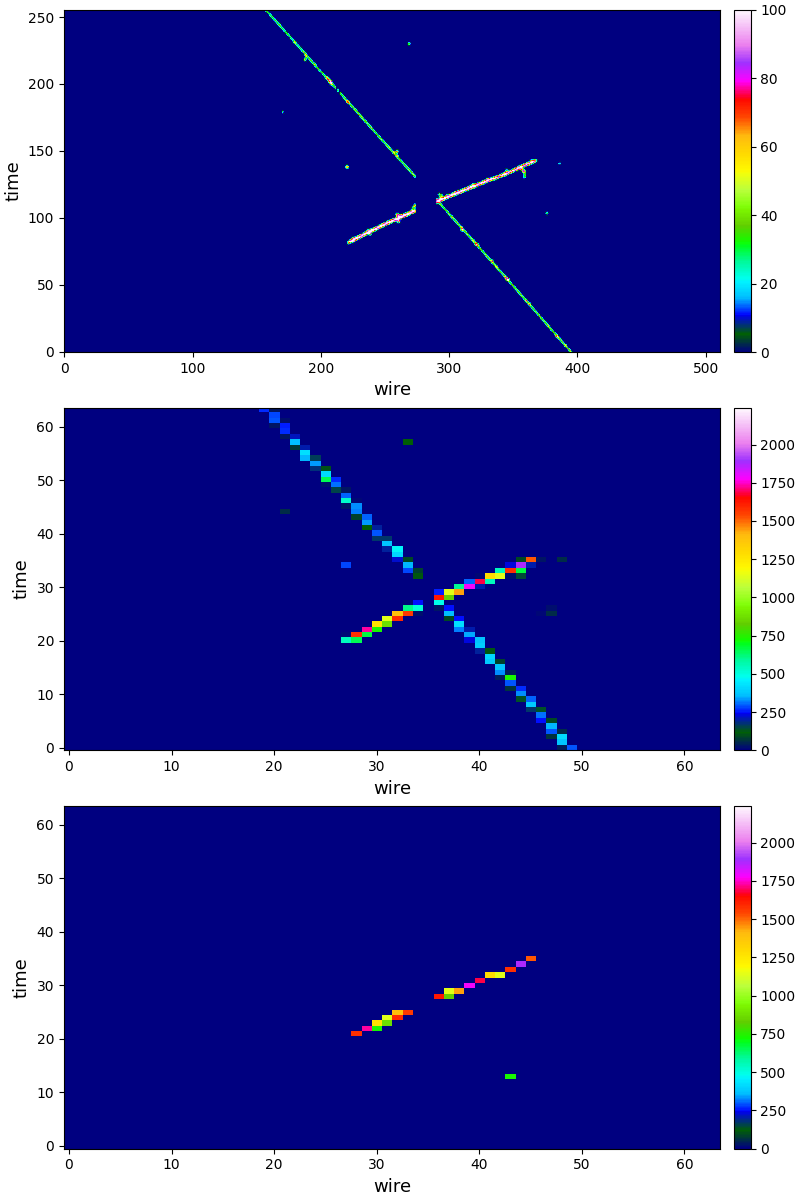}
    \caption{Example MicroBooNE neutrino image. Top: original view with signal (red) and background (orange) windows indicated. Bottom: $256\times 512$ windows downsampled to $63\times63$ and denoised by thresholding.}
    \label{fig:neu_input}
\end{figure}

For the neutrino experiment, we use the MicroBooNE Open Samples dataset~\cite{MicroBooNE:2016pwy,abratenko_2022_7262009,abratenko_2022_7262140}.
MicroBooNE uses a LArTPC to record neutrino interaction events.
In a LArTPC, an incoming neutrino (or a cosmic-ray particle) interacts with an argon nucleus, producing charged particles that ionize argon along their trajectories.
Ionization electrons drift to the anode under an applied electric field and induce signals on wire sensor planes.
The arrival time and wire position are used to reconstruct 2D images containing particle trajectories.
The dataset we use contains simulated neutrino interactions overlaid on real cosmic-ray background.
Originally, each LArTPC image has 6400 time bins and 3456 wire bins (or 2400, depending on plane), and pixel values are sensor amplitudes.
An example is shown at the top of Fig.~\ref{fig:neu_input}, where the time axis is compressed by a factor of 6, and the pixel intensity is saturated at 100 and cut at 10.
The objective is to detect whether an image contains a neutrino interaction.
We extract windows of size $256\times 512$, which are large enough to contain nearly all of the activity from a neutrino interaction.
Windows containing a simulated neutrino interaction (based on truth labels) are labeled as signal, while background windows are randomly cropped elsewhere.
To make the computation manageable, we downsample the windows to $63\times 63$ by summing the bins, then denoise by zeroing pixels below 700, as shown in Fig.~\ref{fig:neu_input}.
Signal windows tend to contain intersecting tracks forming a vertex while background windows contain fewer active pixels or disconnected tracks with low intensity.
Finally, features are normalized by a factor of 3200 to unit order.
The task is supervised binary classification.

\subsubsection{Experimental setup}
\label{sec:exp-eval-setup}

\begin{figure}[!t]
    \centering
    \includegraphics[width=0.9\textwidth]{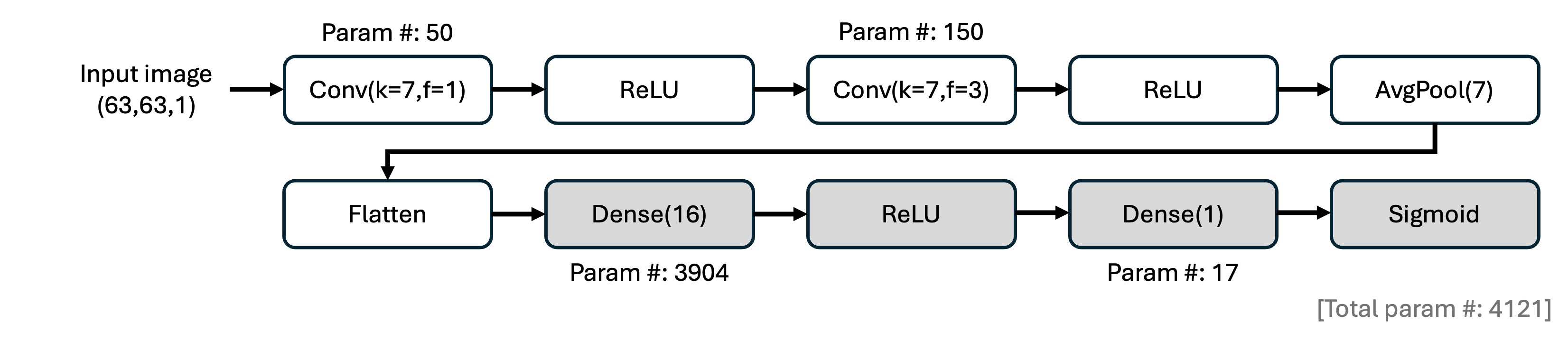}
    \caption{Base architecture for constructing both standard and sparse CNNs for the neutrino experiment. All have two convolutional blocks followed by a 2-layer MLP, with a compact model size of around 4k parameters.}
    \label{fig:models_neutrino}
\end{figure}

For a fair comparison, we use the same base architecture to construct both the standard CNN and the sparse CNN in each experiment.
The base architecture is shown in Fig.~\ref{fig:models_neutrino}.
Each model consists of two convolutional blocks (a convolutional layer followed by ReLU), uses average pooling for downsampling, then flattens to feed a 2-layer multilayer perceptron (MLP) classifier.
These models are compact (approx. 4k parameters), which is typical for low-latency FPGA applications.
The only architectural difference between the standard and sparse models is that the sparse CNN uses sparse convolution to compute on at most $N_{\text{active}}^{\text{max}}$ pixels.
For the sparse CNNs, we test $N_{\text{active}}^{\text{max}}\in\{8,12,16,20\}$, referred to as $\textit{sparse-\{tiny,small,medium,large\}}$.
In practice, a proper $N_{\text{active}}^{\text{max}}$ can be chosen by studying the occupancy distribution of the data to pick a high percentile value after thresholding, then scanning around this value and checking the performance; this empirical process can be optimized by making $N_{\text{active}}^{\text{max}}$ a trainable parameter, which is a planned future upgrade (Sec.~\ref{sec:future}).
We compare accuracy and FPGA cost between the standard CNN and these sparse variants, training with our $\tt{SparsePixels}$ library using $\tt{QKeras}$ as the QAT backend, evaluating two fixed total bit-widths: 8 and 16.

We use 160k windows with balanced signal and background, split 70/10/20\% for train/val/test.
Models are trained with batch size 128 for up to 200 epochs with early stopping, using Adam~\cite{kingma2014adam} with a learning rate of $10^{-3}$ to optimize binary cross-entropy.

After training, we generate HLS for all models with $\tt{hls4ml}$~\cite{Aarrestad:2021zos,fastml_hls4ml,Duarte:2018ite}.
Standard CNNs are implemented with stream I/O (they are too large with parallel I/O) and synthesized with Vivado HLS (2020.1)~\cite{vivado}, which is the currently stable version for convolutional architectures.
Sparse CNNs use parallel I/O and are synthesized with Vitis HLS (2023.1)~\cite{vitis}.
The reported resources are obtained after the logic synthesis step.
As in the scaling studies, the target device is an AMD/Xilinx Alveo FPGA with part number xcu250-figd2104-2L-e, and the clock frequency is set to 200 MHz (5 ns clock period).

\subsubsection{Results}
\label{sec:exp-eval-results}

\begin{figure}[!t]
    \centering
    \includegraphics[width=0.256\textwidth]{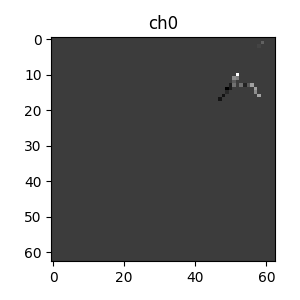}\\
    \includegraphics[width=0.7\textwidth]{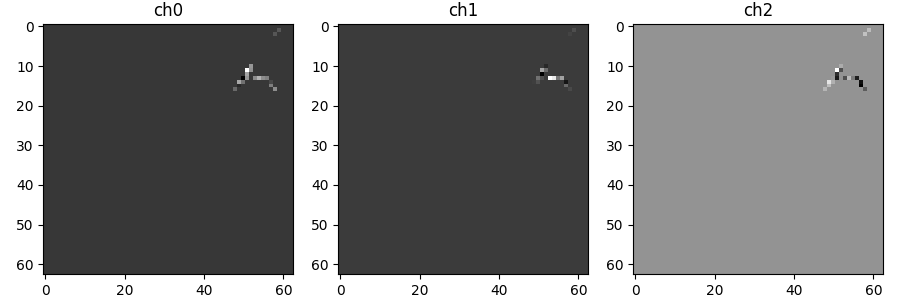}
    \caption{Illustration of sparse convolution on an example neutrino image from the MicroBooNE dataset. Top: single-channel input. Bottom: three output channels after one sparse convolution.}
    \label{fig:neu_conv}
\end{figure}

Fig.~\ref{fig:neu_conv} illustrates sparse convolution on an example neutrino signal image from the MicroBooNE dataset, showing the input and the output after one sparse convolutional layer.
As designed, the active pixel locations are unchanged.
Sparse convolution is a constrained operation, where feature updates occur only at the active pixel locations.
This limits the spread of information to neighboring locations and narrows how pixels communicate across layers.
Since the number of active pixels never increases (sparse convolution preserves the active set and pooling typically reduces it), one can trade computation for accuracy by increasing the number of output channels per layer.
Likewise, a larger kernel can increase the receptive field among active pixels, as our HLS implementation does not directly scale the iteration count with the kernel size (see Sec.~\ref{sec:method-hls-conv}).

\begin{figure}[!t]
    \centering
    \includegraphics[width=0.55\textwidth]{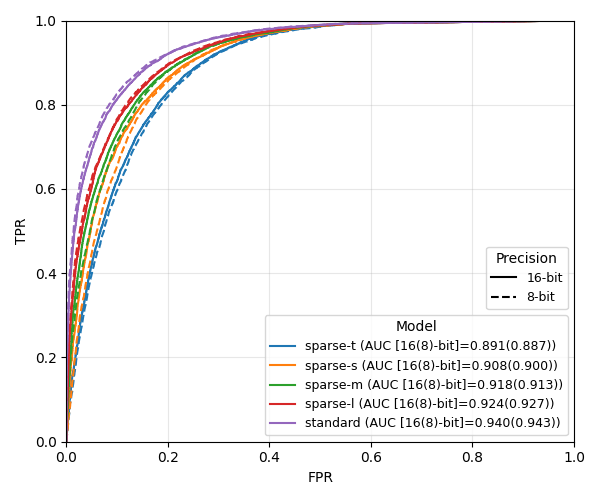}
    \caption{Model performance comparing the standard CNNs and the sparse CNNs for the neutrino dataset.}
    \label{fig:performance_neutrino}
\end{figure}

Fig.~\ref{fig:performance_neutrino} shows the model performance.
As expected, sparse CNN accuracy improves with larger $N_{\text{active}}^{\text{max}}$ because more semantic information is retained.
The sparse-large models ($N_{\text{active}}^{\text{max}}=20$) are the most accurate among the sparse variants and are close to the standard CNNs.
A small performance loss is expected, since sparse CNNs compute on a subset of pixels and enforce a sparsity-preserving convolution, but these small losses are traded for significant FPGA inference speedups.

Tab.~\ref{tab:results_neutrino} summarizes latency, speedup, and performance relative to the standard CNNs.
The sparse-large model gains more than an order-of-magnitude speedup with comparable or slightly lower accuracy.
The high latency of standard CNNs comes from computing on all pixels and, given the large image sizes, the need to use stream I/O (one pixel streaming per cycle plus additional buffering).
In contrast, sparse CNNs selectively compute on a small fraction of pixels, for example, the sparse-large model consumes 20 of $63\times 63$ input pixels (0.5\%), and the compact representation enables parallel processing on an FPGA, where all active pixels can be processed concurrently.
Concretely, the sparse-large model at 8-bit achieves a $\times 73$ speedup over the 8-bit standard CNN (133 vs. 9733 clock cycles, i.e., 0.665 vs. 48.665 $\mu$s) with $<2$\% ROC AUC drop (0.927 vs. 0.943).
As designed, the latencies and IIs of the sparse CNNs are frozen at synthesis, exact constants for every input regardless of the input sparsity.
Similar results are observed from the two synthetic datasets in Appendix~\ref{sec:apd-mnist-jet}.

\begin{table*}[!t]
\centering
\caption{Comparison of FPGA latency and model performance between standard and sparse CNNs. The sparse-\{t,s,m,l\} variants correspond to $N_{\text{active}}^{\text{max}}=\{8,12,16,20\}$. Latency and initiation interval (II) are measured in clock cycles (cc) with 5 ns clock period. Speedups for sparse CNNs are measured with respect to the standard CNN at the same total bit-width. $\epsilon_{\mathrm{s}}$ and $\epsilon_{\mathrm{b}}$ refer to signal and background efficiency, respectively.}
\label{tab:results_neutrino}

\begin{subtable}
\centering
\small
\scalebox{1}{
\begin{tabular}{lccccc}
    \multicolumn{6}{c}{\bfseries MicroBooNE neutrino detection, 2 classes, $63\times63=3969$ input pixels} \\
    \toprule
    \textbf{CNN model} & \textbf{Latency [cc]} & \textbf{Speedup} & \textbf{II [cc]} &
    $\boldsymbol{\epsilon_{\mathrm{s}}(\epsilon_{\mathrm{b}}=0.2)}$ & \textbf{AUC} \\
    \midrule
    Sparse-t 8b & 69 (0.345 $\mu$s) & $\times 141$ & 35 (0.175 $\mu$s) & 0.819 & 0.887 \\
    Sparse-t 16b & 71 (0.355 $\mu$s) & $\times 137$ & 35 (0.175 $\mu$s) & 0.830 & 0.891 \\ \hline
    
    Sparse-s 8b & 90 (0.45 $\mu$s) & $\times 108$ & 52 (0.26 $\mu$s) & 0.856 & 0.900 \\
    Sparse-s 16b & 94 (0.47 $\mu$s) & $\times 104$ & 52 (0.26 $\mu$s) & 0.864 & 0.908 \\ \hline
    
    Sparse-m 8b & 111 (0.555 $\mu$s) & $\times 88$ & 67 (0.335 $\mu$s) & 0.879 &  0.913\\
    Sparse-m 16b & 116 (0.58 $\mu$s) & $\times 84$ & 67 (0.335 $\mu$s) & 0.881 &  0.918\\ \hline
    
    Sparse-l 8b & 133 (0.665 $\mu$s) & $\times 73$ & 84 (0.42 $\mu$s) & 0.896 & 0.927 \\
    Sparse-l 16b & 136 (0.68 $\mu$s) & $\times 72$ & 84 (0.42 $\mu$s) & 0.894 & 0.924 \\ \hline\hline
    
    Standard 8b & 9733 (48.665 $\mu$s) & $-$ & 5029 (25.145 $\mu$s) & 0.921 & 0.943 \\
    Standard 16b & 9736 (48.68 $\mu$s) & $-$ & 5029 (25.145 $\mu$s) & 0.921 & 0.940 \\
    \bottomrule
\end{tabular}
}
\end{subtable}

\end{table*}

\begin{figure}[!t]
    \centering
    \includegraphics[width=0.95\textwidth]{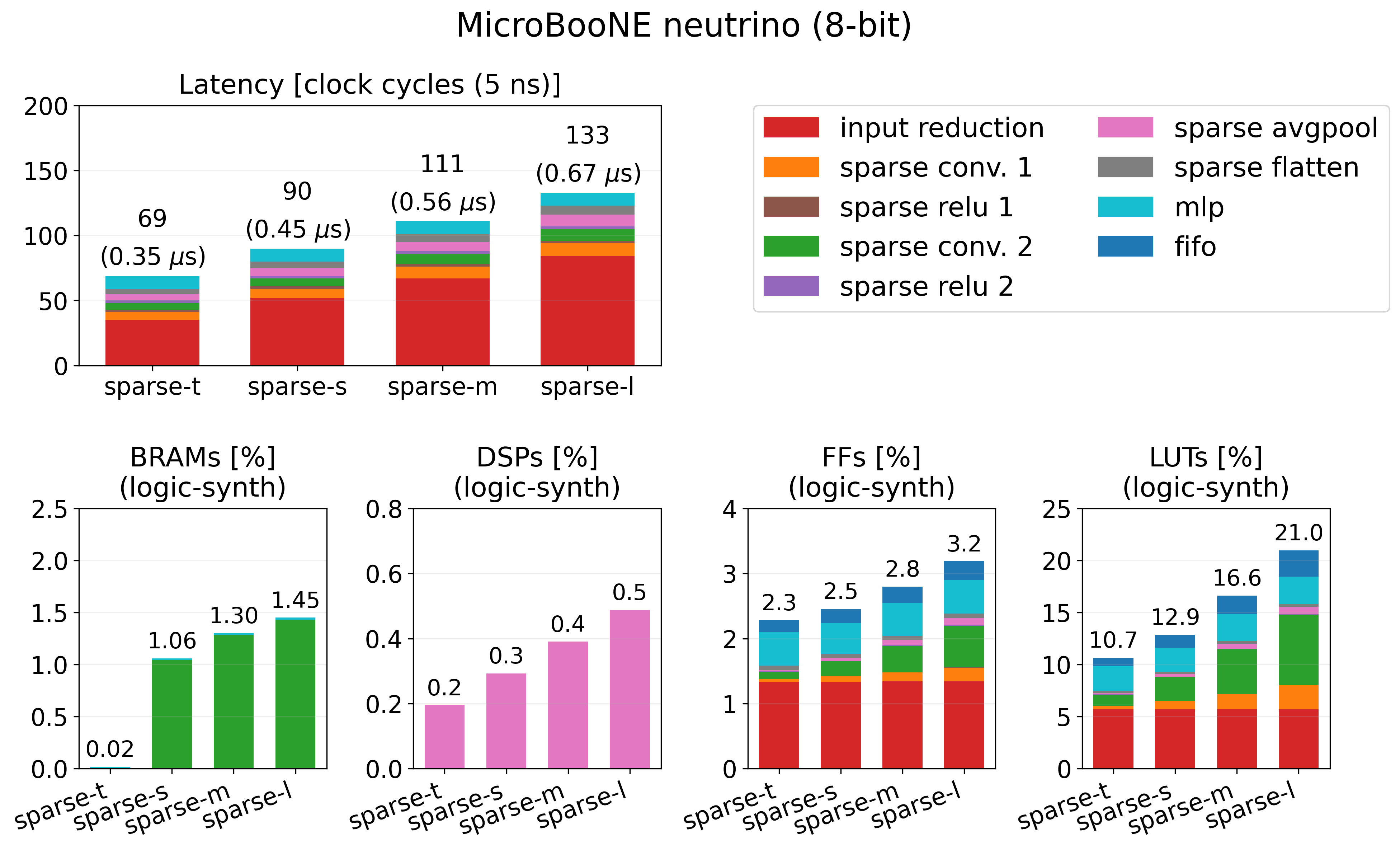}
    \caption{Per-layer resource breakdown for 8-bit sparse CNNs for the neutrino dataset. Top: latency. Bottom: BRAMs, DSPs, FFs, and LUTs. Latency is measured in clock cycles (cc) with 5 ns clock period. Resource utilization is obtained after the logic synthesis step.}
    \label{fig:breakdown8b}
\end{figure}

For all sparse CNNs, the initiation interval (II)---the number of clock cycles between accepting successive inputs---equals the latency of the sparse input reduction layer (the first layer in sparse models), which can be seen in the per-layer latency breakdown in Fig.~\ref{fig:breakdown8b} (8-bit) and Fig.~\ref{fig:breakdown16b} (16-bit).
This is because the recursive tree splitting requires buffering and processing the current input array until scanning completes before the next array can enter.
Subsequent sparse layers operate on sparse arrays and their operations are maximally parallelized.

\begin{figure}[!t]
    \centering
    \includegraphics[width=0.95\textwidth]{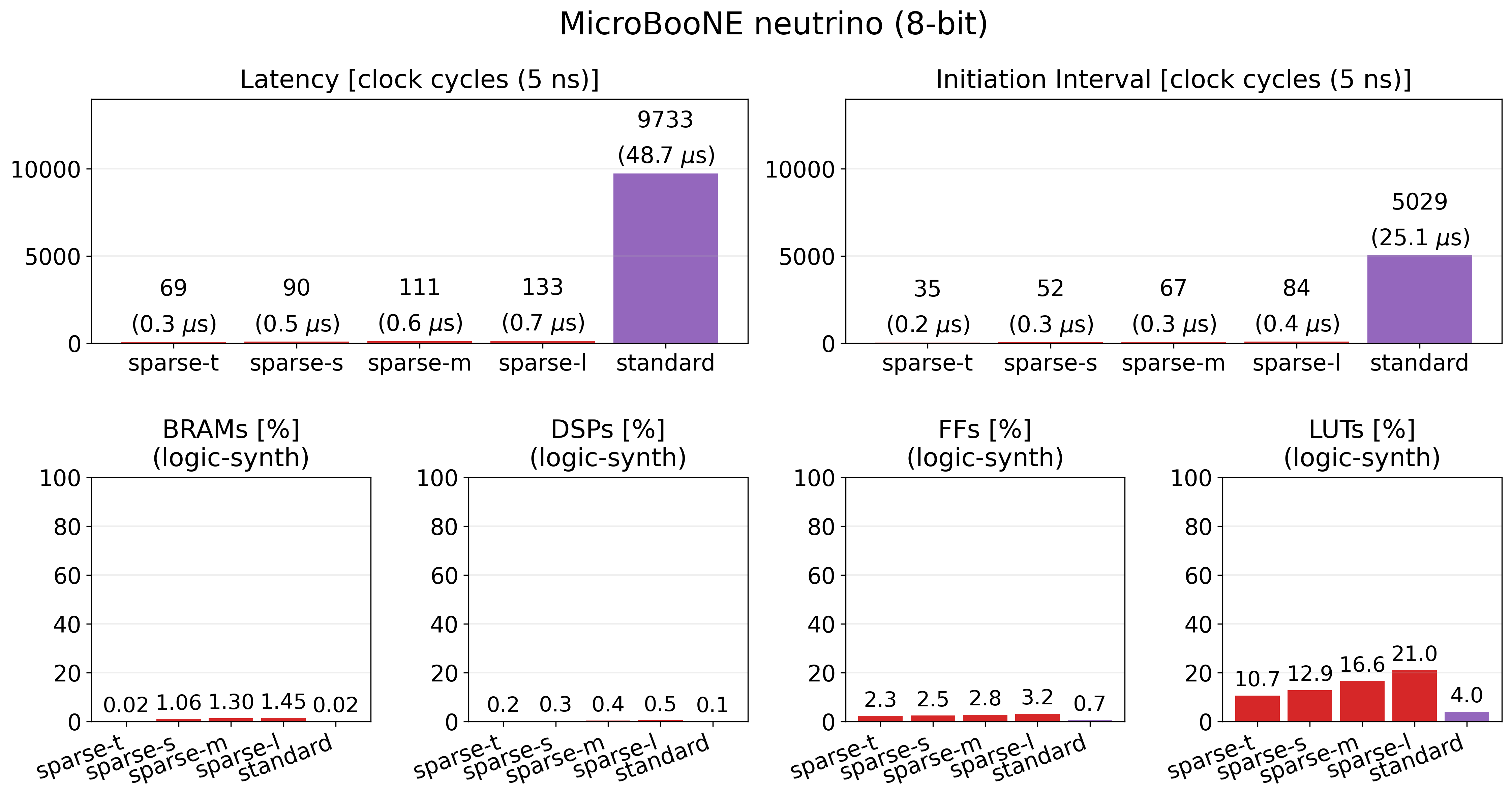}
    \caption{Synthesis results of standard (purple) and sparse (red) CNNs at 8-bit for the neutrino dataset. Top: latency and II. Bottom: BRAMs, DSPs, FFs, and LUTs. Latency and II are measured in clock cycles (cc) with 5 ns clock period. Resource utilization is obtained after the logic synthesis step.}
    \label{fig:bar8b}
\end{figure}

Fig.~\ref{fig:bar8b} (Fig.~\ref{fig:bar16b}) shows resource usage as a fraction of the total available on device for 8-bit (16-bit) models.
As discussed, standard CNNs, implemented with stream I/O, tend to use fewer resources due to sequential processing, while sparse CNNs, implemented with parallel I/O, consume more resources in exchange for much lower latency.
For example, LUT usage is several times higher in sparse CNNs because loops are largely unrolled to execute concurrently.
This increase is less concerning since even the sparse-large models consume only $\mathcal{O}(20\%)$ of total LUTs on device, but this is traded for much larger speedups.
When needed, the degree of parallelization can be adjusted for the sparse CNNs to reduce resources at the cost of higher latency, which is a trade-off that is acceptable given the observed substantial baseline speedups.
The per-layer resource breakdowns shown in Fig.~\ref{fig:breakdown8b} (8-bit) and Fig.~\ref{fig:breakdown16b} (16-bit) are consistent with the scaling behaviors in Sec.~\ref{sec:exp-scaling}.

\section{Limitations and future work}
\label{sec:future}

With the results presented above as a proof of concept, we summarize the main limitations of the current design, with concrete improvements planned for a future upgrade which aims to make the framework readily usable in broader experiments.

\textbf{Full parallelization in HLS limits input and model sizes.} The current HLS implementation unrolls most loops and fully parallelizes computation wherever possible across all sparse layers, prioritizing minimal latency at the cost of higher resource usage. This is why we have kept the model sizes small in both the scaling studies and the experiments, since doing otherwise would quickly exhaust on-chip resources. However, unlike conventional images with full occupancy and multiple channels that encode more complex semantics (e.g., dog vs. cat), our method targets inputs with high spatial sparsity, which typically carry much simpler semantic patterns (e.g., the neutrino images contain single-channel 1D tracks in a vastly empty 2D space), so models with larger depth and number of filters would not gain enough performance to justify the extra hardware costs in constrained environments. Nonetheless, while full parallelization is the simplest design to validate first, the obvious next step is to enable flexible control over the degree of parallelization in every sparse layer so one can trade between latency and resource utilization. This would make a wider range of model sizes and larger inputs feasible for practical real-world experiments.

\textbf{Choosing a proper $N_{\text{active}}^{\text{max}}$ is currently empirical.} The model hyperparameter $N_{\text{active}}^{\text{max}}$ fixes the maximum number of active pixels the model processes from the input: too small a value can mislead the model since most of the relevant pixels are masked out (see the MNIST digit example in Fig.~\ref{fig:mnist_leak}), while too large a value retains sufficient semantic information but significantly increases the model size and resource usage. We currently find a proper value by an empirical scan. A possible improvement is to make $N_{\text{active}}^{\text{max}}$, and potentially the threshold, trainable hyperparameters in the training library, so they can be learned directly from the data for a given model configuration.

\textbf{A fixed $N_{\text{active}}^{\text{max}}$ bounds the maximal information a model can process.} While this is the core design choice to maximize hardware efficiency with a constant runtime for sparse input data, it could in fact reduce the semantic information processable by the model when the chosen value is too small, in contrast to a standard CNN which always processes the full input image. This could raise concern since, in practice, the number of active pixels can vary between inputs at runtime, but such expected fluctuations can be absorbed by allowing a margin when choosing $N_{\text{active}}^{\text{max}}$ and tuning the threshold, based on the occupancy distribution of the training data. However, in trigger applications, a sharp increase beyond the training occupancy distribution would correspond to a change in normal run conditions, which would affect all trigger algorithms systematically and would require retuning in any case.

\textbf{Coarse quantization.} The current training library uses $\tt{QKeras}$ as the backend for QAT, which quantizes the precision uniformly per layer.
The future upgrade is to adopt HGQ~\cite{Sun:2024soe} for finer-grained per-weight quantization, further reducing resource usage at fixed accuracy.

\section{Summary}
\label{sec:conclusion}

We have introduced $\tt{SparsePixels}$, a framework for sub-microsecond FPGA inference of CNNs using sparse convolution on spatially sparse 2D data.
The key features in our HLS implementation are: (1) dynamically identifying and retaining a small subset of active pixels from a large input into compact arrays, then (2) performing all subsequent computation only on those retained pixels.
By selectively computing on the informative subset of pixels, the design avoids computation on empty regions, yielding significant inference speedups and resource savings, especially when only a percent-level fraction of pixels or fewer is active.
Our design operates on a fixed pixel budget that does not change between inputs, so the hardware runtime is input-independent and constant by construction.
On a representative example of MicroBooNE neutrino images with 4k input pixels, a sparse CNN that computes on at most 20 active pixels achieves a $\times 73$ inference speedup on an FPGA over a standard CNN with the same base architecture (133 vs. 9733 clock cycles; 0.665 vs. 48.665 $\mu$s), with $<2\%$ ROC AUC drop, and resource utilization well within device budgets.
For easy adoption, we provide a Python library supporting QAT of sparse CNNs and HLS C++ modules for FPGA deployment.
We hope this work benefits experiments operating under stringent sub-microsecond latency constraints, such as modern particle experiments, where standard CNNs are often infeasible due to their high latency, which would otherwise force substantial performance compromises.

\section*{Acknowledgments}
HFT and DSR are supported by the U.S. Department of Energy (DOE), Office of Science, Office of High Energy Physics subprogram on Computational High Energy Physics under Award No. DE-SC0026801.
DSR is also supported by the U.S. Department of Energy (DOE), Office of Science, Office of High Energy Physics Early Career Research program under Award No. DE-SC0025324.
PH is supported by the U.S. National Science Foundation (NSF) Harnessing the Data Revolution (HDR) Institute for Accelerated AI Algorithms for Data-Driven Discovery (A3D3) under Cooperative Agreement PHY-2117997.
PH is also supported by the Institute for Artificial Intelligence and Fundamental Interactions (IAIFI) under the NSF grant \#PHY-2019786.
We acknowledge the MicroBooNE Collaboration for making publicly available the datasets~\cite{abratenko_2022_7262009,abratenko_2022_7262140} employed in this work.
These data sets consist of simulated neutrino interactions from the Booster Neutrino Beamline overlaid on top of cosmic data collected with the MicroBooNE detector~\cite{MicroBooNE:2016pwy}.
This work used resources available through the National Research Platform (NRP) at the University of California, San Diego~\cite{10.1145/3708035.3736060}. NRP has been developed, and is supported in part, by funding from National Science Foundation, from awards 1730158, 1540112, 1541349, 1826967, 2112167, 2100237, and 2120019, as well as additional funding from community partners.
%ANONYMIZE

\bibliographystyle{naturemag}
\bibliography{references}

\clearpage

\appendix

\section{Appendix}

\subsection{Sparse input reduction and sparse convolution in HLS}
\label{sec:apd-input-conv}

The algorithmic details of the sparse input reduction and sparse convolution are given in Alg.~\ref{alg:sparse_input_reduce} and Alg.~\ref{alg:sparse_conv}, respectively.

\begin{algorithm}[!t]
    \caption{Sparse Input Reduction in HLS}
    \label{alg:sparse_input_reduce}
    
    \SetKwInOut{Input}{Inputs}
    \SetKwInOut{Output}{Outputs}
    \SetKwFunction{OpActive}{OpActive}
    \SetKwFunction{FindActive}{FindActive}
    \SetKwProg{Fn}{Function}{:}{end}
    \SetKwProg{Proc}{Procedure}{:}{end}
    
    \Input{
    flat channel-last input $X[0{..}HWC - 1]$,
    active pixel threshold $t$,
    max number of active pixels $N_{\text{active}}^{\text{max}}$
    }
    \Output{
    sparse feature array $a_{\text{feat}}[0{..}N_{\text{active}}^{\text{max}}\cdot C - 1]$,
    sparse hash array $a_{\text{hash}}[0{..}N_{\text{active}}^{\text{max}}\cdot 2 - 1]$
    }
    
    \Fn{\OpActive{$a,b;t$}\Hdr{$a,b$ are pairs $(\mathrm{getValue},\,\mathrm{getIndex})$}}{
      \If{$a.\mathrm{getValue} > t$}{\Return $a$}
      \ElseIf{$b.\mathrm{getValue} > t$}{\Return $b$}
      \Else{\Return $(0,0)$}
    }
    
    \Fn{\FindActive{$P[0{..}N-1],\, t$}\Hdr{returns leftmost active by recursive reduction}}{
      \If{$N = 1$}{\Return $P[0]$}
      \ElseIf{$N = 2$}{\Return \OpActive{$P[0],\, P[1],\, t$}}
      \Else{$N_{\text{left}} \gets 2^{\lfloor \log_2(N-1)\rfloor}$ \RC{largest power-of-two $< N$}
      $u \gets$ \FindActive{$P[0{..}N_{\mathrm{left}}-1],\, t$}\;
      $v \gets$ \FindActive{$P[N_{\mathrm{left}}{..}N-1],\, t$}\;
      \Return \OpActive{$u,\, v,\, t$}}
    }
    
    \Proc{SparseInputReduce\Hdr{keep up to $N_{\mathrm{active}}^{\text{max}}$ active pixels by scanning channel-0}}{
      \For{$j \gets 0$ \KwTo $HW - 1$}{
        $P[j].\mathrm{getValue} \gets X[C\cdot j]$;\quad
        $P[j].\mathrm{getIndex} \gets j$ \RC{prepare channel-0 data for scan}
        %$h[j] \gets 1 + \left\lfloor j/W \right\rfloor$;\quad
        %$w[j] \gets 1 + (j \bmod W)$ \RC{height,width coord. start from 1}
        $h[j] \gets \text{height for $j$-th pixel}$;\quad
        $w[j] \gets \text{width for $j$-th pixel}$
      }
    
      \For{$i \gets 0$ \KwTo $N_{\mathrm{active}}^{\text{max}} - 1$}{
        $p \gets$ \FindActive{$P[0{..}HW{-}1],\, t$}\;
          $a_{\text{feat}}[C\cdot i] \gets p.\mathrm{getValue}$ \RC{fill feature for first channel}
          \For{$c \gets 1$ \KwTo $C - 1$}{
            $a_{\text{feat}}[C\cdot i + c] \gets X[C\cdot p.\mathrm{getIndex} + c]$ \RC{fill feature for other channels}
          }
          $a_{\text{hash}}[2i] \gets h[p.\mathrm{getIndex}]$\RC{fill height}
          $a_{\text{hash}}[2i+1] \gets w[p.\mathrm{getIndex}]$\RC{fill width}
          $P[p.\mathrm{getIndex}].\mathrm{getValue} \gets 0$ \RC{after consuming, mask to reveal the next active}
      }
    }
\end{algorithm}

\begin{algorithm}[!t]
    \caption{Sparse 2D Convolution in HLS}
    \label{alg:sparse_conv}
    
    \SetKwInOut{Input}{Inputs}
    \SetKwInOut{Output}{Outputs}
    \SetKwFunction{MultAtOffset}{MultAtOffset}
    \SetKwProg{Fn}{Function}{:}{end}
    \SetKwProg{Proc}{Procedure}{:}{end}
    
    \Input{
    sparse feature array $a_{\text{feat}}^{\text{in}}[0{..}N_{\text{active}}^{\max}\cdot C_{\text{in}} - 1]$,
    sparse hash array $a_{\text{hash}}[0{..}N_{\text{active}}^{\max}\cdot 2 - 1]$,
    weights $W[0{..}K^2 C_{\text{in}} C_{\text{out}} - 1]$,
    bias $b[0{..}C_{\text{out}} - 1]$,
    odd kernel size $K$
    }
    \Output{
    sparse feature array $a_{\text{feat}}^{\text{out}}[0{..}N_{\text{active}}^{\max}\cdot C_{\text{out}} - 1]$,
    $same$ $a_{\text{hash}}$ \RC{unchanged active pixel set}
    }
    
    \Fn{\MultAtOffset{$\Delta h,\ \Delta w,\ c_{\mathrm{out}},\ p_{\mathrm{in}}$}\Hdr{Mult-accum based on offset in $K^2$ field}}{
      $R \gets (K-1)/2$\;
      \If{$|\Delta h|>R$ \textbf{or} $|\Delta w|>R$}{
        \Return $0$ \RC{offset is outside kernel field}
      }
      \Else{
      \LC{map offset to weight array index}
      $pos \gets (R - \Delta h)\cdot K + (R - \Delta w)$ \RC{flattened spatial position $0{..}K^2-1$}
      $w\_\text{idx} \gets C_{\text{out}}\cdot C_{\text{in}}\cdot pos + C_{\text{in}}\cdot c_{\text{out}}$ \RC{weight array ordering: pixel\_pos$\rightarrow$ch\_out$\rightarrow$ch\_in}
      $M \gets 0$\;
      \For{$c_{\text{in}} \gets 0$ \KwTo $C_{\text{in}}-1$}{
        $M \gets M + W[w\_\text{idx} + c_{\text{in}}]\cdot a_{\text{feat}}^{\text{in}}[C_{\text{in}}\cdot p_{\text{in}} + c_{\text{in}}]$ \RC{dot over input channels}
      }
      \Return $M$}
    }
    
    \Proc{SparseConv\Hdr{convolve on sparse arrays ($N_{\mathrm{active}}^{\max}\cdot N_{\mathrm{active}}^{\max}\cdot C_{\mathrm{in}}\cdot C_{\mathrm{out}}$ total iterations)}}{
        \LC{loop over output pixels}
      \For{$p_{\text{out}} \gets 0$ \KwTo $N_{\mathrm{active}}^{\max}-1$}{
      \LC{loop over output channels}
        \For{$c_{\mathrm{out}} \gets 0$ \KwTo $C_{\mathrm{out}}-1$}{
          $A \gets 0$\;
          \LC{loop over input pixels}
          \For{$p_{\text{in}} \gets 0$ \KwTo $N_{\mathrm{active}}^{\max}-1$}{
            $\Delta h \gets a_{\text{hash}}[2\,p_{\text{out}}]   - a_{\text{hash}}[2\,p_{\text{in}}]$\RC{compute offset in height}
            $\Delta w \gets a_{\text{hash}}[2\,p_{\text{out}}+1] - a_{\text{hash}}[2\,p_{\text{in}}+1]$\RC{compute offset in width}
            $A \gets A \,+ $ \MultAtOffset{$\Delta h,\, \Delta w,\, c_{\mathrm{out}},\, p_{\mathrm{in}}$}\RC{offset-based mult-accum}
          }
          $A \gets A + b[c_{\text{out}}]$\;
          \If{$\mathrm{pixel\,}p_{\mathrm{out}}\mathrm{\,is\,padded}$}{$A \gets 0$}
          $a_{\text{feat}}^{\text{out}}[C_{\text{out}}\cdot p_{\text{out}} + c_{\text{out}}] \gets A$
        }
      }
    }
\end{algorithm}

\subsection{Sparse activation in HLS and scaling}
\label{sec:apd-activation}

\begin{figure}[!t]
    \centering
    \includegraphics[width=0.35\textwidth]{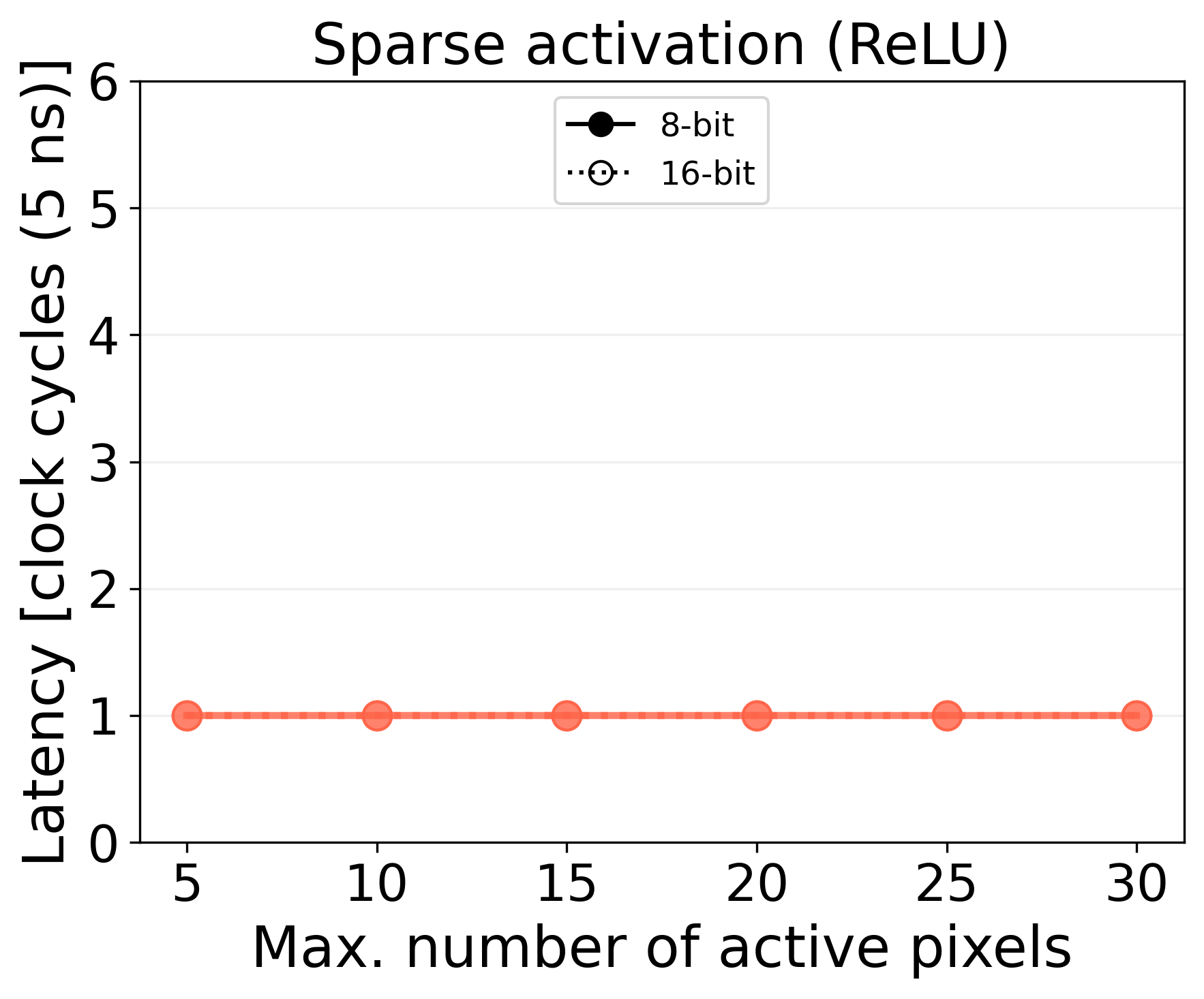}
    \includegraphics[width=0.35\textwidth]{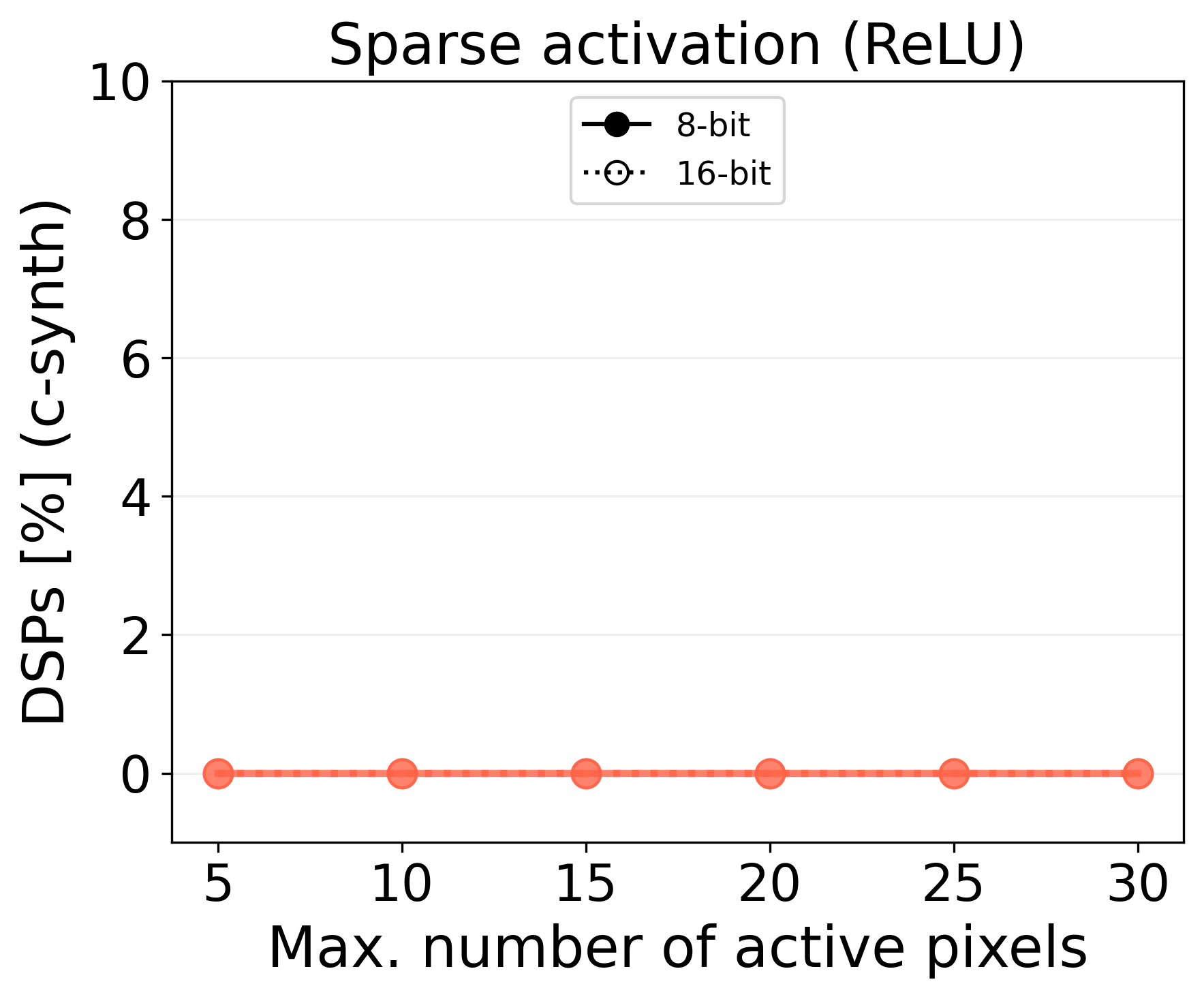}\\
    \includegraphics[width=0.35\textwidth]{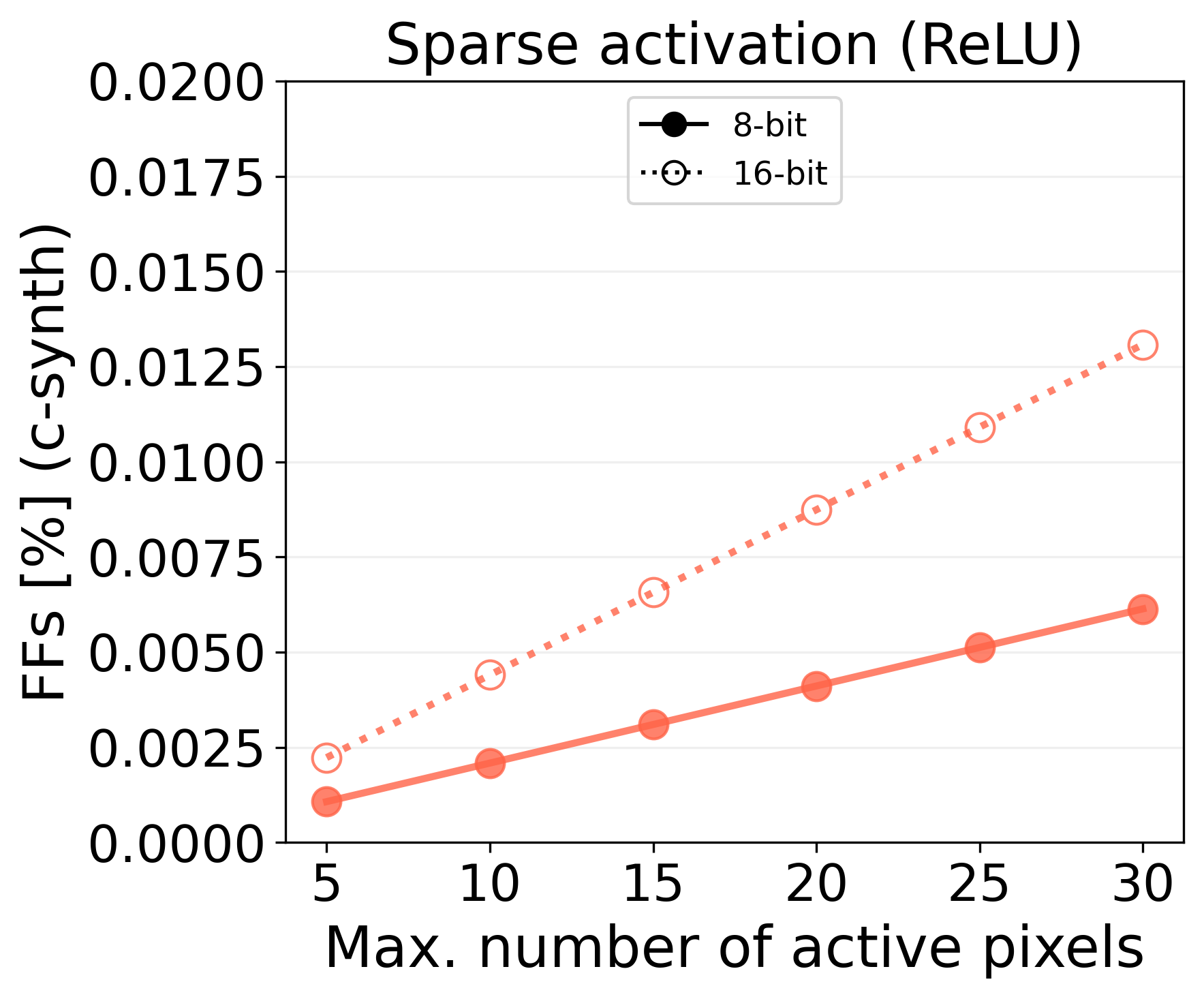}
    \includegraphics[width=0.35\textwidth]{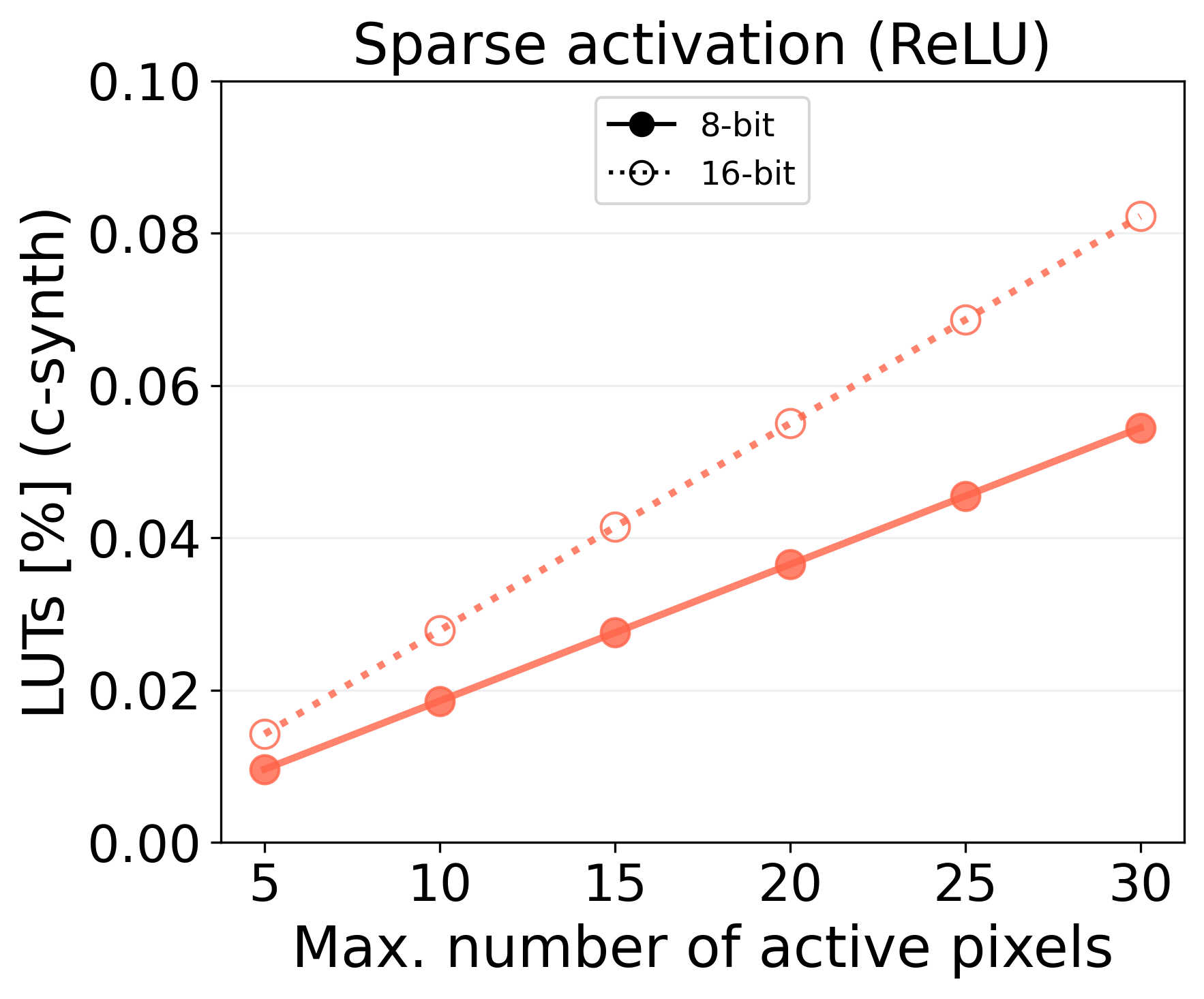}
    \caption{Scaling of the sparse activation layer on an FPGA for different $N_{\text{active}}^{\text{max}}$: latency (upper left), DSPs (upper right), FFs (lower left), and LUTs (lower right). For latency, 1 clock cycle is equivalent to 5 ns. Resource utilization is obtained from the HLS C-synthesis step.}
    \label{fig:scaling_relu}
\end{figure}

The algorithmic details are given in Alg.~\ref{alg:sparse_activation}.
For scaling, we use ReLU and scan $N_{\text{active}}^{\text{max}}$ from 5 to 30 in steps of 5.
Fig.~\ref{fig:scaling_relu} reports latency and resource utilization.
The activation loop is fully unrolled, so the layer finishes in one clock cycle and uses no DSPs. 
FF/LUT usage increases with $N_{\text{active}}^{\text{max}}$ because more elements are processed in parallel.

\begin{algorithm}[!t]
\caption{Sparse Activation in HLS}
\label{alg:sparse_activation}

\SetKwInOut{Input}{Inputs}
\SetKwInOut{Output}{Outputs}
\SetKwFunction{Act}{Act}
\SetKwProg{Fn}{Function}{:}{end}
\SetKwProg{Proc}{Procedure}{:}{end}

\Input{
sparse feature array $a_{\text{feat}}^{\text{in}}[0{..}N_{\text{active}}^{\max}\cdot C - 1]$
}
\Output{
sparse feature array $a_{\text{feat}}^{\text{out}}[0{..}N_{\text{active}}^{\max}\cdot C - 1]$
}

\Proc{SparseActivation\Hdr{element-wise activation on sparse features}}{
  \For{$i \gets 0$ \KwTo $N_{\mathrm{active}}^{\max}\cdot C-1$}{
    $a_{\text{feat}}^{\text{out}}[i] \gets$ \Act{$a_{\mathrm{feat}}^{\mathrm{in}}[i]$}
  }
}
\end{algorithm}

\subsection{Sparse pooling in HLS and scaling}
\label{sec:apd-pooling}

\begin{figure}[!t]
    \centering
    \includegraphics[width=0.35\textwidth]{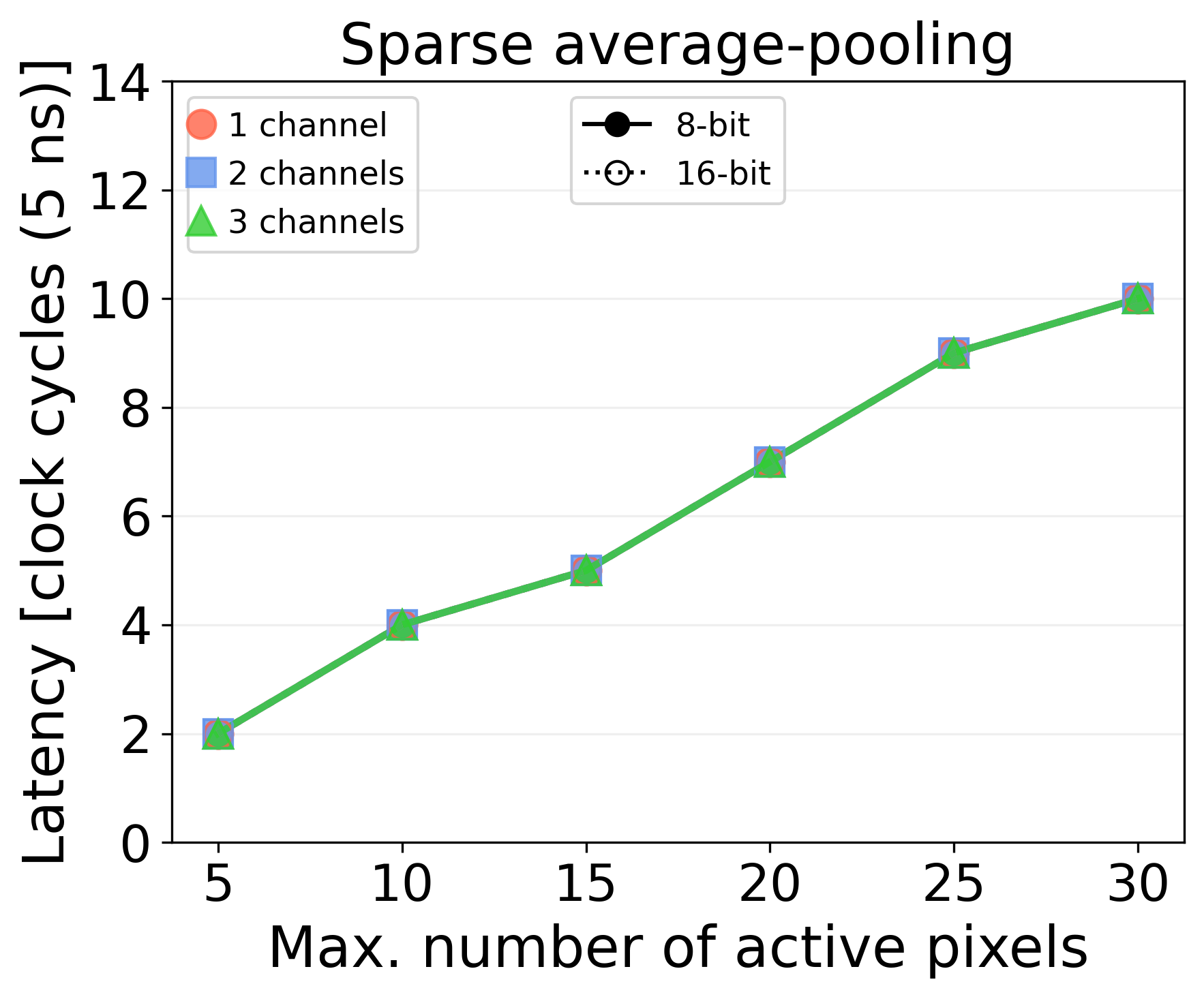}
    \includegraphics[width=0.35\textwidth]{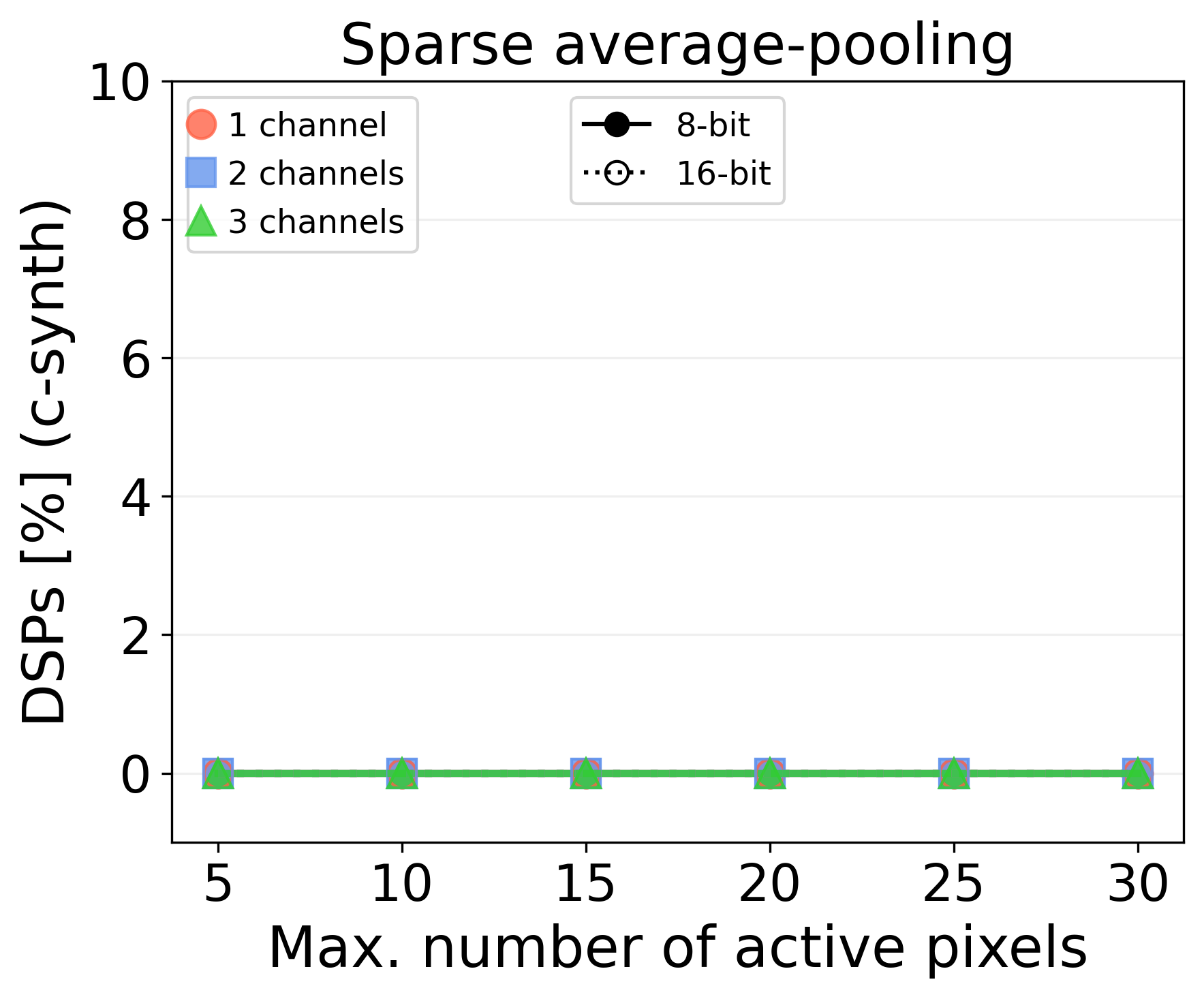}\\
    \includegraphics[width=0.35\textwidth]{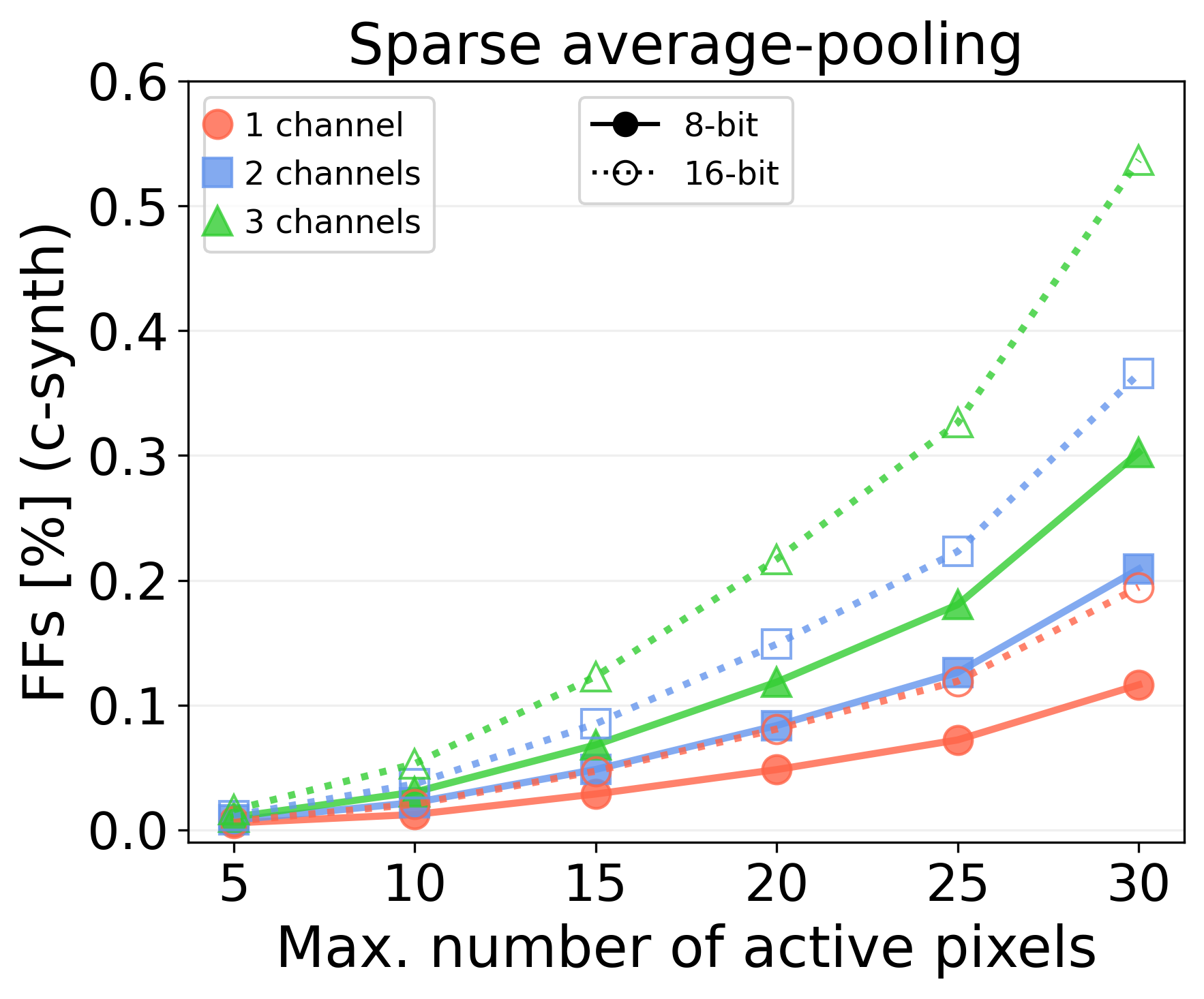}
    \includegraphics[width=0.35\textwidth]{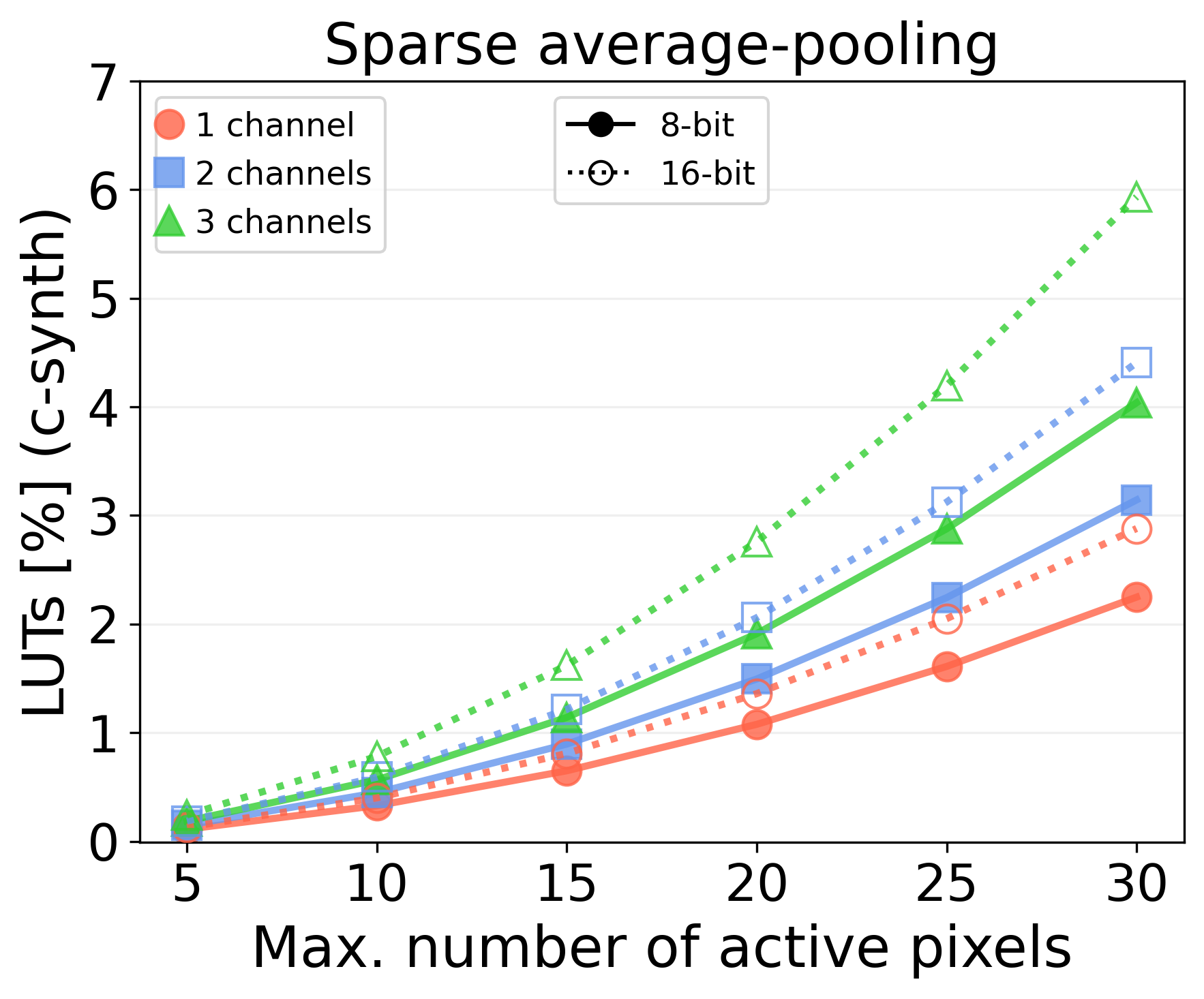}
    \caption{Scaling of the sparse pooling layer on an FPGA for different number of channels and $N_{\text{active}}^{\text{max}}$: latency (upper left), DSPs (upper right), FFs (lower left), and LUTs (lower right). For latency, 1 clock cycle is equivalent to 5 ns. Resource utilization is obtained from the HLS C-synthesis step.}
    \label{fig:scaling_pooling}
\end{figure}

The algorithmic details are given in Alg.~\ref{alg:sparse_pool}.
For scaling, we use average pooling as an example, test $C_{\text{in}}\in\{1,2,3\}$, and scan $N_{\text{active}}^{\text{max}}$ from 5 to 30 in steps of 5.
Fig.~\ref{fig:scaling_pooling} shows latency and resource utilization.
The latency curves across $C_{\text{in}}$ coincide because the channel loop is unrolled.
No DSPs are used; FF/LUT usage increases with $N_{\text{active}}^{\text{max}}$ and $C_{\text{in}}$ due to more parallel comparisons and accumulators.

\begin{algorithm}[!t]
    \caption{Sparse Pooling (Avg) in HLS}
    \label{alg:sparse_pool}
    
    \SetKwInOut{Input}{Inputs}
    \SetKwInOut{Output}{Outputs}
    \SetKwFunction{PoolMap}{PoolMap}
    \SetKwProg{Fn}{Function}{:}{end}
    \SetKwProg{Proc}{Procedure}{:}{end}
    
    \Input{
    sparse feature array $a_{\text{feat}}^{\text{in}}[0{..}N_{\text{active}}^{\max}\cdot C - 1]$,
    sparse hash array $a_{\text{hash}}^{\text{in}}[0{..}N_{\text{active}}^{\max}\cdot 2 - 1]$,
    pool size $P$
    }
    \Output{
    sparse feature array $a_{\text{feat}}^{\text{out}}[0{..}N_{\text{active}}^{\max}\cdot C - 1]$,
    sparse hash array $a_{\text{hash}}^{\text{out}}[0{..}N_{\text{active}}^{\max}\cdot 2 - 1]$
    }

    \Proc{SparseAvgPool}{
      \LC{compute pooled coords and fill output hash}
      \For{$i \gets 0$ \KwTo $N_{\text{active}}^{\max}-1$}{
        $a_{\mathrm{hash}}^{\text{out}}[2i] \gets \lfloor (a_{\mathrm{hash}}^{\text{in}}[2i]-1)/P \rfloor + 1$\;
        $a_{\mathrm{hash}}^{\text{out}}[2i+1] \gets \lfloor (a_{\mathrm{hash}}^{\text{in}}[2i+1]-1)/P \rfloor + 1$
      }
    
      \LC{loop over output hash}
      \For{$i \gets 0$ \KwTo $N_{\text{active}}^{\max}-1$}{
        \LC{loop over channels}
        \For{$c \gets 0$ \KwTo $C-1$}{
          $A \gets 0$\;
          \LC{collect pixels that map to same pool}
          \For{$j \gets 0$ \KwTo $N_{\text{active}}^{\max}-1$}{
            \If{$a_{\mathrm{hash}}^{\mathrm{out}}[2j] = a_{\mathrm{hash}}^{\mathrm{out}}[2i]$ \textbf{and} $a_{\mathrm{hash}}^{\mathrm{out}}[2j+1] = a_{\mathrm{hash}}^{\mathrm{out}}[2i+1]$}{
              $A \gets A + a_{\text{feat}}^{\text{in}}[C\cdot j + c]$\;
              $a_{\text{feat}}^{\text{in}}[C\cdot j + c] \gets 0$ \RC{after consuming, mask to avoid double counting}
            }
          }
          $a_{\text{feat}}^{\text{out}}[C\cdot i + c] \gets A / P^2$
        }
      }
    }
\end{algorithm}

\subsection{Sparse flatten in HLS and scaling}
\label{sec:apd-flatten}

\begin{figure}[!t]
    \centering
    \includegraphics[width=0.35\textwidth]{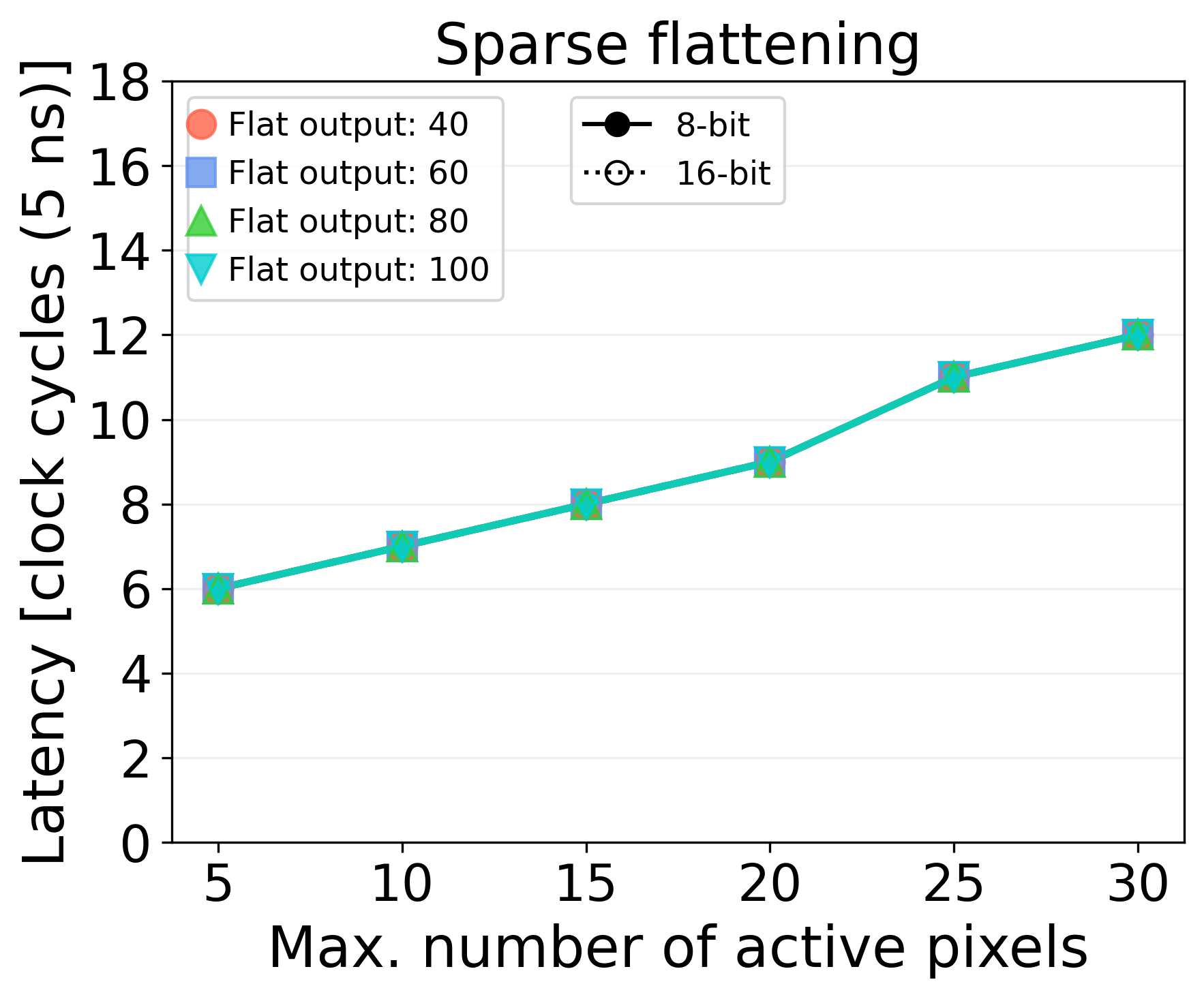}
    \includegraphics[width=0.35\textwidth]{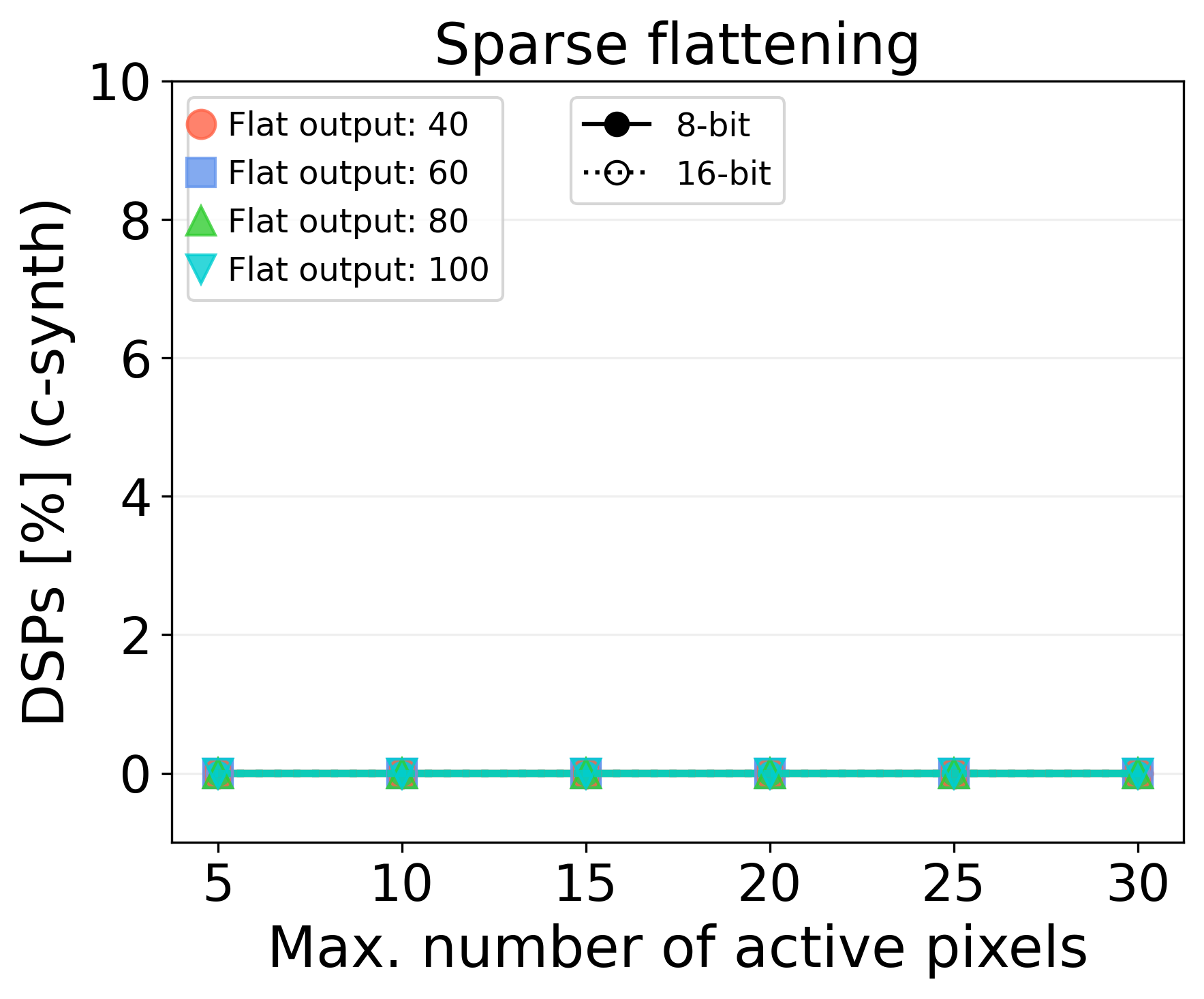}\\
    \includegraphics[width=0.35\textwidth]{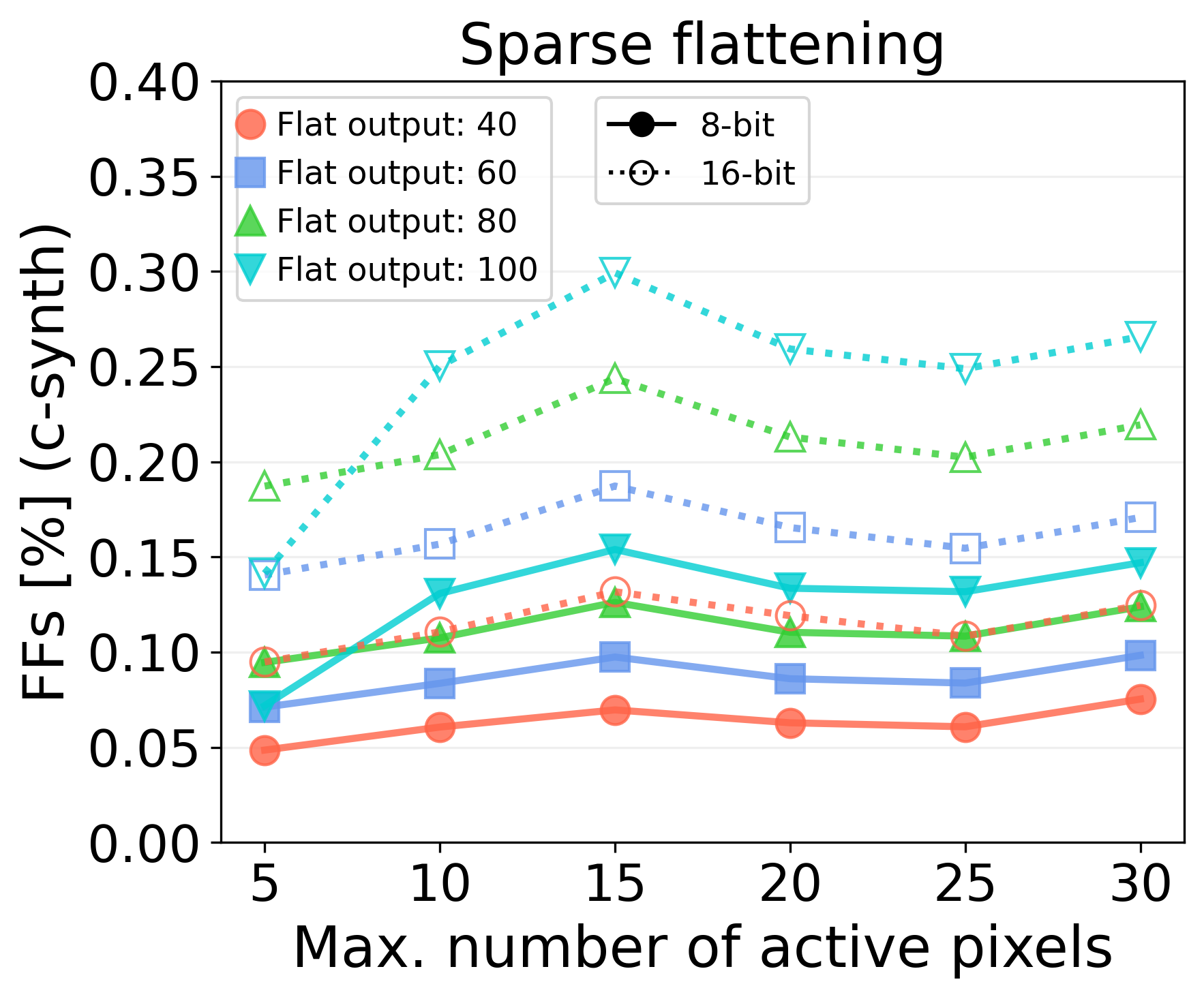}
    \includegraphics[width=0.35\textwidth]{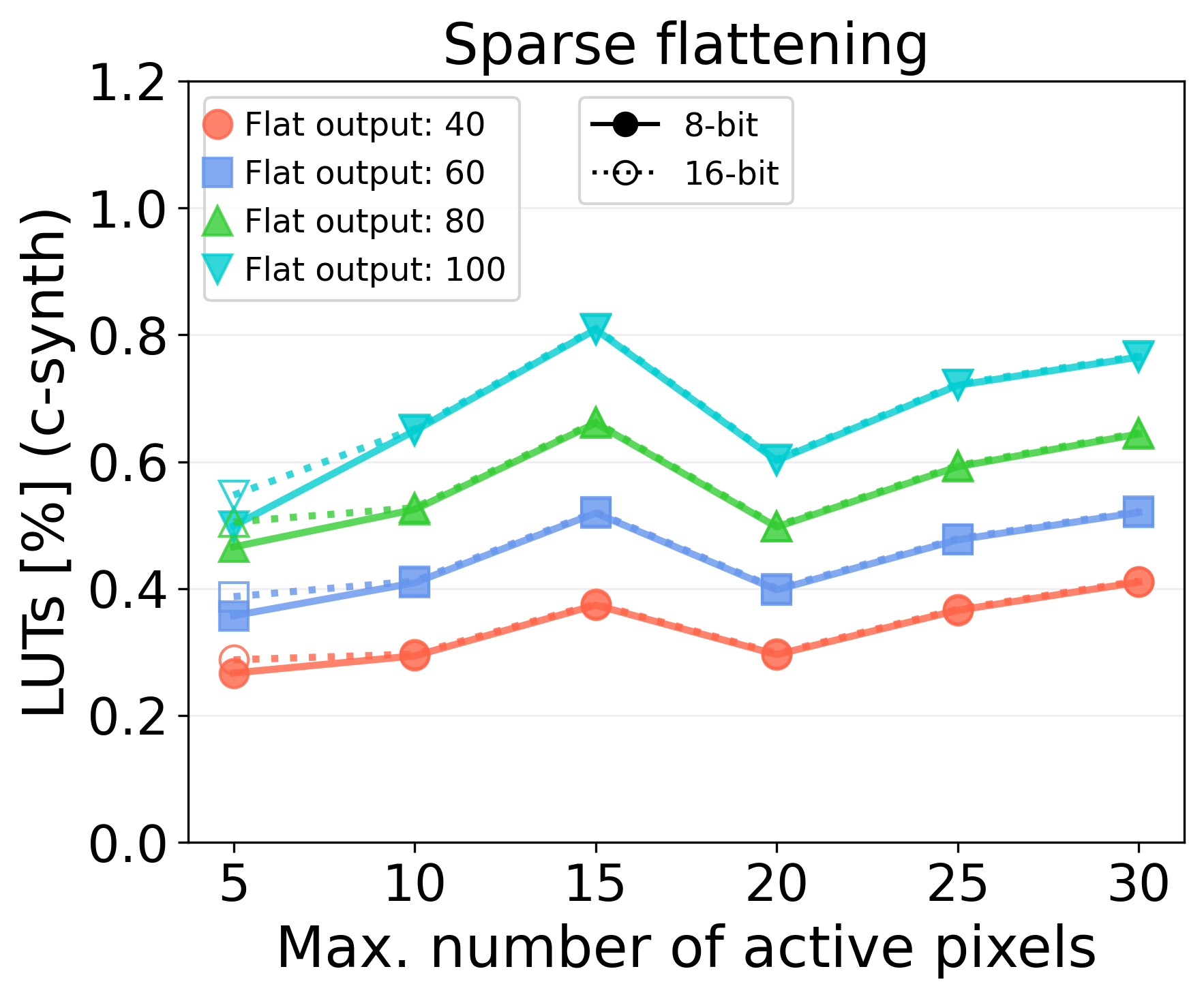}
    \caption{Scaling of the sparse flattening layer on an FPGA for different flat output array size and $N_{\text{active}}^{\text{max}}$: latency (upper left), DSPs (upper right), FFs (lower left), and LUTs (lower right). For latency, 1 clock cycle is equivalent to 5 ns. Resource utilization is obtained from the HLS C-synthesis step.}
    \label{fig:scaling_flatten}
\end{figure}

The algorithmic details are given in Alg.~\ref{alg:sparse_flatten}.
For scaling, we test flat output sizes $HWC\in\{40, 60, 80, 100\}$ and scan $N_{\text{active}}^{\text{max}}$ from 5 to 30 in steps of 5.
Fig.~\ref{fig:scaling_flatten} shows latency and resource utilization.
The latency is essentially independent of the small flat output size because $HWC$ initialization is fully unrolled.
Because this is address generation plus writes, DSP usage is 0, while FF/LUT usage increases with $N_{\text{active}}^{\text{max}}$ and the flat output size $HWC$.

\begin{algorithm}[!t]
    \caption{Sparse Flatten in HLS}
    \label{alg:sparse_flatten}
    
    \SetKwInOut{Input}{Inputs}
    \SetKwInOut{Output}{Outputs}
    \SetKwFunction{PixIndex}{PixIndex}
    \SetKwProg{Fn}{Function}{:}{end}
    \SetKwProg{Proc}{Procedure}{:}{end}
    
    \Input{
    sparse feature array $a_{\text{feat}}^{\text{in}}[0{..}N_{\text{active}}^{\max}\cdot C - 1]$,
    sparse hash array $a_{\text{hash}}^{\text{in}}[0{..}N_{\text{active}}^{\max}\cdot 2 - 1]$,
    dim $(H,W,C)$
    }
    \Output{
    flat dense array $X[0{..}HWC - 1]$
    }
    
    \Proc{SparseFlatten\Hdr{scatter sparse pixels into flat dense array (channel-last)}}{
      \For{$j \gets 0$ \KwTo $HWC - 1$}{$X[j] \gets 0$ \RC{zero-initialize}}
    
      \For{$i \gets 0$ \KwTo $N_{\text{active}}^{\max} - 1$}{
        $\text{pix} \gets (a_{\text{hash}}^{\text{in}}[2i] - 1)\cdot W + (a_{\text{hash}}^{\text{in}}[2i+1] - 1)$\RC{map coord to flat dense position}
        \For{$c \gets 0$ \KwTo $C - 1$}{
          $X[C\cdot \text{pix} + c] \gets a_{\text{feat}}^{\text{in}}[C\cdot i + c]$ \RC{scatter}
        }
      }
    }
\end{algorithm}

\subsection{Evaluation on synthetic datasets (MNIST and jet tagging)}
\label{sec:apd-mnist-jet}

In this section we evaluate on two synthetic datasets following the same setup from the neutrino evaluation in Sec.~\ref{sec:exp-eval-setup}.
The datasets here are artificially sparsified for demonstration purposes.

\begin{figure}[!t]
    \centering
    \includegraphics[width=0.45\textwidth]{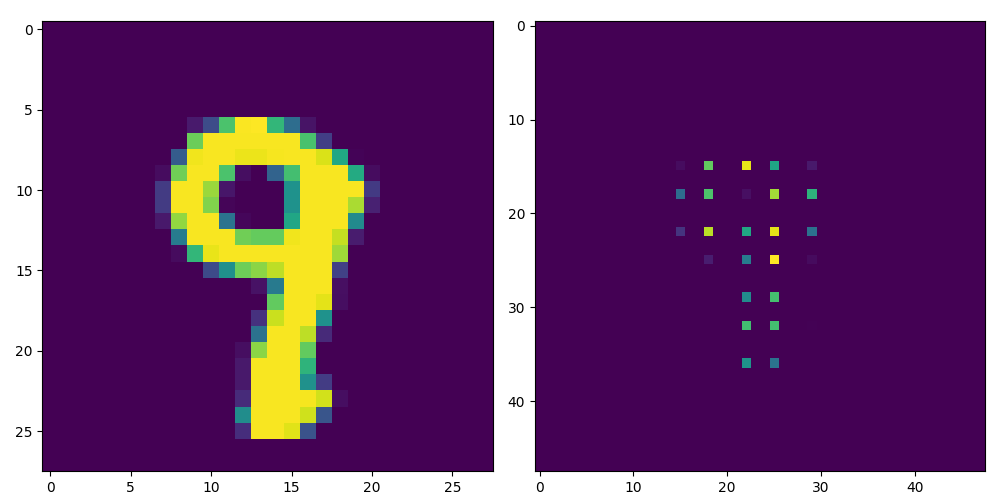}
    \caption{Example MNIST image before (left) and after (right) preprocessing.}
    \label{fig:mnist_input}
\end{figure}

\begin{figure}[!t]
    \centering
    \includegraphics[width=0.8\textwidth]{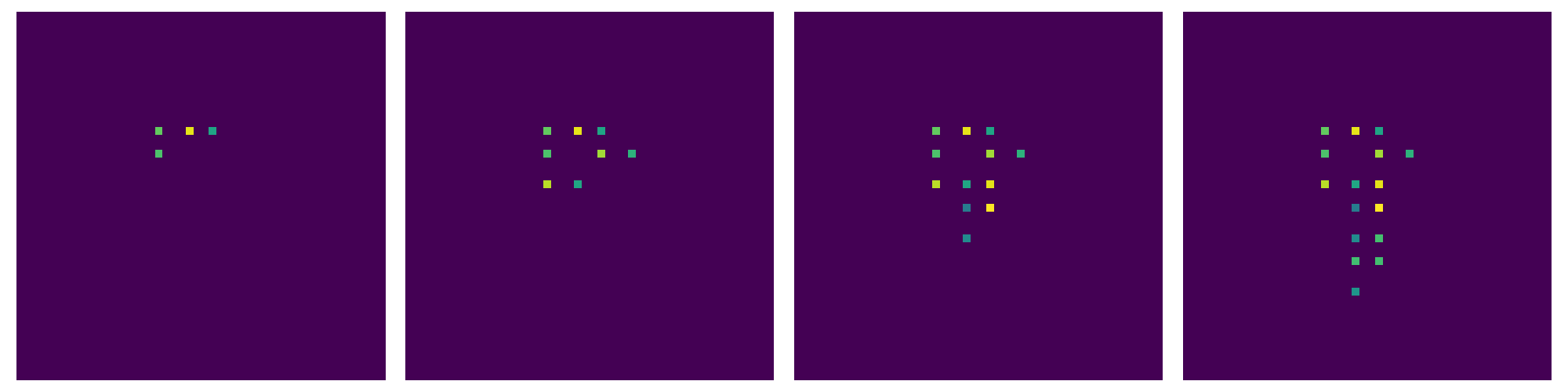}
    \caption{Effect of $N_{\text{active}}^{\text{max}}$ on retained semantics. From left to right: $N_{\text{active}}^{\text{max}}=4,8,12,16$.}
    \label{fig:mnist_leak}
\end{figure}

We first inspect a toy dataset derived from MNIST~\cite{lecun-mnisthandwrittendigit-2010}.
The original images have $28\times 28$ pixels and are already somewhat sparse, but we target sparsity at the few-percent level or below.
For demonstration, we further sparsify the features with the following transformations: (1) apply average pooling with pool size 3 to reduce the number of nonzeros; (2) zero-pad the borders to resize to $48\times 48$; (3) radially ``inflate'' pixels outward proportional to their distance from the image center, so the active pixels are more separated spatially (this avoids over-compression by later pooling); (4) zero out low-valued pixels with a threshold at 0.4.
An example image is shown in Fig.~\ref{fig:mnist_input}.
The task we consider is supervised multi-class classification of the processed images.
We also illustrate the importance of choosing a proper $N_{\text{active}}^{\text{max}}$, a hyperparameter of our sparse CNNs.
As described in Sec.~\ref{sec:method-hls-input}, $N_{\text{active}}^{\text{max}}$ caps the number of active pixels retained in sparse arrays which are subsequently processed by sparse convolution.
Pixels are scanned in a row-major order (top left to bottom right).
A larger $N_{\text{active}}^{\text{max}}$ preserves more information but increases hardware cost, whereas too small a value leads to semantic information loss in a way that depends on the spatial sparsity of a given dataset.
In Fig.~\ref{fig:mnist_leak}, using the digit ``9'' as an example, $N_{\text{active}}^{\text{max}}=4$ is insufficient to resemble any class; 8 captures only the upper half which resembles the digit ``0''; 12 begins to include the lower part; and 16 recovers the full ``9''.
Thus an $N_{\text{active}}^{\text{max}}$ scan is necessary to balance accuracy and hardware cost for a given dataset.

\begin{figure}[!t]
    \centering
    \includegraphics[width=0.85\textwidth]{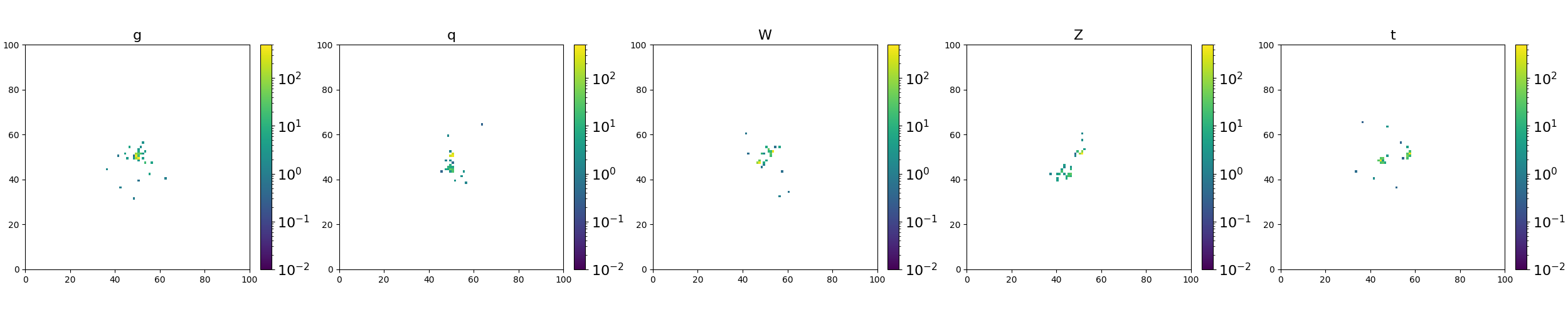}\\
    \includegraphics[width=0.85\textwidth]{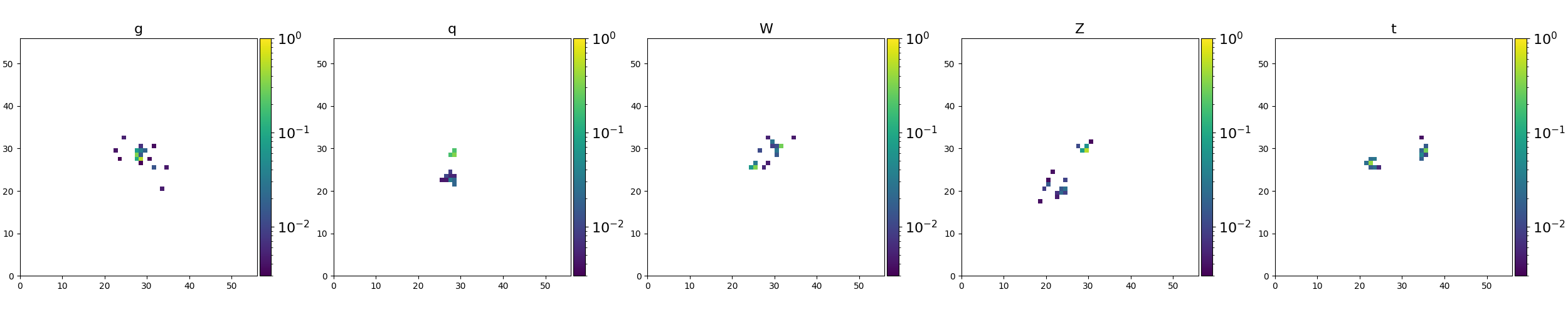}\\
    \includegraphics[width=0.85\textwidth]{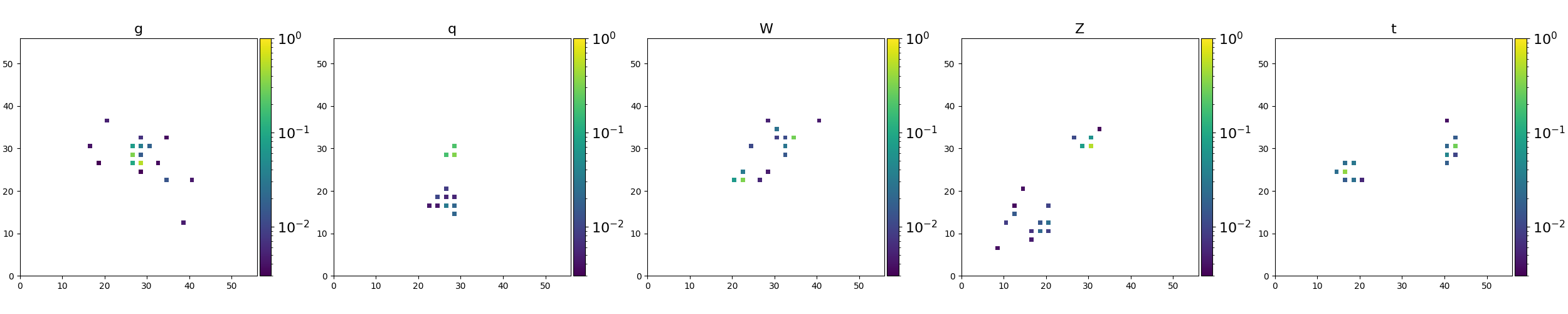}
    \caption{Example jet images for classes g, q, W, Z, and t (left to right). Top: original $100\times 100$ images. Middle: cropped to $56\times 56$ with low-energy hits zeroed. Bottom: sparsified by radial inflation.}
    \label{fig:jet_input}
\end{figure}

Lastly, we consider jet flavor tagging at the LHC~\cite{Moreno:2019bmu,Duarte:2018ite,pierini_2020_3602260}.
In proton-proton collisions, unstable heavy particles are produced and they decay promptly into sprays of particles recorded by the detector.
When the parent particle has high momentum, the constituent particles appear as a collimated cone, called a jet.
Jet substructure (particle content, energy, and geometry) encodes the identity of the parent particle, and identifying the originating particle is called jet tagging.
The dataset we use contains simulated jets with transverse momentum ($p_{\text{T}}$) around 1 TeV produced from proton-proton collisions at the center-of-mass energy of 13 TeV.
Jets are clustered with the anti-$k_{\text{T}}$ algorithm~\cite{Cacciari:2008gp} using the distance parameter at $R=0.8$.
Five jet classes are simulated: gluon (g), light quark (q), W boson (W), Z boson (Z), and top quark (t).
Each jet is represented as a $100\times 100$ image in pseudorapidity $\eta$ (transformed polar angle) and azimuthal angle $\phi$ relative to the jet axis; each pixel stores the scalar sum of particle $p_{\text{T}}$ measured in calorimeters, and up to the 150 highest-$p_{\text{T}}$ constituents are retained per jet.
Example images are shown in the top row of Fig.~\ref{fig:jet_input}.
The task is supervised multi-class classification.
Because the original images are too large for full parallelization on an FPGA, we preprocess the images for demonstration: (1) crop from $100\times 100$ to $56\times 56$ by removing border pixels; (2) normalize pixel values by a factor of 1200, saturate at 1, and cut at 0.003 (middle row in Fig.~\ref{fig:jet_input}); (3) as with MNIST, apply radial inflation to spread active pixels outward from the center so that subsequent pooling does not easily collapse information (bottom row in Fig.~\ref{fig:jet_input}).

\begin{figure}[!t]
    \centering
    \includegraphics[width=0.9\textwidth]{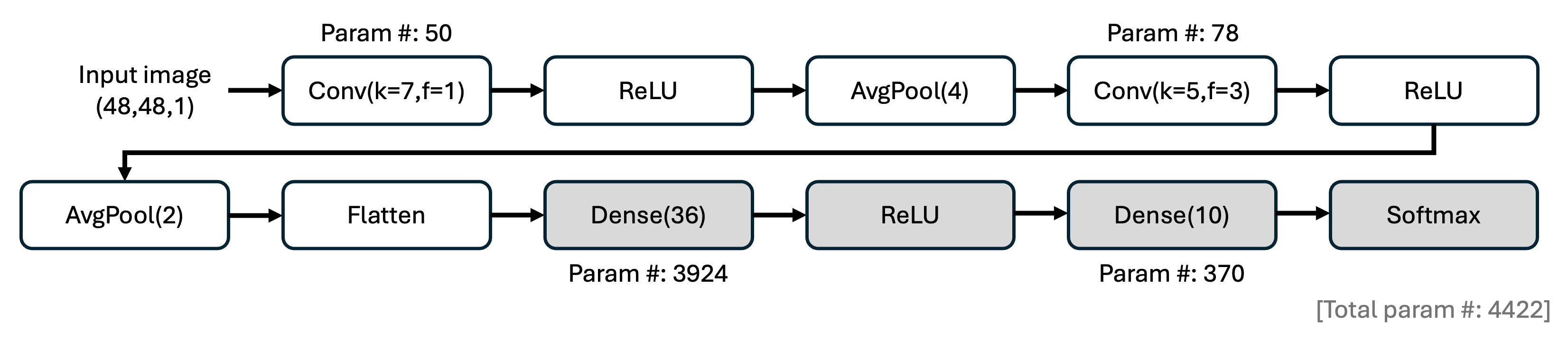}
    \includegraphics[width=0.9\textwidth]{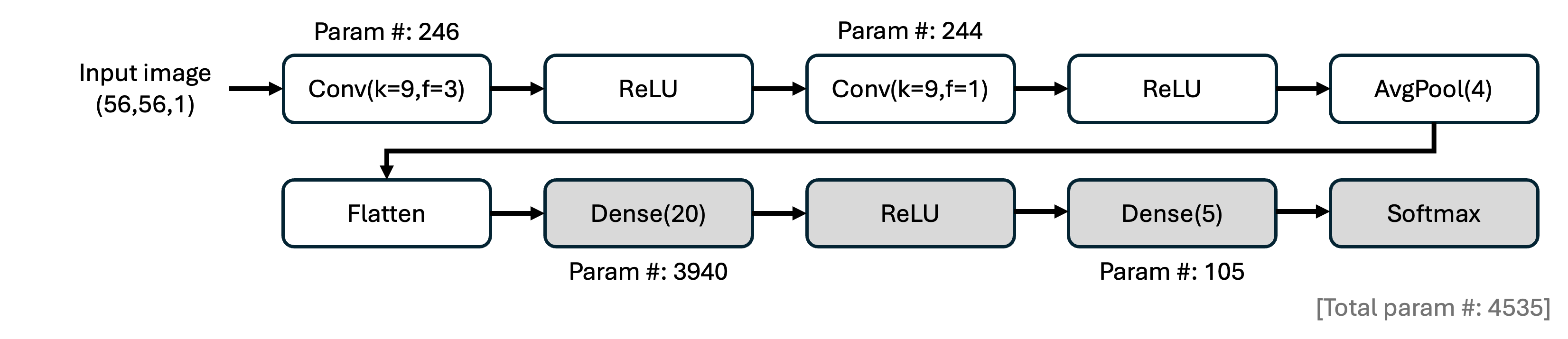}
    \caption{Base architectures for constructing both standard and sparse CNNs for the MNIST (top) and jet tagging (bottom) experiments. All have two convolutional blocks followed by a 2-layer MLP, with a compact model size of around 4k parameters.}
    \label{fig:models_mnist_jet}
\end{figure}

\begin{figure}[!t]
    \centering
    \includegraphics[width=0.45\textwidth]{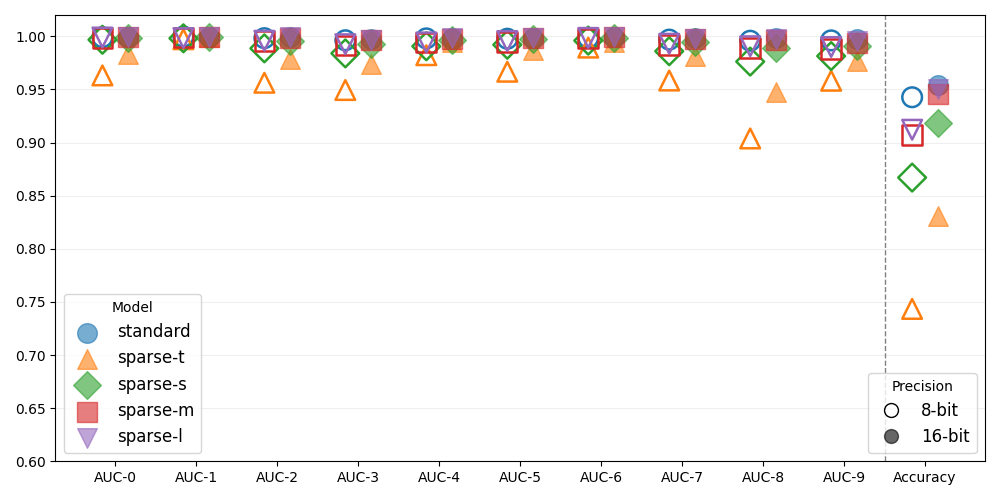}
    \includegraphics[width=0.45\textwidth]{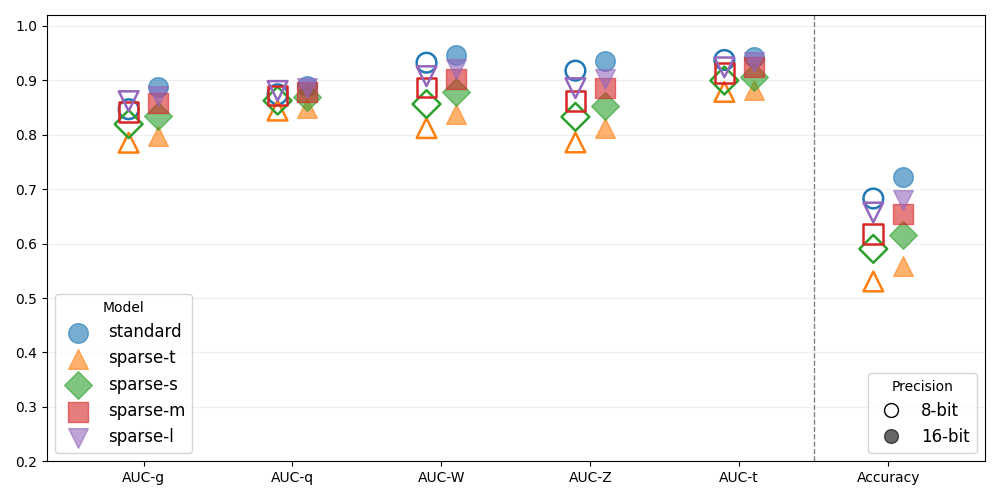}
    \caption{Model performance comparing the standard CNNs and the sparse CNNs for MNIST (left) and jet tagging (right).}
    \label{fig:performance_mnist_jet}
\end{figure}

\begin{figure}[!t]
    \centering
    \includegraphics[width=0.95\textwidth]{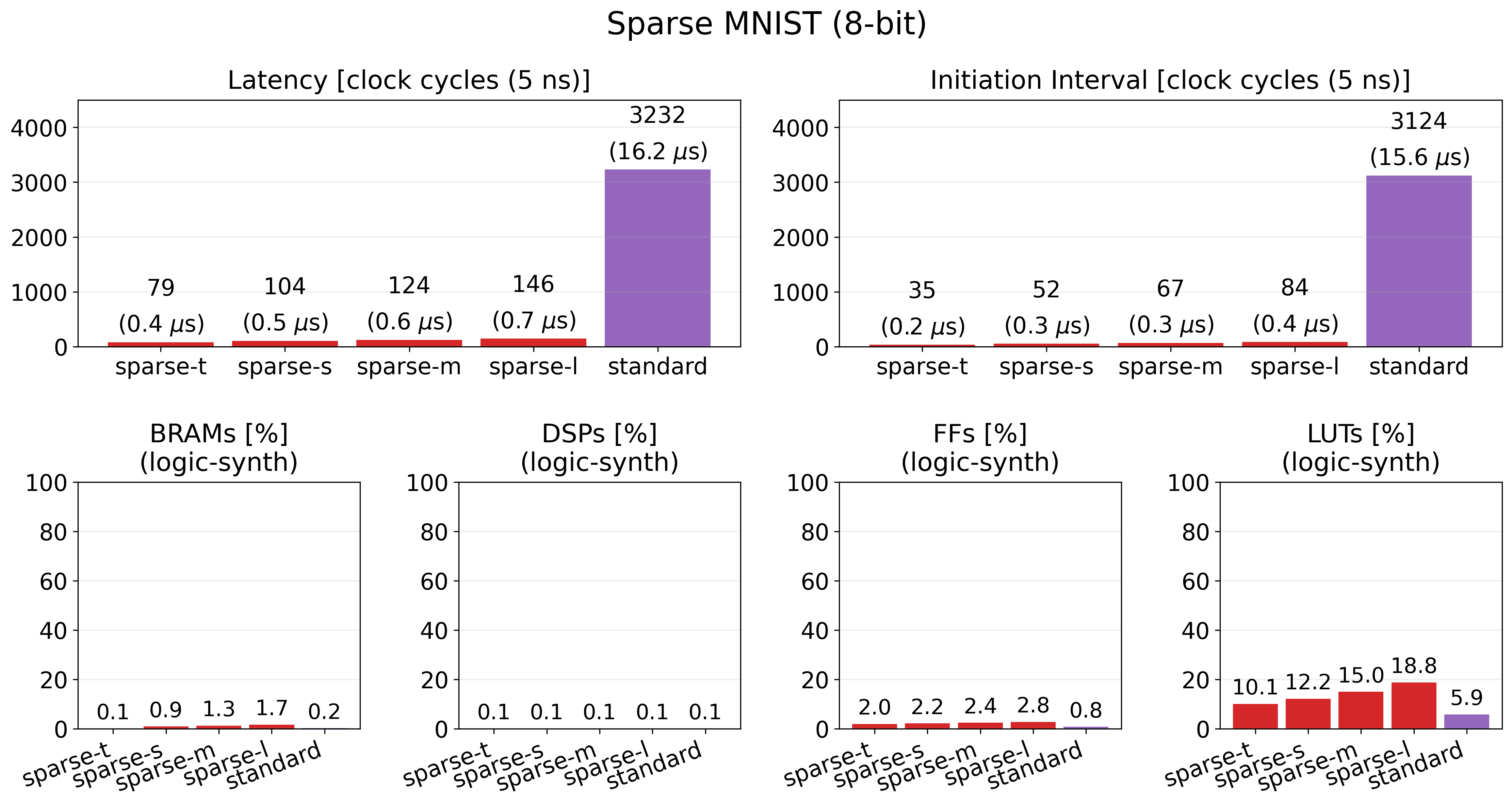}\\
    \includegraphics[width=0.95\textwidth]{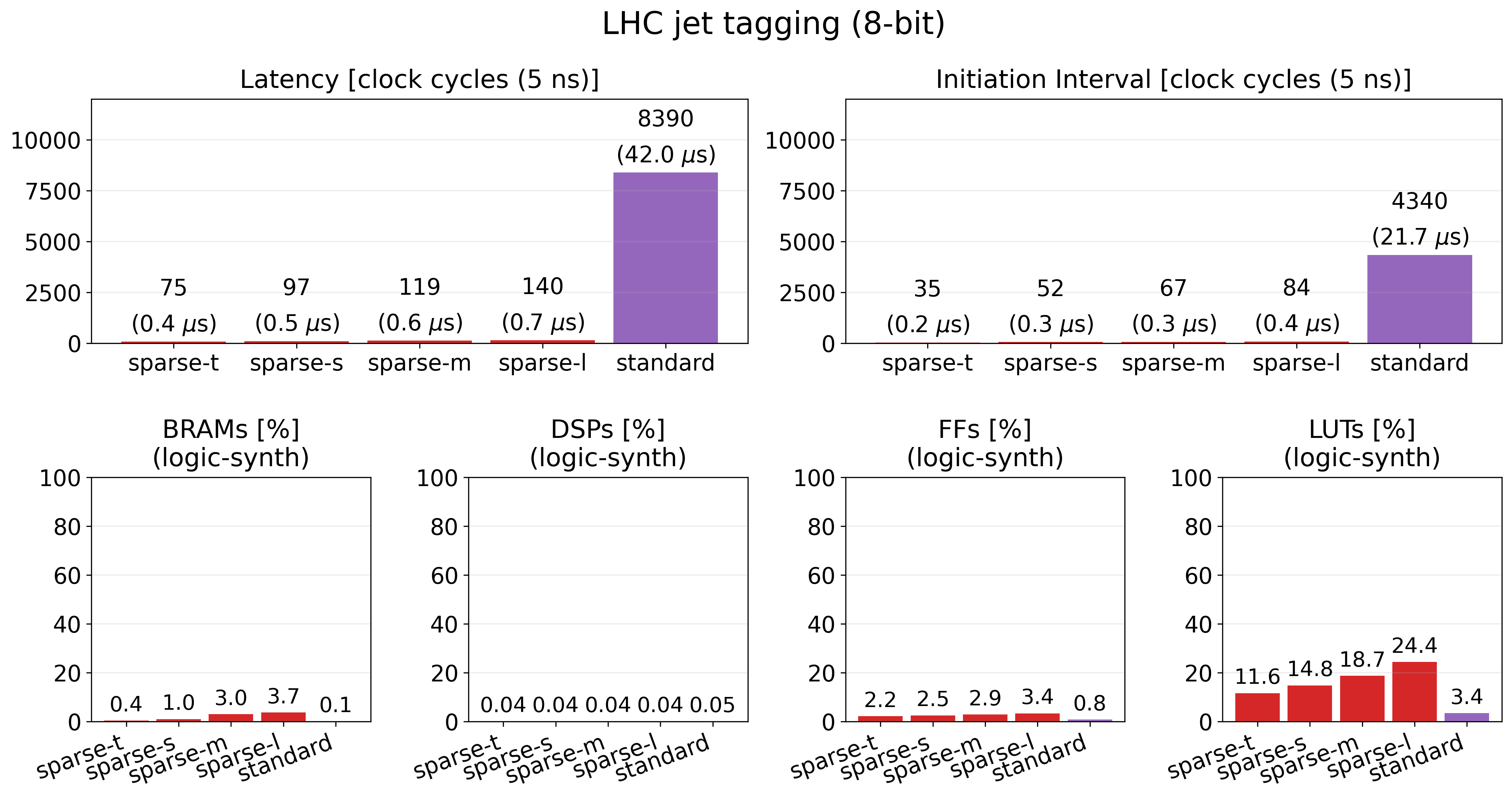}
    \caption{Synthesis results of standard (purple) and sparse (red) CNNs at 8-bit for MNIST (top) and jet tagging (bottom). Latency and II are measured in clock cycles (cc) with 5 ns clock period. Resource utilization is obtained after the logic synthesis step.}
    \label{fig:bar8b_mnist_jet}
\end{figure}

\begin{figure}[!t]
    \centering
    \includegraphics[width=0.95\textwidth]{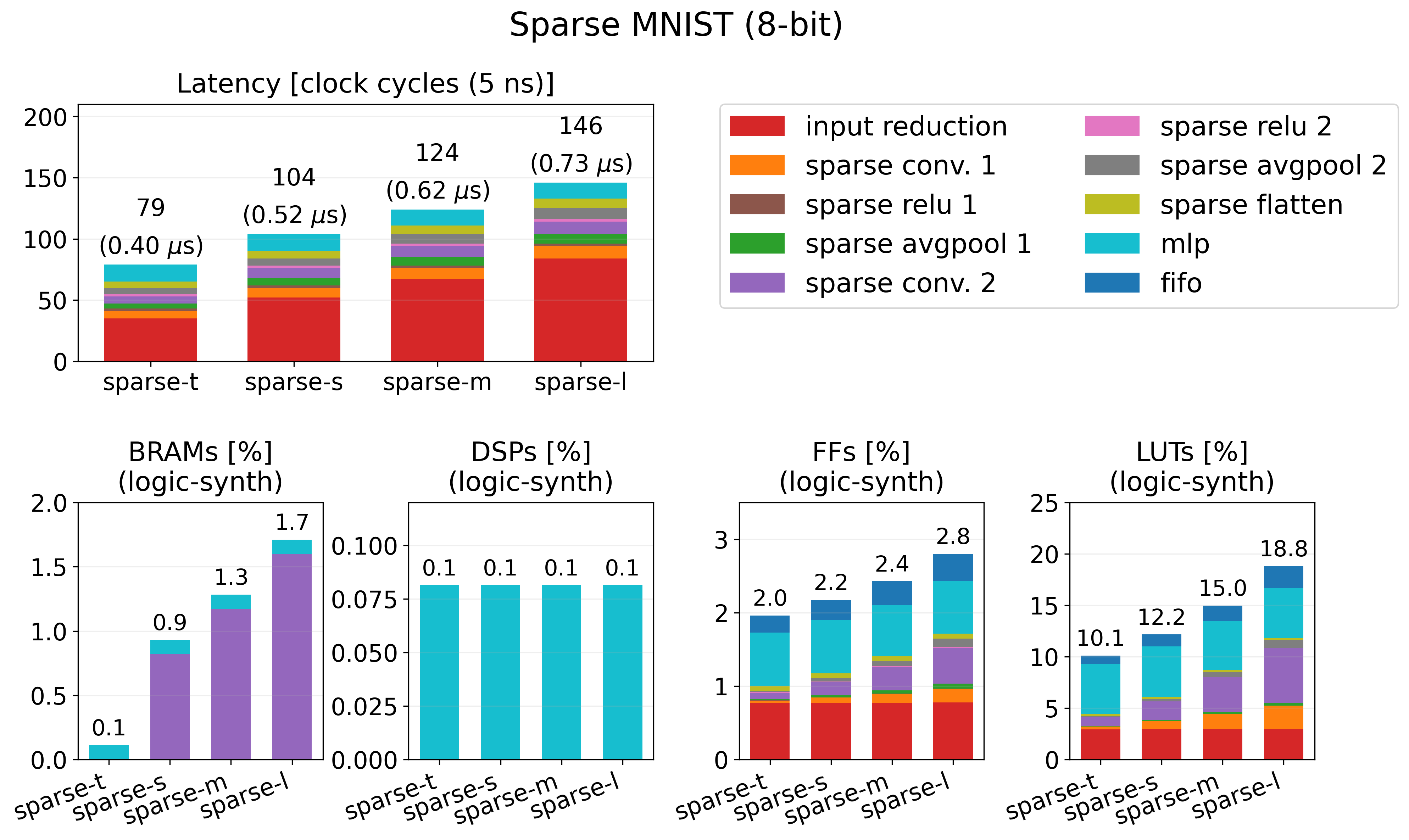}\\
    \includegraphics[width=0.95\textwidth]{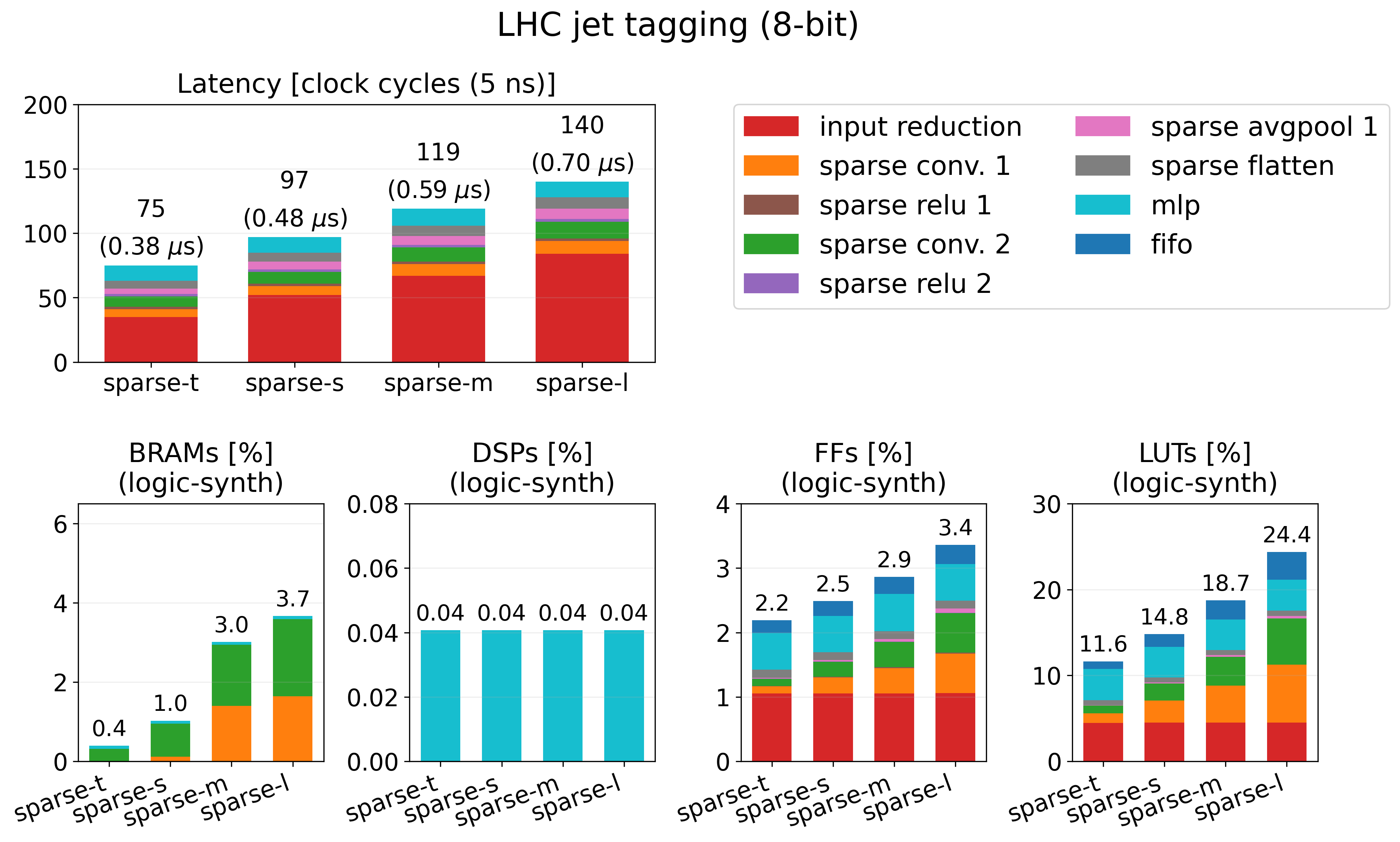}
    \caption{Per-layer resource breakdown for 8-bit sparse CNNs for MNIST (top) and jet tagging (bottom).}
    \label{fig:breakdown8b_mnist_jet}
\end{figure}

For MNIST, we use the 70k examples split into 50k/10k/10k for train/val/test.
For the LHC jet dataset, we use 300k jets split 70/10/20\% for train/val/test sets.
Training settings are the same as in Sec.~\ref{sec:exp-eval-setup} except the loss is categorical cross-entropy.
The base architectures are shown in Fig.~\ref{fig:models_mnist_jet} and the model performance in Fig.~\ref{fig:performance_mnist_jet}.
Tab.~\ref{tab:results_mnist_jet} summarizes latency, speedup, and performance relative to the standard CNNs.
Fig.~\ref{fig:bar8b_mnist_jet} and Fig.~\ref{fig:breakdown8b_mnist_jet} show the resource usage and per-layer breakdowns.
The sparse-large model at 8-bit achieves a $\times 22$ speedup on MNIST and a $\times 60$ speedup on jet tagging over the 8-bit standard CNNs, with accuracy drops within a few percent.

\begin{table*}[!t]
\centering
\caption{Comparison of FPGA latency and model performance between standard and sparse CNNs. The sparse-\{t,s,m,l\} variants correspond to $N_{\text{active}}^{\text{max}}=\{8,12,16,20\}$. Latency and initiation interval (II) are measured in clock cycles (cc) with 5 ns clock period. Speedups for sparse CNNs are measured with respect to the standard CNN at the same total bit-width.}
\label{tab:results_mnist_jet}

\begin{subtable}
\centering
\small
\scalebox{0.605}{
\begin{tabular}{lcccccccccccccc}
    \multicolumn{15}{c}{\bfseries Sparse MNIST, 10 classes, $48\times48=2304$ input pixels} \\
    \toprule
    \textbf{CNN model} & \textbf{Latency} [cc] & \textbf{Speedup} & \textbf{II} [cc] & \textbf{Acc.} [\%]
    & \textbf{AUC(0)} & \textbf{AUC(1)} & \textbf{AUC(2)} & \textbf{AUC(3)} & \textbf{AUC(4)}
    & \textbf{AUC(5)} & \textbf{AUC(6)} & \textbf{AUC(7)} & \textbf{AUC(8)} & \textbf{AUC(9)} \\
    \midrule
    Sparse-t 8b & 79 (0.395 $\mu$s) & $\times 41$ & 35 (0.175 $\mu$s) & 74.3 & 0.9631 & 0.9972 & 0.9562 & 0.9494 & 0.9822 & 0.9666 & 0.9893 & 0.9583 & 0.9038 & 0.9581 \\
    Sparse-t 16b & 81 (0.405 $\mu$s) & $\times 40$ & 35 (0.175 $\mu$s) & 83.1 & 0.9834 & 0.9990 & 0.9788 & 0.9735 & 0.9941 & 0.9872 & 0.9948 & 0.9817 & 0.9476 & 0.9770 \\ \hline
    
    Sparse-s 8b & 104 (0.52 $\mu$s) & $\times 31$ & 52 (0.26 $\mu$s) & 86.7 & 0.9967 & 0.9981 & 0.9886 & 0.9839 & 0.9906 & 0.9920 & 0.9959 & 0.9859 & 0.9762 & 0.9813 \\
    Sparse-s 16b & 106 (0.53 $\mu$s) & $\times 30$ & 52 (0.26 $\mu$s) & 91.8 & 0.9987 & 0.9994 & 0.9956 & 0.9922 & 0.9969 & 0.9971 & 0.9982 & 0.9943 & 0.9893 & 0.9908 \\ \hline
    
    Sparse-m 8b & 124 (0.62 $\mu$s) & $\times 26$ & 67 (0.335 $\mu$s) & 90.7 & 0.9984 & 0.9987 & 0.9956 & 0.9914 & 0.9942 & 0.9946 & 0.9980 & 0.9919 & 0.9892 & 0.9882 \\
    Sparse-m 16b & 127 (0.635 $\mu$s) & $\times 25$ & 67 (0.335 $\mu$s) & 94.6 & 0.9994 & 0.9996 & 0.9984 & 0.9963 & 0.9978 & 0.9983 & 0.9991 & 0.9975 & 0.9966 & 0.9940 \\ \hline
    
    Sparse-l 8b & 146 (0.73 $\mu$s) & $\times 22$ & 84 (0.42 $\mu$s) & 91.2 & 0.9988 & 0.9982 & 0.9957 & 0.9925 & 0.9942 & 0.9955 & 0.9984 & 0.9929 & 0.9906 & 0.9899 \\
    Sparse-l 16b & 150 (0.75 $\mu$s) & $\times 22$ & 84 (0.42 $\mu$s) & 95.0 & 0.9991 & 0.9997 & 0.9989 & 0.9969 & 0.9987 & 0.9979 & 0.9991 & 0.9971 & 0.9974 & 0.9952 \\ \hline\hline
    
    Standard 8b & 3232 (16.16 $\mu$s) & $-$ & 3124 (15.62 $\mu$s) & 94.3 & 0.9991 & 0.9997 & 0.9982 & 0.9962 & 0.9980 & 0.9977 & 0.9985 & 0.9968 & 0.9958 & 0.9961 \\
    Standard 16b & 3232 (16.16 $\mu$s) & $-$ & 3124 (15.62 $\mu$s) & 95.4 & 0.9996 & 0.9997 & 0.9989 & 0.9978 & 0.9988 & 0.9981 & 0.9989 & 0.9985 & 0.9979 & 0.9976 \\
    \bottomrule
\end{tabular}
}
\end{subtable}

\vspace{0.6em}

\begin{subtable}
\centering
\small
\scalebox{0.605}{
\begin{tabular}{lccccccccc}
    \multicolumn{10}{c}{\bfseries LHC jet tagging, 5 classes, $56\times56=3136$ input pixels} \\
    \toprule
    \textbf{CNN model} & \textbf{Latency [cc]} & \textbf{Speedup} & \textbf{II [cc]} & \textbf{Acc. [\%]}
    & \textbf{AUC(g)} & \textbf{AUC(q)} & \textbf{AUC(W)} & \textbf{AUC(Z)} & \textbf{AUC(t)} \\
    \midrule
    Sparse-t 8b & 75 (0.375 $\mu$s) & $\times 112$ & 35 (0.175 $\mu$s) & 53.0 & 0.7853 & 0.8443 & 0.8122 & 0.7858 & 0.8787 \\
    Sparse-t 16b & 78 (0.39 $\mu$s) & $\times 108$ & 35 (0.175 $\mu$s) & 55.8 & 0.7976 & 0.8494 & 0.8388 & 0.8117 & 0.8822 \\ \hline
    
    Sparse-s 8b & 97 (0.485 $\mu$s) & $\times 86$ & 52 (0.26 $\mu$s) & 59.0 & 0.8195 & 0.8626 & 0.8563 & 0.8328 & 0.8996 \\
    Sparse-s 16b & 101 (0.505 $\mu$s) & $\times 83$ & 52 (0.26 $\mu$s) & 61.5 & 0.8340 & 0.8685 & 0.8779 & 0.8529 & 0.9053 \\ \hline
    
    Sparse-m 8b & 119 (0.595 $\mu$s) & $\times 71$ & 67 (0.335 $\mu$s) & 61.8 & 0.8418 & 0.8722 & 0.8866 & 0.8619 & 0.9130 \\
    Sparse-m 16b & 120 (0.6 $\mu$s) & $\times 70$ & 67 (0.335 $\mu$s) & 65.5 & 0.8580 & 0.8786 & 0.9029 & 0.8856 & 0.9238 \\ \hline
    
    Sparse-l 8b & 140 (0.7 $\mu$s) & $\times 60$ & 84 (0.42 $\mu$s) & 65.7 & 0.8617 & 0.8805 & 0.9077 & 0.8856 & 0.9238 \\
    Sparse-l 16b & 143 (0.715 $\mu$s) & $\times 59$ & 84 (0.42 $\mu$s) & 68.1 & 0.8710 & 0.8852 & 0.9208 & 0.9032 & 0.9328 \\ \hline\hline
    
    Standard 8b & 8390 (41.95 $\mu$s) & $-$ & 4340 (21.7 $\mu$s) & 68.3 & 0.8468 & 0.8743 & 0.9324 & 0.9178 & 0.9374 \\
    Standard 16b & 8390 (41.95 $\mu$s) & $-$ & 4340 (21.7 $\mu$s) & 72.2 & 0.8874 & 0.8893 & 0.9463 & 0.9361 & 0.9426 \\
    \bottomrule
\end{tabular}
}
\end{subtable}

\end{table*}

\subsection{Synthesis results for 16-bit models}
\label{sec:apd-results}

%\begin{figure}[!t]
%    \centering
%    \includegraphics[width=0.22\textwidth]{figs/mnist_conv_in.png}\\
%    \includegraphics[width=0.6\textwidth]{figs/mnist_conv_out.png}
%    \caption{Sparse convolution on an example image preserving the sparsity pattern. Top: input image with one channel. Bottom: output image with three channels.}
%    \label{fig:mnist_conv}
%\end{figure}

%\begin{figure}[!t]
%    \centering
%    \includegraphics[width=0.25\textwidth]{figs/jet_conv_in.png}\\
%    \includegraphics[width=0.7\textwidth]{figs/jet_conv_out.png}
%    \caption{f}
%    \label{fig:jet_conv}
%\end{figure}

\begin{figure}[!t]
    \centering
    \includegraphics[width=0.75\textwidth]{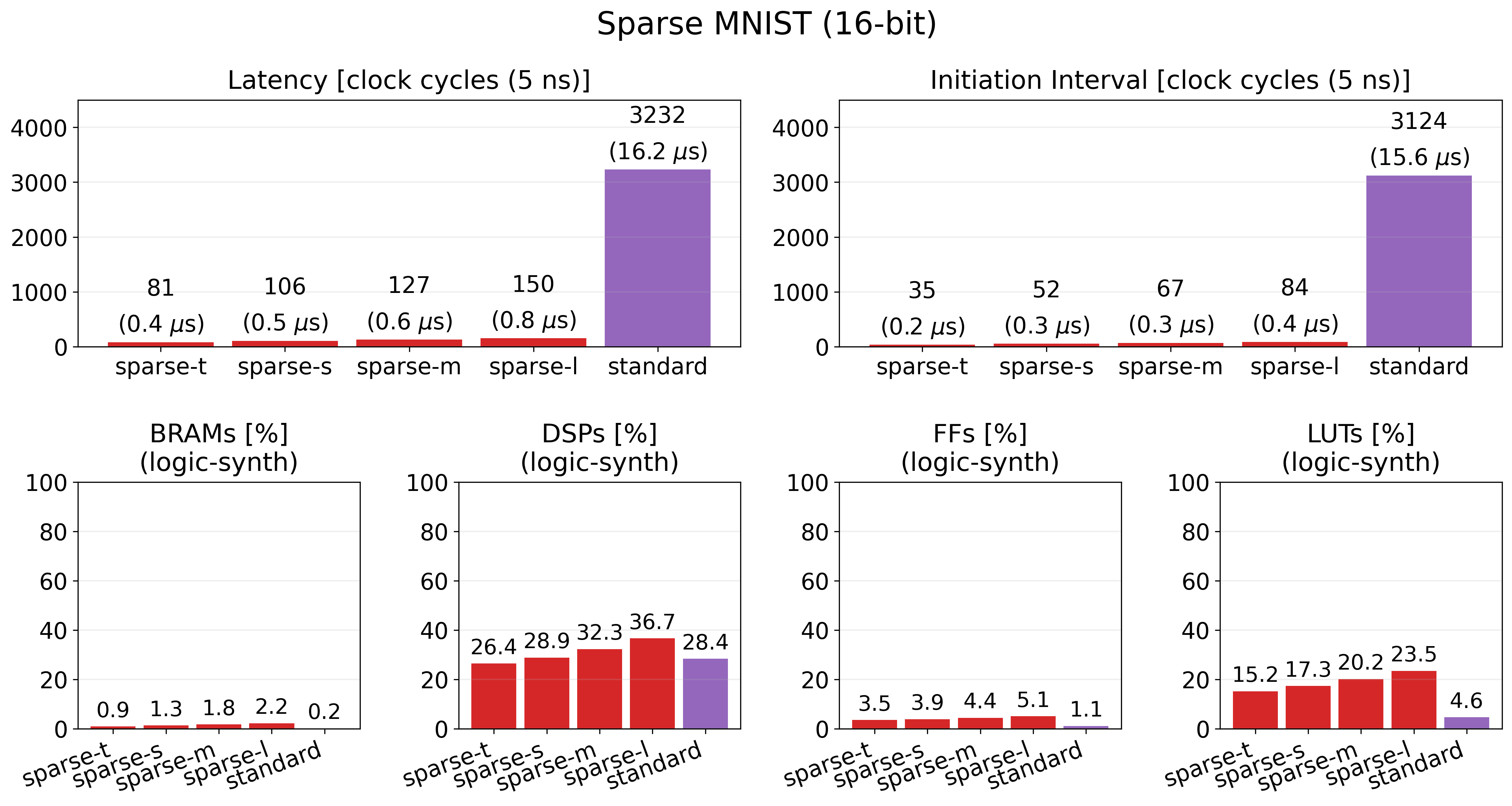}\\
    \includegraphics[width=0.75\textwidth]{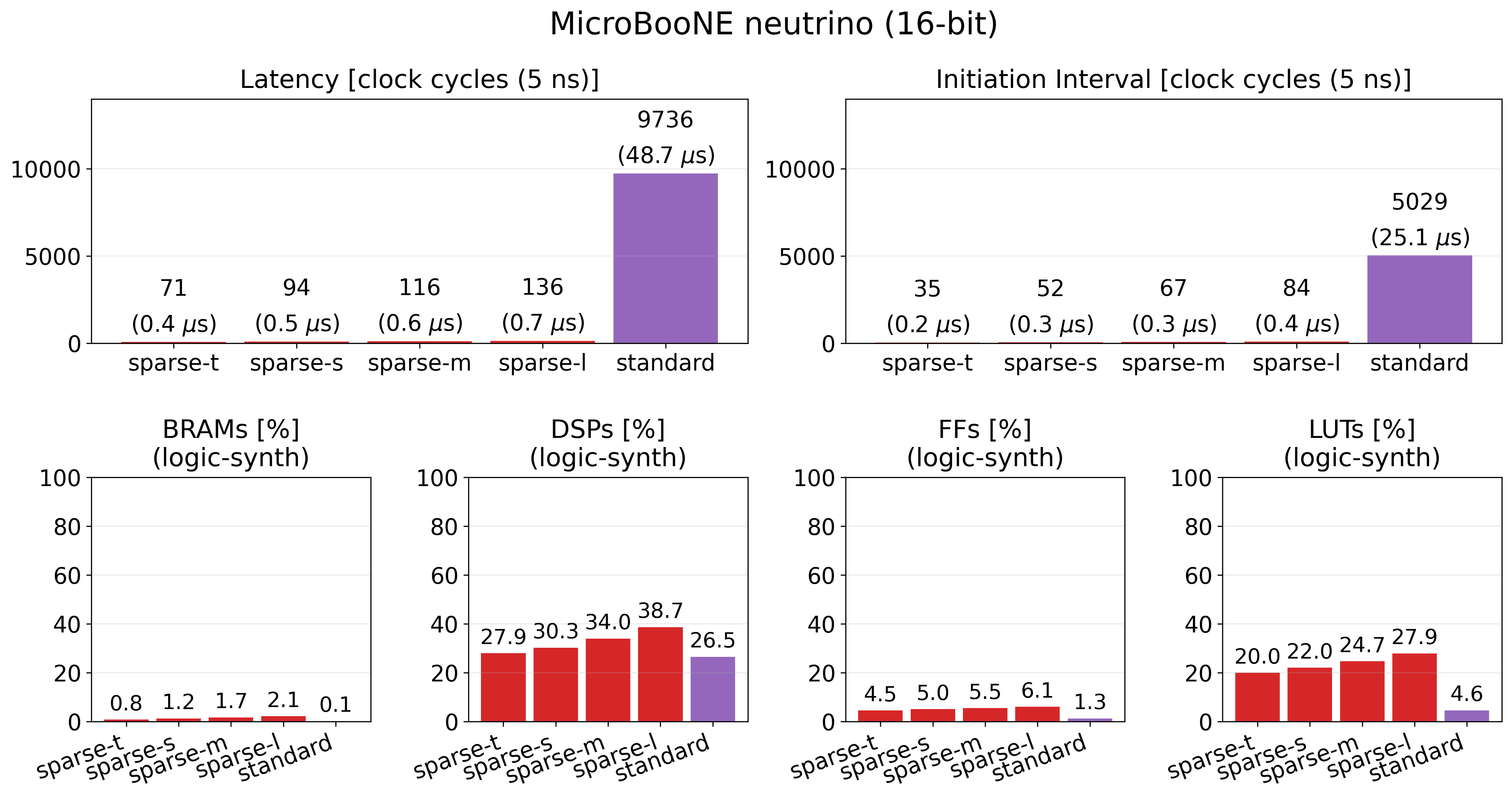}\\
    \includegraphics[width=0.75\textwidth]{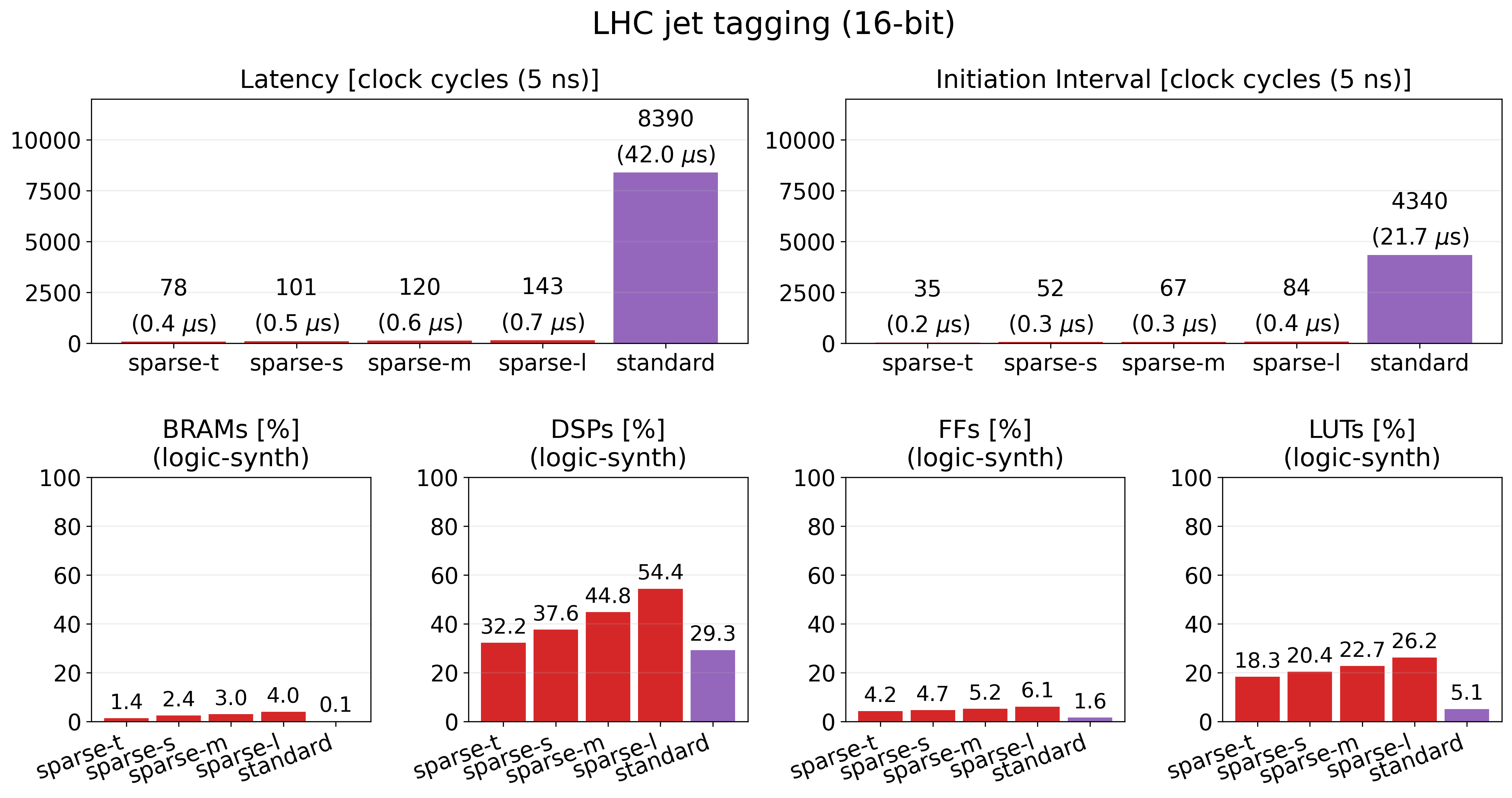}
    \caption{Synthesis results of standard (purple) and sparse (red) CNNs at 16-bit for MNIST (top), neutrino (middle), and jet tagging (bottom). Latency and II are measured in clock cycles (cc) with 5 ns clock period. Resource utilization is obtained after the logic synthesis step.}
    \label{fig:bar16b}
\end{figure}

\begin{figure}[!t]
    \centering
    \includegraphics[width=0.7\textwidth]{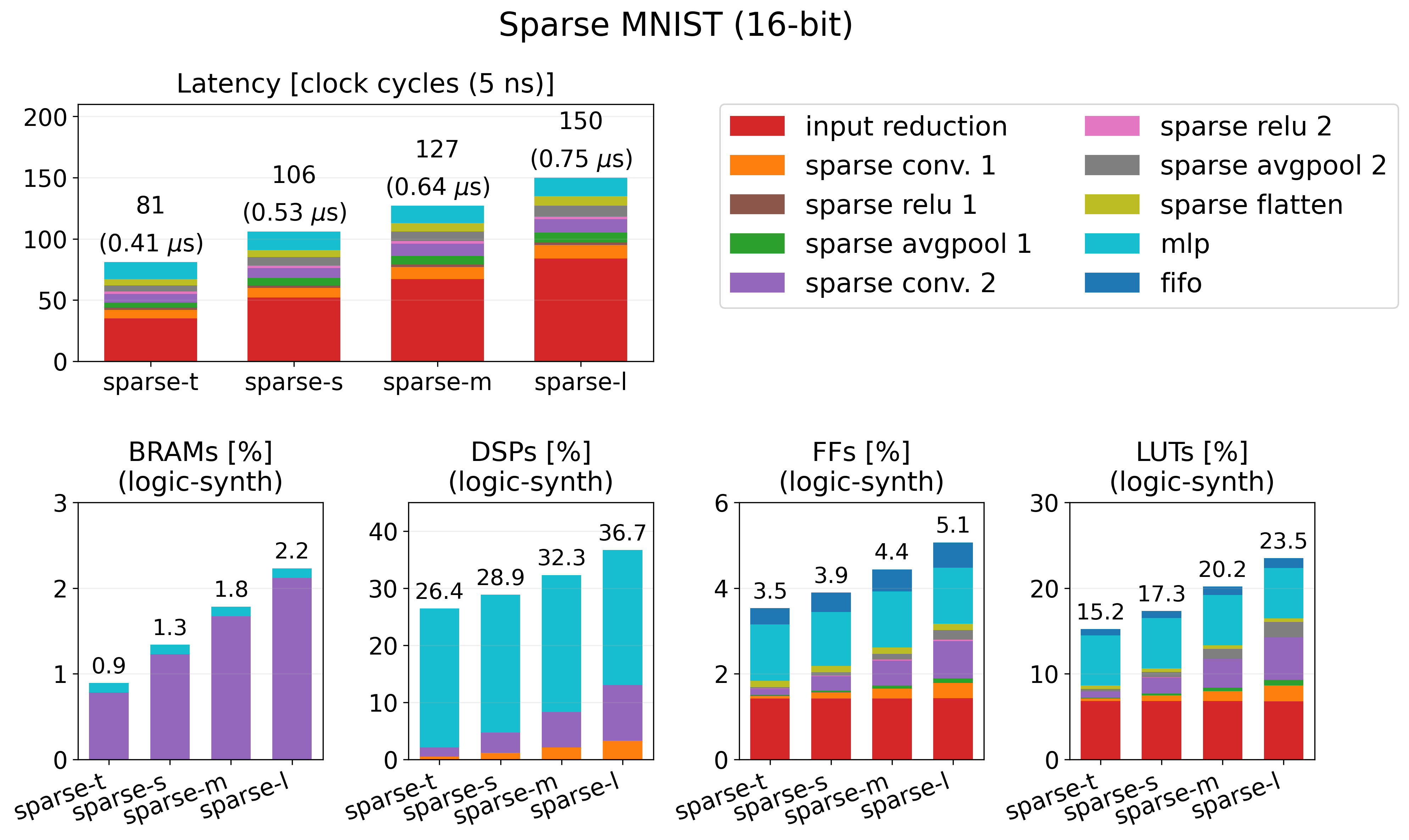}\\
    \includegraphics[width=0.7\textwidth]{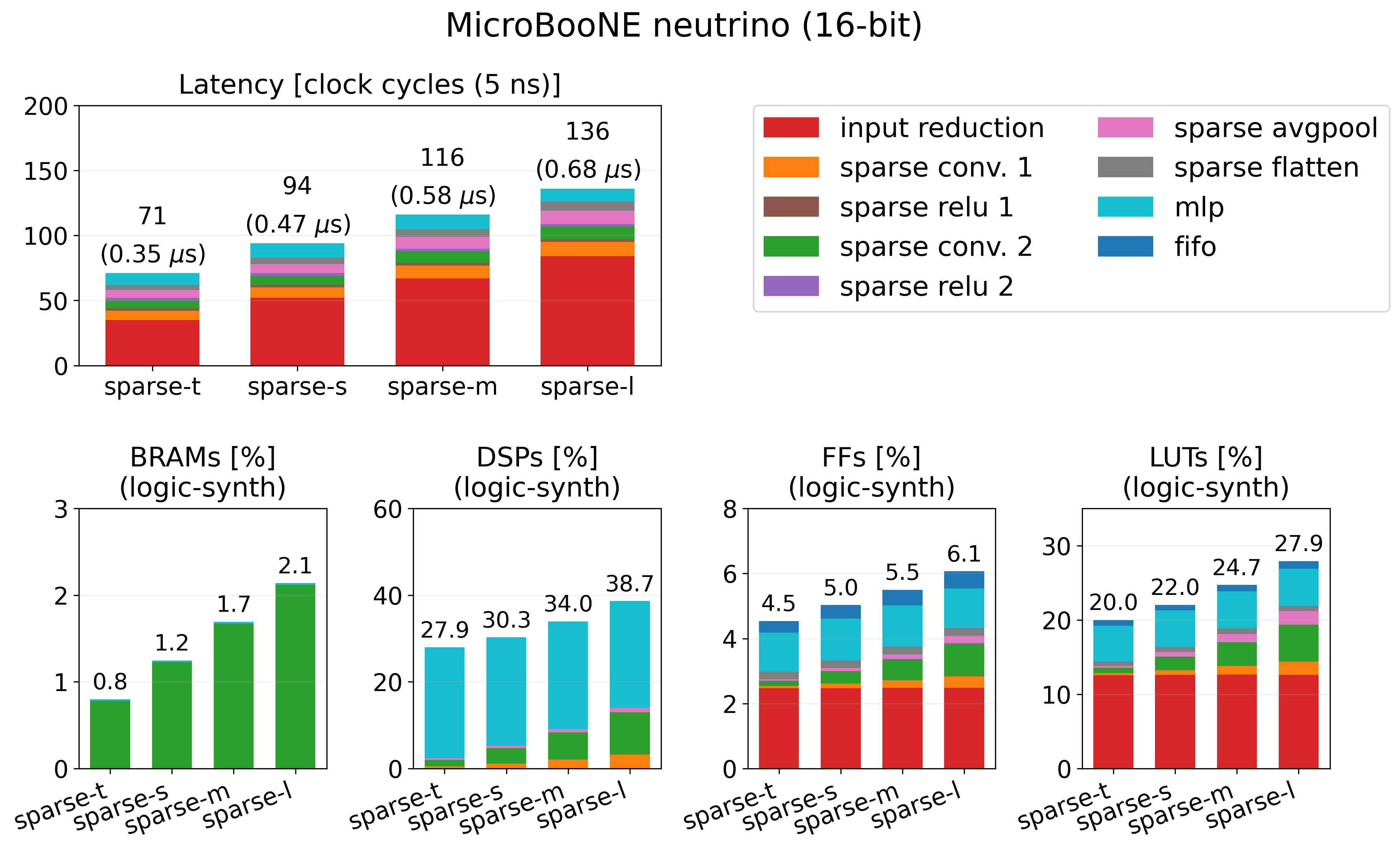}\\
    \includegraphics[width=0.7\textwidth]{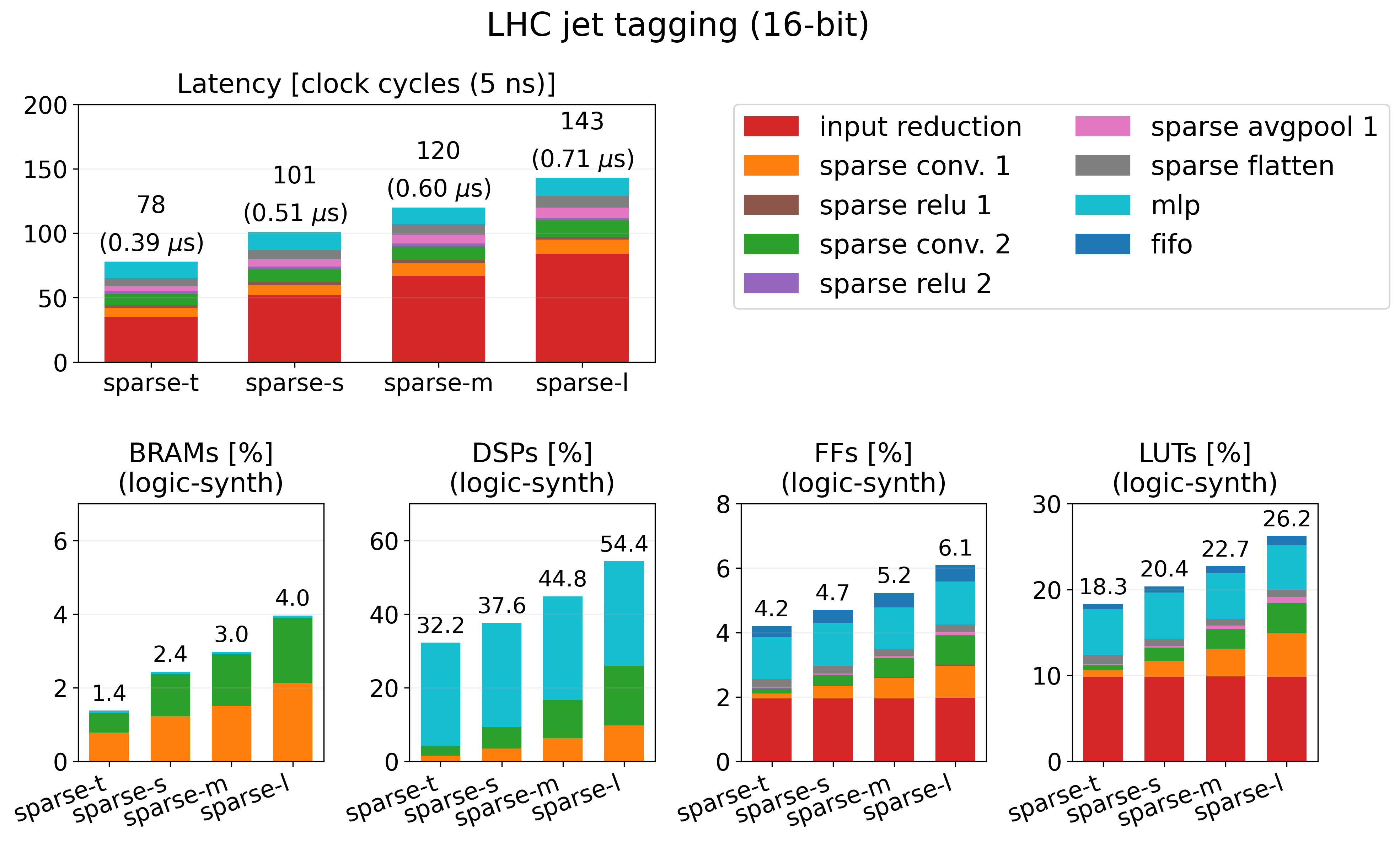}
    \caption{Per-layer resource breakdown for 16-bit sparse CNNs on MNIST (top), neutrino (middle), and jet tagging (bottom). Latency is measured in clock cycles (cc) with 5 ns clock period. Resource utilization is obtained after the logic synthesis step.}
    \label{fig:breakdown16b}
\end{figure}

Fig.~\ref{fig:bar16b} and Fig.~\ref{fig:breakdown16b} summarize results of the 16-bit models.
Latencies are similar to those of the 8-bit models because of the same loop structures in the architectural design.
DSP usage increases primarily because arithmetic operations such as multiplications are preferentially mapped to DSPs at higher bit-width by the HLS tool.

\end{document}